\documentclass[12pt,a4paper,twoside,final]{mathias}

\usepackage[dvips]{graphicx}

\usepackage{latexsym,isolatin1,deluxetable}

\newcounter{numrom}
{\begin{list}{\Roman{numrom}.}
 {\usecounter{numrom}}}{\end{list}}

\renewcommand{\arraystretch}{1}

\begin{document}

\newcommand{\mathbold}[1]{\mbox{\protect\boldmath $#1$}}

\def\rme{{\rm e}}
\def\rmi{{\rm i}}
\def\rmx{{\rm x}}

\def\lra{\leftrightarrow}

\def\be{\begin{equation}}
\def\ee{\end{equation}}
\def\bea{\begin{eqnarray}}
\def\eea{\end{eqnarray}}

\newcommand{\abs}[1]{\left | #1 \right |}

\def\re{{\rm e}}
\def\elpo{\re^\pm}
\def\rp{{\rm p}}
\def\rn{{\rm n}}
\def\rN{{\rm N}}
\def\nue{\nu_{\rm e}}
\def\numu{\nu_\mu}
\def\Ye{Y_{\rm e}}
\def\GF{G_{\rm F}}

\newcommand{\lsim}{\; ^< \!\!\!\! _\sim \;}
\newcommand{\gsim}{\; ^> \!\!\!\! _\sim \;}

\def\dmsq{\Delta m^2}
\def\msqa{\Delta m^2_{\rm atm}}
\def\msqs{\Delta m^2_{\odot}}
\def\thsol{\theta_\odot}
\def\ebar{{\bar{\re}}}


\newfont{\headfo}{cmssbx10 scaled 3000}
\pagestyle{empty}


\begin{center}

{\sc  Dissertation}

\vspace{1.3cm}

{\headfo Supernova Neutrino Spectra \\ \vspace{3mm} and Applications
to \\ \vspace{3mm} Flavor Oscillations}\\ 

\vspace{2.2cm}
by

\vspace{0.3cm}
{\LARGE \sc Mathias Thorsten Keil}

\vfill

\columnwidth=5cm
\plotone{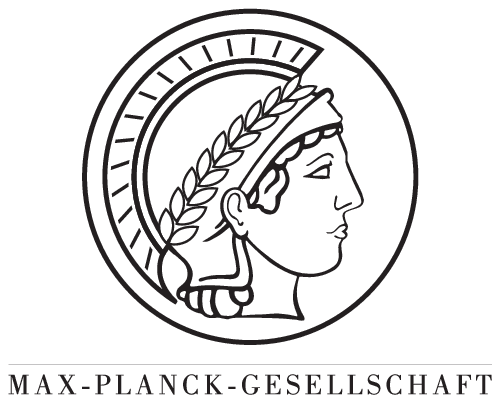}

\end{center}

\cleardoublepage
\begin{center}

Institut f\"ur Theoretische Physik T30, Univ.-Prof.~Dr.~Manfred Lindner

\vspace{1.3cm}
{\headfo Supernova Neutrino Spectra \\ \vspace{3mm} and Applications to \\ \vspace{3mm} Flavor Oscillations}\\
\vspace{3.2cm}

{\LARGE\sf\sc Mathias Thorsten Keil}
\vfill

{\raggedright
Vollst\"andiger Abdruck der von der Fakult\"at f\"ur Physik der Technischen Universit\"at M\"unchen zur Erlangung des akademischen Grades eines
{\bf \sf Doktors der Naturwissenschaften (Dr. rer. nat.)}
genehmigten Dissertation.

\begin{center}
\begin{tabular}{lll}
Vorsitzender:              &    & Univ.-Prof.~Dr.~F.~v.~Feilitzsch\\[0.3cm]
Pr\"ufer der Dissertation: & 1. & Univ.-Prof.~Dr.~M.~Lindner\\[0.1cm]
		  	   & 2. & Univ.-Prof.~Dr.~M.~Drees\\
\end{tabular}
\end{center}

\noindent
Die Dissertation wurde am 27.05.2003 bei der Technischen
Universit\"at M\"unchen eingereicht und durch die Fakult\"at f\"ur
Physik am 25.06.2003 angenommen.
}
\end{center} 
\cleardoublepage

\pagenumbering{roman}
\pagestyle{plain}

\vspace*{-1.5cm}
\noindent
\hfill\parbox[h]{4cm}{
\huge \bfseries Summary 

}\hfill
\vspace{.5cm}

\noindent
We study the flavor-dependent neutrino spectra formation in the core
of a supernova (SN) by means of Monte Carlo simulations.  Several
neutrino detectors around the world are able to detect a
high-statistics signal from a galactic SN.  From such a signal one may
extract information that severely constrains the parameter space for
neutrino oscillations.  Therefore, reliable predictions for
flavor-dependent fluxes and spectra are urgently needed.

In all hydrodynamic simulations the treatment of $\nu_{\mu,\tau}$ and
$\bar\nu_{\mu,\tau}$ is rather schematic, with the exception of the
most recent Garching-Group simulation, where the interaction processes
were updated partly based on our results.  The interactions commonly
included in traditional simulations are iso-energetic scattering on
nucleons $\rN\nu\to\nu\rN$, scattering on electrons and positrons
$\re^\pm\nu\to\nu\re^\pm$, and electron-positron pair annihilation
$\re^+\re^-\to\nu_{\mu,\tau}\bar\nu_{\mu,\tau}$.  In our Monte Carlo 
simulations we vary the static stellar background models in addition
to systematically switching on and off our set of interaction
processes, i.e., recoil and weak magnetism in $\rN\nu\to\nu\rN$,
scattering on $\re^\pm$ and $\nue$/$\bar\nue$, $\re^+\re^-$ and
$\nue\bar\nue$ annihilation into $\nu_{\mu,\tau}\bar\nu_{\mu,\tau}$
pairs, and neutrino bremsstrahlung off nucleons
$\rN\rN\to\rN\rN\nu\bar\nu$.  As $\nu_{\mu,\tau}$ and
$\bar\nu_{\mu,\tau}$ sources, $\rN\rN\to\rN\rN\nu\bar\nu$ and
$\nue\bar\nue\to\nu_{\mu,\tau}\bar\nu_{\mu,\tau}$ dominate.  The
latter process has never been studied before in the context of SNe and
turns out to be always more important than the traditional
$\re^+\re^-$ annihilation process by a factor of 2--3.  For energy
transfer, the most important reactions are $\rN\nu\to\nu\rN$ with
recoil, and scattering on $\re^\pm$.  Weak magnetism has a very small
effect and scattering on $\nue$ and $\bar\nue$ is negligible.

The charged current reactions $\re^-\rp\to\nue\rn$ and
$\re^+\rn\to\bar\nue\rp$ dominate for $\nue$ and $\bar\nue$.  In
comparing our numerical results for all flavors we find the standard
hierarchy of mean energies $\langle\epsilon_{\nue}\rangle <
\langle\epsilon_{\bar\nue}\rangle \lsim
\langle\epsilon_{\nu_{\mu,\tau}}\rangle$, with, however, very similar
values for $\nu_{\mu,\tau}$ and $\bar\nue$.  The luminosities of
$\nu_{\mu,\tau}$ and $\bar\nue$ can differ by up to a factor of 2 from
$L_{\bar\nue}\approx L_{\nue}$.  The Garching Group obtains similar
results from their self-consistent simulation with the full set of
interactions.  These results are almost orthogonal to the previous
standard picture of exactly equal luminosities of all flavors and
differences in mean energies of up to a factor of 2.

Existing concepts for identifying oscillation effects in a SN neutrino
signal need to be revised.  We present two methods for detecting the
earth-matter effect that are rather independent of predictions from SN
simulations.  An earth-induced flux difference can be measured by the
future IceCube detector in Antarctica and a co-detector like Super- or
Hyper-Kamiokande.  At a single detector with high energy
resolution the Fourier transform of the inverse-energy spectrum can
reveal the modulations of the spectrum.  Both methods are sensitive to
the small mixing angle $\theta_{13}$ and the neutrino mass hierarchy.

\cleardoublepage

\vspace*{\fill}
\begin{center}
\hfill F\"ur meine Eltern
\end{center}
\vspace*{8cm}
\vspace*{\fill}

\cleardoublepage

\begin{center}
{\huge \bfseries Acknowledgments}
\end{center}

\noindent
Many people contributed to this dissertation with their advice,
support, patience, and kindness.  I express my gratitude to everyone
who helped making this work come true.

Foremost I am indebted to my advisor, Georg Raffelt, not only for
proposing the project and thereby getting me into Astroparticle
Physics, but even more for his great support and advice in so many
ways.  I benefited very much from our close and very inspiring
collaboration.  Also for the proofreading and uncountable valuable
suggestions on the manuscript I am very grateful.

The numerical part of this work was only possible with the code that
Hans-Thomas Janka had developed and kindly introduced me to.  I
enjoyed the collaboration with Thomas and appreciate his patience with
my questions and his exhaustive answers on all kinds of code- and
supernova-related questions.  Thomas and his Garching Supernova Group,
i.e., Robert Buras and Markus Rampp, contributed very valuable results
from their hydrodynamic simulations.  I also thank Robert for our many
long discussions and his valuable suggestions on parts of the
manuscript.

With Amol Dighe I had a wonderful collaboration on the
neutrino-oscillation part.  But also for all other questions and 
discussions I found Amol's door wide open.  In addition, I thank Amol
for improving parts of the manuscript by his proofreading and
suggestions.  Also for  discussions with Michael Kachelrie\ss, Sergio
Pastor, Dimitry Semikoz, and Ricard Tom\'as I am very grateful.

As important as the support at the institute were all those friends
and my family who supported and encouraged me throughout all the years,
even when I was far away.  For their love and support in any possible
way I owe my deepest thanks to my parents as well as my brother and
sister.  Christoph even fitted some proofreading into his busy
conference schedule, thank you.

For your love, support, and encouragement, your patience, your smiles,
and everything you did for me, words cannot express my gratitude and
love for you, Meike.

\vspace*{3cm}
\vspace*{\fill}

\newpage

\vspace*{\fill}

\noindent
Parts of this dissertation have already been published as:
\begin{itemize}

\item M.~T.~Keil, G.~G.~Raffelt, and H.~T.~Janka,\\
``Monte Carlo study of supernova neutrino spectra formation,''\\
Astrophys.\ J.\  in press,\\
astro-ph/0208035.

\item R.~Buras, H.~T.~Janka, M.~T.~Keil, G.~G.~Raffelt, and M.~Rampp,\\
``Electron-neutrino pair annihilation: A new source for muon and tau
neutrinos in supernovae,''\\ 
Astrophys.\ J.\  {\bf 587} (2003) 320 [astro-ph/0205006].

\item  M.~T.~Keil, G.~G.~Raffelt, and H.~T.~Janka,\\
``Supernova Neutrino Spectra Formation,''\\
Nucl.\ Phys.\ B Proc.\ Suppl.\ {\bf 118} (2003) 506.

\item G.~G.~Raffelt, M.~T.~Keil, R.~Buras, H.~T.~Janka, and M.~Rampp,\\
``Supernova neutrinos: Flavor-dependent fluxes and spectra,''\\
astro-ph/0303226.

\item A.~S.~Dighe, M.~T.~Keil, and G.~G.~Raffelt,\\
``Detecting the neutrino mass hierarchy with a supernova at
IceCube,''\\ 
JCAP, in press (2003), hep-ph/0303210.

\item A.~S.~Dighe, M.~T.~Keil, and G.~G.~Raffelt,\\
``Identifying earth matter effects on supernova neutrinos at a single
detector,''\\ 
hep-ph/0304150.

\end{itemize}

\newpage

\tableofcontents
\cleardoublepage


\pagestyle{headings}
\pagenumbering{arabic}

\cleardoublepage

\chapter{Introduction}

The physics of neutrino oscillations has received tremendous attention
during the past years and today oscillations are thought to be
established.  A large number of experiments has contributed to our present
knowledge of the mixing schemes and parameters.  For several of the mixing
parameters the allowed ranges are known.  Within the next few
decades the magnitude of the small mixing angle $\theta_{13}$ and the
question whether the neutrino mass hierarchy is normal or inverted will
remain open and can be settled only by future precision measurements
at dedicated long-baseline oscillation experiments (Barger et~al.\
2001, Cervera et~al.\ 2000, Freund, Huber, \& Lindner 2001) or the
observation of a future galactic supernova (SN).

Along with the interest in neutrino oscillations developed the branch
of neutrino astronomy.  For the early experiments Raymond Davis
Jr.\ and Masatoshi Koshiba received last year's Physics Nobel Prize
``for pioneering contributions to astrophysics, in particular for the
detection of cosmic neutrinos'' (Nobel e-Museum, 2002).  Current and
planned experiments will be able to record some 10,000--100,000
neutrinos from a future galactic SN and therefore would provide
valuable information on SN physics and neutrino properties (Barger,
Marfatia, \& Wood 2001, Minakata et~al.~2002).

There exists a large number of publications addressing the
question, what the neutrino signal of a galactic SN will tell us about
neutrino oscillations.  With the measurement of such a neutrino
signal it would be possible to differentiate between existing
oscillation scenarios (Chiu \& Kuo 2000, Dighe \& Smirnov 2000, Dutta
et~al.\ 2000, Fuller, Haxton, \& McLaughlin 1999, Lunardini \& Smirnov
2001b, 2003, Minakata \& Nunokawa 2001, Takahashi \& Sato 2002).
At the same time, an understanding of neutrino oscillations is crucial
for interpreting  a recorded signal.

In 1987 neutrinos from a SN were detected for the first and only time
(Bionta et al.~1987, Hirata et al.~1987).  This signal from SN1987A
in the Large Magellanic Cloud was analyzed taking neutrino
oscillations into account (Jegerlehner, Neubig, \&  Raffelt 1996,
Kachelriess et~al.\ 2002, Lunardini \& Smirnov 2001a, Smirnov,
Spergel, \& Bahcall 1994).  However, the neutrino detectors that were
operational at that time measured only about 20 events connected to
SN1987A.  With such low statistics, determining SN and neutrino
parameters was difficult, but some limits could be obtained.

For the analysis of a SN neutrino signal it would be helpful to have a
firm prediction for the neutrino spectra emitted by the core of the
SN.  Oscillation effects in a neutrino signal depend on the spectral
differences between the various flavors.  Such spectral differences
are obtained by neutrino-transport simulations coupled to self
consistent hydrodynamic calculations.  In order to get reliable
neutrino spectra the way neutrinos are transported in these simulations
plays a crucial role.  In the past, the transport of $\numu$ was very
schematic (here and in the following $\numu$ stands for either muon or
tau neutrinos, or anti-neutrinos, unless stated differently).  Spectra
obtained for $\numu$ in these simulations therefore inherit large 
uncertainties.  However, these uncertainties barely affect the
dynamics of the explosion.  From the SN model builders point of view,
details of $\numu$ spectra formation do therefore  not play a crucial
role.  The greatest challenge for numerical models still is to
understand how SNe explode.  The most elaborate simulations agree
rather well on the early stages of the SN mechanism, but do not show
explosions.

According to the commonly accepted ``delayed-explosion scenario'' a 
massive star with a mass greater than about $8 M_\odot$ eventually
becomes a core-collapse SN.  At the end of its life, such a star has
accumulated an iron core of about 1.5 $M_\odot$, close to the
Chandrasekhar limit.  Once the core mass is greater than the
stability limit it collapses within some hundred milliseconds to a
proto-neutron star.  Due to this collapse a large amount of
gravitational binding energy is liberated, but particles cannot escape
the dense medium.  A shock wave that formed at the edge of the
proto-neutron star travels outwards and plows through the still
infalling outer layers of the iron core.  The disintegration of these
infalling iron nuclei consumes the kinetic energy of the shock wave
and finally causes it to stall after a few 100 ms.  Neutrino emission
drains energy from the proto-neutron star and at the same time heats
up the region behind the stalled shock wave.  This energy transfer
causes the shock wave to regain velocity and finally blow off the
envelope of the star.

In a crude approximation the proto-neutron star is a blackbody source
for neutrinos of all flavors.  The flavor dependent differences, that
are most important for oscillations, arise due to the flavor
dependence of the neutrino-matter interactions.  For the case of
electron neutrinos ($\nue$) and electron anti-neutrinos ($\bar\nue$)
the dominant reactions are the charged-current interactions with
nucleons $\re^-\rp\leftrightarrow\nue\rn$ and
$\re^+\rn\leftrightarrow\bar\nue\rp$.  Since there are more neutrons
than protons in a proto-neutron star, the interactions are more
efficient for $\nue$ and keep $\nue$ in local thermal equilibrium
(LTE) up to larger radii than $\bar\nue$.  Therefore the emerging
$\bar\nue$ are hotter than the $\nue$.  The other flavors undergo
various types of neutral-current interactions that each are of
comparable importance.  The detailed interplay of these reactions in
the context of SNe has been studied for the first time in the present
work.

Traditionally, the $\numu$-matter interactions that are implemented
in numerical simulations are iso-energetic scattering with nucleons,
scattering on electrons and positrons, and electron-positron pair
annihilation into $\numu\bar\numu$ pairs.  For several years there
have been suggestions on improving these rates. Burrows \& Sawyer
(1998) and Reddy et al.\ (1999) calculated the effect of many-body
correlations that change the neutrino-nucleon interactions.  Weak
magnetism even causes different rates for $\numu$ and $\bar\numu$
(Horowitz \& Li 2000, Horowitz 2002, Vogel \& Beacom 1999).  A
standard simplification in numerical simulations was to take the
nucleon mass to be infinite for neutrino-nucleon scattering.  However,
it has been known for some time that the influence of nucleon recoils
in these scattering reactions was not negligible (Janka et~al.\ 1996,
Raffelt 2001).  An important source term for $\numu$ is nucleon
bremsstrahlung ${\rm NN}\to {\rm NN}\nu_{\mu}\bar\nu_{\mu}$ that even
dominates for low energies (Hannestad \& Raffelt 1998, Suzuki
1991,1993, Thompson, Burrows, \& Horvath 2000).  The traditionally
included source for $\numu$ in hydrodynamic simulations, $\re^+\re^-
\to \nu_{\mu}\bar\nu_{\mu}$, turns out to be rather unimportant.  For
the first time we showed that electron-neutrino pair annihilation
$\nue\bar\nue \to \numu\bar\numu$ is always a factor of 2--3 more
important than this traditional process.

In order to study the relative importance of all possible interaction
processes and their dependence on a reasonable range of stellar
background models we have adapted the Monte Carlo code of
Janka (1987, 1991) and added new microphysics to it.  We go beyond the
work of Janka \& Hillebrandt (1989a,b) in that we include the
bremsstrahlung process, nucleon recoils and weak magnetism, $\nue
\bar\nue$ pair annihilation into $\nu_\mu\bar\nu_\mu$, and scattering
of $\nu_{\mu}$ on $\nue$ and $\bar\nue$. With these extensions we
investigate the neutrino transport systematically for a variety of
medium profiles that are representative for different SN phases.
Raffelt (2001) has recently studied the $\nu_\mu$ spectra-formation
problem with the limitation to nucleonic processes (elastic and
inelastic scattering, recoils, bremsstrahlung), to Maxwell-Boltzmann
statistics for the neutrinos, and plane-parallel geometry.  Our
present study complements this more schematic work by including the
leptonic processes, Fermi-Dirac statistics, and spherical geometry.
In addition we apply our Monte Carlo code to the transport of $\nue$
and $\bar\nue$ and thus are able to compare the flavor-dependent
fluxes and spectra.

With the complete set of neutrino interactions we find a situation
that is almost orthogonal to what has been assumed so far.  The
standard picture concerning SN neutrino fluxes and spectra
was an exact equipartition of the neutrino luminosity among all
flavors and a strong hierarchy of the mean energies of the different
neutrino flavors $\langle\epsilon_{\nu_\mu}\rangle >
\langle\epsilon_{\bar\nue}\rangle > \langle\epsilon_{\nue}\rangle$
with a difference of up to a factor of 2 between
$\langle\epsilon_{\bar\nue}\rangle$ and
$\langle\epsilon_{\nu_\mu}\rangle$.  These assumptions in turn lead to
strong oscillation effects in the predicted signal of a galactic SN.
Our findings suggest that the difference between
$\langle\epsilon_{\bar\nue}\rangle$ and
$\langle\epsilon_{\nu_\mu}\rangle$ is about 10\% or less, but the
luminosities can differ by a factor of 2 in either direction,
depending on the explosion phase.

This new picture is also supported by all hydrodynamic simulations
including Boltzmann solvers for the transport of neutrinos.  If these
simulations use the traditional set of neutrino interactions they find
differences between $\langle\epsilon_{\bar\nue}\rangle$ and
$\langle\epsilon_{\numu}\rangle$ around 20--30\%.  Very recently the
Garching group implemented what we found to be the relevant set of
interactions in their two-dimensional hydrodynamic simulation (Buras
et al.\ 2003b).  Their findings are equivalent to the results of our
Monte Carlo simulation.

For studies of neutrino-oscillation effects in SN neutrino spectra
these findings mean a fundamental change of paradigm, because the main
focus of these studies are the effects on $\bar\nue$ and $\bar\numu$.
Available detectors will record only $\bar\nue$ with high statistics.
The basic idea was to identify an oscillation effect by comparing
characteristics of the observed and the predicted $\bar\nue$
spectrum.  If oscillations were present, one would observe deviations,
because the $\bar\nue$ and $\bar\numu$ spectra would mix.  In addition
to the fact that the predictions have large uncertainties, these
concepts fail once the new findings are taken into account.
Therefore different methods are needed.

According to our findings and the most recent results of self-consistent
SN simulations, a concept for inferring neutrino-oscillation
parameters from the signal of a galactic SN should be based on the
flux differences between $\bar\nue$ and $\bar\numu$ rather than
differences in mean energies.  With a SN rate in our galaxy of a few
per century the concept needs to involve detectors that will be
operational for several decades. 

With such methods one might be able to determine oscillation
parameters that are difficult to obtain by earth-based experiments.
If the currently discussed ``neutrino factories'' and ``super beams''
are built they will be able to determine neutrino parameters in great
detail (Apollonio et al.~2002, Barger et al.~2001, Cervera et
al.~2000, Freund, Huber, \& Lindner, 2001, Huber et al.~2003, Huber,
Lindner, \& Winter 2002).  Today, it is not clear whether such
experiments will become reality.  Both concepts can benefit from one
another, because detectors being built for earth bases experiments can
also serve as SN detectors and results obtained from one concept can
be used to improve the other.

In view of our findings, we developed two methods for
identifying the earth matter effect in a SN neutrino signal that only
depend on robust features of simulation results rather than detailed
predictions.  
The first method involves two neutrino detectors at different sites,
such that for one the neutrino path goes through the earth and the
other sees the signal ``from above.''  The energy deposited
in one detector differs from that in the other if the earth-matter
effect is present, due to different original $\bar\nue$ and
$\bar\numu$ fluxes.  Surprisingly, the future high-energy neutrino
telescope IceCube at the South Pole, that will begin taking data from
2005, will yield the most accurate signal (Dighe, Keil, \& Raffelt
2003a).  Due to its unique location, with a co-detector on the
northern hemisphere, e.g., in Japan, the setup covers most of the
sky.  Alternatively, with a single detector one can identify the
modulations of the energy spectrum that are induced by the
earth-matter effect with the help of a Fourier transform (Dighe, Keil,
\& Raffelt 2003b).  This is a difficult task with the size and energy
resolution of available detectors.  However, with a detector like
Hyper-Kamiokande or a large scintillation detector this is a powerful
procedure.

This dissertation is structured as follows.
In Chapter~\ref{chap:paradigm} we explain the current explosion
paradigm that we 
believe is realized in nature.  We then discuss the neutrino-matter
interactions that are relevant inside the proto-neutron star in
Chapter~\ref{chap:rates}.  We put special emphasis on the
$\nue\bar\nue$ process that has never been studied before in the
context of SNe.  In Chapter~\ref{chap:SetStage} we present the
stellar profiles that we use in our numerical simulations.  We
introduce the concept of a ``thermalization depth'' and apply it to
our stellar profiles in order to estimate the relative importance of
the interaction processes.  For describing the spectra that we obtain
from our calculations we use the characteristic quantities and fits
that we explicate in Chapter~\ref{chap:specchar}.  With all the
necessary tools in hand we then first describe our numerical results
for $\numu$ in Chapter~\ref{chap:importance} and then move on to the
comparison of all flavors in Chapter~\ref{chap:AllFlavComp}, where we
also give an overview of the previous literature.  As an application
of our findings we show in Chapter~\ref{chap:EarthMatter} how to
identify the earth-matter effect in a detected SN neutrino signal. And
finally we summarize our findings in Chapter~\ref{chapter:conclusions}.

\cleardoublepage

\chapter{The Core-Collapse Paradigm}
\label{chap:paradigm}

Massive stars accumulate iron in their center after several stages of
nuclear burning.  The iron core becomes unstable once it reaches a
certain mass limit and collapses.  According to the so called
``delayed-explosion mechanism,'' the collapse is inverted into an
explosion and a shock wave travels outwards powered by the energy
deposition of the neutrino flux.  The neutrinos are emitted by the
proto-neutron star that formed in the center.

There exist a number of numerical simulations.  Even the most
elaborate among these simulations, however, fail to produce
explosions.  It is an unsettled issue what the missing ingredients
are.

With a focus on the neutrino emission, we discuss the various stages
of the explosion.  Flavor dependent differences in the emitted
neutrino spectra arise due to neutrino-matter
interactions. Electron neutrinos and electron anti-neutrinos are
dominantly produced by charged-current processes.  Several neutral
current processes govern the transport of the other flavors.

\section{The Life of a Star: Balancing Forces}
\label{sec:LifeStar}

During its whole life a star has to balance two possibly fatal
effects that work in opposite directions.  On one side, there is the
gravitational pull, that tries to collapse the star, on the other side
the thermal pressure that expands the star.  The connection of thermal
and gravitational energy is given by the virial theorem.  For the
time-averaged kinetic energy $\langle E_{\rm kin}\rangle$ and
gravitational energy $\langle E_{\rm G}\rangle$ of an atom inside
the star the virial theorem is
\be
\langle E_{\rm kin} \rangle = - \frac{1}{2} \langle E_{\rm
G}\rangle \;.
\ee
By making the star more compact, $\langle E_{\rm G}\rangle$ becomes
more negative.  Since the temperature is proportional to $\langle
E_{\rm kin}\rangle$, it rises as the star's radius shrinks.  Shrinking
the star by heating it up, or vice versa, corresponds to a negative
heat capacity and stabilizes the star.  For a certain stellar radius
the gravitational pull is balanced by the thermal pressure.

A star radiates photons and
therefore loses energy.  By burning just enough of its fuel in order to
produce the energy radiated away, the star becomes a stable object.
Initially, this fuel is hydrogen, that
produces energy in nuclear-fusion reactions to helium.  The stronger
the gravitational pull, the hotter the star's interior and the more
energy is radiated away.  Therefore more massive stars burn fuel at a
higher rate.

Due to the nuclear-fusion reactions more and more helium collects in
the central region.  When hydrogen is depleted at the center, the
helium eventually reaches high enough density and temperature
in order to ignite.  Due to the strong temperature dependence of
nuclear fusion reactions, helium burning in the central region is
spatially well separated from the hydrogen burning that can still
continue at the edge of the helium core.  The helium fusion process
produces again heavier elements that can also start fusion reactions
at a later stage.  Since the star has to be denser and hotter to
ignite the burning of heavier elements it becomes brighter and loses
more energy at this stage.  Additionally, these burning processes
liberate less energy in each reaction.  The reaction rate rises
drastically.  While hydrogen burning can last millions or even
billions of years the silicon burning stage usually takes less than a
day (see e.g.\ Herant et al.~1997). 

\begin{figure}[t]
\columnwidth=11cm
\plotone{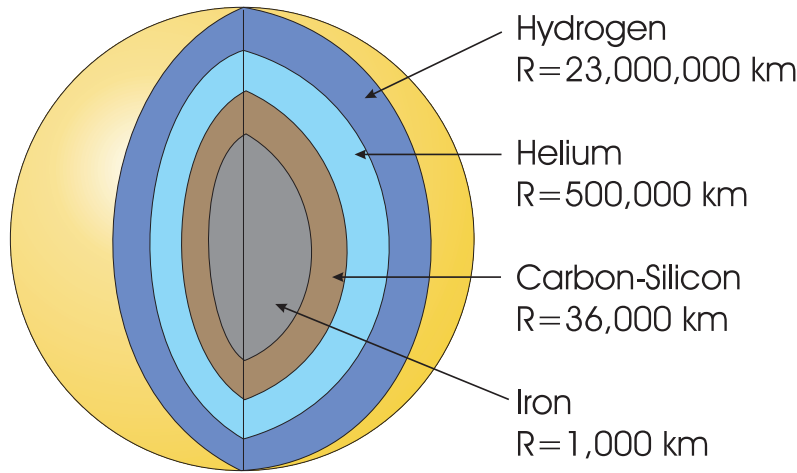}
\caption{\label{fig:onion} Schematic picture of a typical progenitor's
onion structure (not drawn to scale).}
\end{figure}

After several stages of nuclear burning the initial composition of
helium and hydrogen transformed into an onion-shell structure with 
heavier elements up to iron-group nuclei forming the star's core
(Figure~\ref{fig:onion}).  For stars with a mass greater than about 8
$M_\odot$ the core mainly consists of $^{56}$Fe and $^{56}$Ni.  Due to
silicon burning the iron core eventually reaches a mass close to its
Chandrasekhar limit of $1.2$--$1.5 M_\odot$.  Iron cannot ignite and
further fuse to heavier elements since it is already the most tightly
bound element.

The Chandrasekhar limit is a stability criterion for compact objects
like white dwarfs or the iron core of a massive star.  Such objects
are stabilized by electron degeneracy pressure.  The following very basic
explanation of the Chandrasekhar limit can be found in Shapiro \&
Teukolsky (1983).  The gravitational energy of a particular electron
at radius $R$ is $E_{\rm G}\propto - N R^{-1}$ for $N$ gravitating
particles.  By moving to the center the electron gets lower
gravitational energy and therefore occupies an energetically favored
state.  As long as the electron is non-relativistic its Fermi energy
$E_{\rm F}$ is proportional to the Fermi momentum $p_{\rm F}$ squared.
Therefore, when the electron moves towards the center, $E_{\rm F} \propto
p_{\rm F}^2 \propto N^{2/3} R^{-2}$ rises faster than the potential
energy falls.  For some radius the total energy $E_{\rm G}+E_{\rm F}$
assumes a minimum value and the star is in a stable state.  When the
mass rises and thus the gravitational potential gets deeper, also
$p_{\rm F}$ has to rise and more and more electrons become
relativistic.  For relativistic electrons, however, $E_{\rm F} \propto
p_{\rm F} \propto N^{1/3} R^{-1}$. The minimum in the total energy
disappears.  Depending on its mass being above or below the
Chandrasekhar limit at that time the star can have two distinct 
destinies.
\begin{itemize}
\item If the mass, $M\propto N$, of the star is below the Chandrasekhar
limit the total energy is positive. Therefore the star lowers its
energy by expanding, which in turn lowers $p_{\rm F}$, electrons
become non-relativistic and the star is stable again. 
\item For a mass above the Chandrasekhar limit the total energy
becomes more and more negative by lowering the radius and the star
starts to collapse.
\end{itemize}
The number of nucleons determines the mass of the iron core.  Up to
details of the chemical composition of the star, $M=N m_{\rm
nucleon}$.  The Chandrasekhar limit is then obtained by solving
$E_{\rm F} = E_{\rm G}$ for relativistic electrons and yields $1.5
M_\odot$.

\section{The Death of a Massive Star: Core Collapse}
\label{sec:CoreCollapse}

As soon as the iron core of a SN progenitor reaches its Chandrasekhar
limit the gravitational pull wins over the internal pressure and the
core starts to collapse.  The rise in density and temperature due to
the collapse sets the stage for processes speeding up the infall.
Electron captures on nuclei and protons,
\bea
\label{eq:Urca}
\re^- (Z,A) &\to& \nue \; (Z-1,A) \nonumber \\
\re^- \rp &\to& \nue \; \rn \; ,
\eea
become more and more efficient and do not only take away degeneracy
pressure, but also drain energy due to the emitted neutrinos that
escape the iron core unhindered.  Another energy sink is photo
disintegration of iron nuclei.  Due to the rising temperature inside
the core photons start to disintegrate  iron nuclei.  Within
milliseconds the photo disintegration undoes parts of the work of the
nuclear-fusion processes, and therefore takes away energy and thermal
pressure.

\begin{figure}[t]
\columnwidth=14cm
\plotone{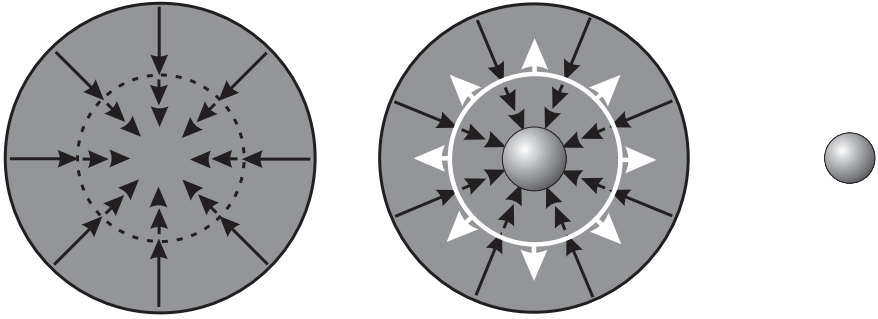}
\caption{\label{fig:collapse} Cartoon of a collapsing star. {\em
Left:} The initial stage, with the inner homologous and the outer
super-sonically infalling region.  {\em Middle:} A hard core has
formed, the shock (white) has bounced and travels outwards. The outer
layers are still infalling, passing through the shock wave, and then
dropping at a slower rate. {\em Right:} The shock wave has blown off
the outer layers of the star; the naked proto-neutron star contracts
and cools by neutrino emission.}
\end{figure}

\begin{figure}[t]
\columnwidth=12cm
\plotone{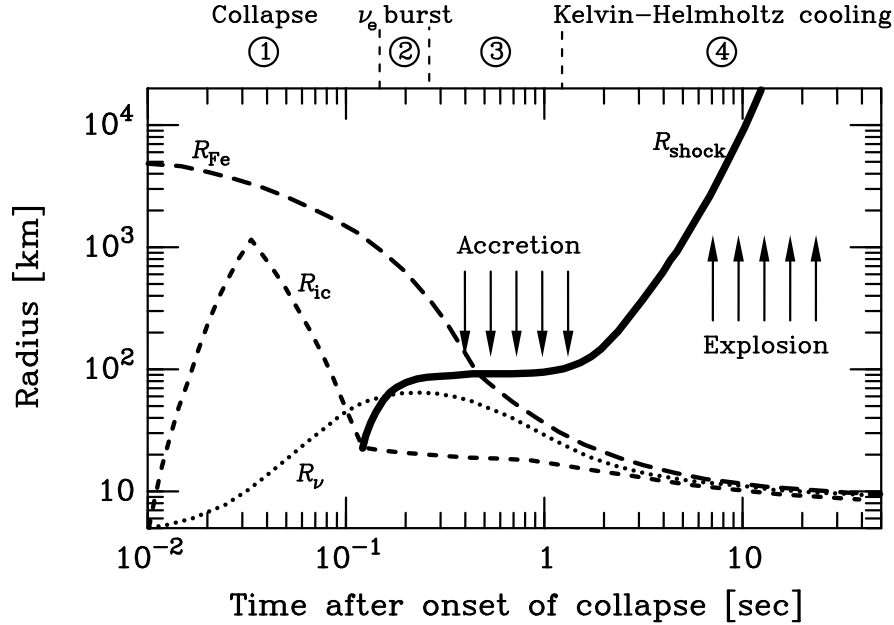}
\caption{\label{fig:schematic} Schematic plot of the time evolution
for various radial zones within the radius of the original iron core.
The evolution is divided in four phases: 1. Collapse of the iron core,
2. prompt $\nue$ burst, 3. accretion phase, and 4. Kelvin-Helmholtz
cooling of the proto-neutron star.  As long dashes the radius of the
iron core ($R_{\rm Fe}$) is shown, short dashes represent the
interface to the inner core ($R_{\rm ic}$), and dots to the radius
below that neutrinos are trapped ($R_\nu$).  The solid line, starting
shortly after 0.1 s shows the position of the shock ($R_{\rm
shock}$). Adapted from Janka (1993).}
\end{figure}

As the collapse of the iron core speeds up, two distinct regions
develop (Figure~\ref{fig:collapse} left, Figure~\ref{fig:schematic}
$R_{\rm ic}$ in phase 1): 
\begin{itemize}
\item the homologously infalling inner part where the speed is
sub-sonic and proportional to radius (Bruenn 1985).  This region
contracts, keeping its relative density profile, until it reaches
nuclear density where the equation of state stiffens drastically.
With its enormous inertia the core contracts even beyond the new
equilibrium and slightly expands again, usually referred to as the
``core bounce.'' 

\item the super-sonically infalling outer part.  Due to the lower
sound speed at larger radii, the matter assumes super-sonic speeds.
The infall corresponds therefore basically to a free fall.
\end {itemize}
A shock wave forms at the interface of the two regions due to the
slight reexpansion of the inner core.  The core bounce triggers the
explosion.  

The collapse happens on a time scale of some 100~ms (first phase in
Figure~\ref{fig:schematic}).  Therefore, the outer part of the iron
core is still infalling and the star's envelope has ``not even
noticed'' that the core collapsed.

\section{Reincarnation: A Proto-Neutron Star Emerges}

The shock wave moves outwards through the infalling outer layers
of the iron core and loses energy by disintegrating the nuclei it
plows through.  The nuclear binding energy of $0.1 M_\odot$ of iron is
about $1.7 \times 10^{51}$ erg, comparable to the explosion energy,
i.e., approximately 1\% of the released $10^{53}$ erg of gravitational
binding energy.  The dissociation of iron also liberates protons that
capture electrons, and in turn emit electron neutrinos.  In the
central region of the core, these neutrinos are trapped by the
surrounding material.  Diffusion is slow compared to the dynamic time
scale.  As shown in Figure~\ref{fig:schematic}, the region of neutrino 
trapping below $R_\nu$ builds up during collapse.  As long as the
shock wave has not passed $R_\nu$ neutrino trapping is very
efficient, because there are still iron nuclei around and due to
coherence effects the cross section for neutrino scattering on nuclei
is large compared to that for scattering on nucleons.

Then we enter the second phase of Figure~\ref{fig:schematic}.  The
shock passes $R_\nu$ and disintegrates the still infalling iron nuclei
in the vicinity of $R_\nu$.  Suddenly, there are no longer iron nuclei
around.  Therefore, instead of neutrinos scattering on nuclei, only
scattering on free nucleons takes place with a much lower cross
section.  At the same time, the free protons capture electrons and
emit electron neutrinos. These neutrinos in the vicinity of the shock
can escape freely and cause a sudden peak in the electron-neutrino
flux.  The luminosity briefly rises to around $10^{54}$~erg~s$^{-1}$,
referred to as the prompt neutrino burst.  This also weakens the
shock.

Lepton number is approximately conserved inside the core because
neutrinos are trapped in the dense matter mainly due to
neutrino-nucleon scattering.  The processes of
Equations~(\ref{eq:Urca}) cause a $\beta$ equilibrium to develop and
a Fermi sea of $\nue$ builds up.  Most of the gravitational binding
energy is now stored in the Fermi seas of electrons and electron
neutrinos.

During the third stage in Figure~\ref{fig:schematic}, at a radius of a
few 100~km, only about a few 100~ms after bounce, the shock wave
stalls due to its energy loss caused by the disintegration of iron
nuclei and neutrino emission.  Still infalling matter from outer
layers of the iron core passes through the standing shock onto the
newly formed proto-neutron star, also shown in the middle panel of
Figure~\ref{fig:collapse}.  The edge of the proto-neutron star is at
$R_{\rm ic}$ in Figure~\ref{fig:schematic}, what used to be the
interface between the homologously and super-sonically infalling
regions of the iron core.

Contraction of the proto-neutron star and the accretion of matter
still cause the temperature to rise and the neutrino luminosity
increases.  In the region between $R_\nu$ and the standing shock wave
convection takes place.  The neutrinos cool the surface of the
proto-neutron star.  In the last stage of Figure~\ref{fig:schematic}
the stalled shock wave is revived by the energy that a fraction of the
outgoing neutrinos deposit in the region just behind the shock.  If
only a few per cent of the liberated $10^{53}$ erg gravitational
binding energy are absorbed by the matter behind the standing shock
wave, the shock will move outwards again.  Finally, this powerful
explosion blows off the whole envelope of the star.  The naked
proto-neutron star remains in the center and cools by neutrino
emission.  The Kelvin-Helmholtz-cooling phase is displayed as the last
phase in Figure~\ref{fig:schematic} and the right panel of
Figure~\ref{fig:collapse}.

Due to the initially stalled and afterwards revived shock wave, this
mechanism is referred to as a delayed explosion.  It was first
suggested by Bethe and Wilson (1985).  Over the years more and more
refined numerical simulations have been developed.  The most elaborate
among them agree in their result that no explosions are obtained
(Buras et~al.~2003a, Liebend\"orfer et al.~2001, Mezzacappa et
al.~2001, Rampp \& Janka 2000, Thompson, Burrows, \& Pinto 2002).
It has been stressed by Buras et al.~(2003a) that some crucial piece
in the game might be missing.

Even if the standard picture is correct, there are also cases when no
neutron star is born by the SN explosion.  If the progenitor star has
a large mass, say greater than 20 $M_\odot$, the evolution of the
proto-neutron star can find a sudden end.  Due to accretion the mass
of the proto-neutron star can exceed its Chandrasekhar limit.  The
proto-neutron star then collapses and ends up as a black hole,
terminating neutrino emission almost at an instant.  The details
depend on the nuclear equation of state and matter accretion, issues
that are not yet settled.  A discussion can be found in Beacom, Boyd,
\& Mezzacappa (2001) and references therein.

\section{Self-Consistent Simulations}
\label{sec:SelfconsinstentSims}

There have been numerical simulations of SNe showing delayed
explosions for almost two decades now (Bethe \& Wilson 1985, Wilson
1985).  These pioneering works obtained robust explosions.  With 
increasing computer power and refined input physics the previous
explosions could not be confirmed by other groups.  Up to now, only
the Livermore group obtains robust explosions (Totani et al.~1998).
However, they have to include convection inside the proto-neutron star
just below the neutrino spheres by a parameterization of
neutron-finger instabilities in order to get explosions (Wilson \&
Mayle 1988, 1993).  This treatment of convection has the effect of
enhancing the early neutrino luminosities, which in turn causes more
energy to be deposited behind the shock.  By neutron-finger convection
the neutron rich outer part of the proto-neutron star mixes with the
inner part in analogy to salt-finger convection in hot salt water and
cold fresh water.  However it is controversial whether convection is
realized in an actual SN (Burrows 1987, Bruenn, Mezzacappa, \& Dineva
1995, Keil, Janka, \& M\"uller 1996, Mezzacappa et al. 1998, Pons et
al.~1999).  In their two-dimensional simulations, Buras et al.~(2003b)
find convection of Ledoux type.

All groups performing computer simulations of the
hydrodynamics of SNe with a state-of-the-art treatment of the
neutrino transport, i.e., Boltzmann solvers, do not get explosions at
all (Liebend\"orfer et al.~2001, Mezzacappa et al.~2001, Rampp \&
Janka 2000, Thompson, Burrows, \& Pinto 2002).  Even the most elaborate
SN simulations with multi-dimensional  hydrodynamics and all
relevant microphysics in the neutrino interactions included fail to
explode (Buras et al. 2003b).  The reason for the models not to
explode is obscure.

The delayed-explosion mechanism described in the previous section is
in principle found by numerical simulations.  The collapse sets in,
a shock wave forms, bounces and moves outwards.  On its way it loses
energy mainly due to the disintegration of nuclei and
finally stalls.  The simulations also show the appearance of a cooling
region above the neutrino sphere where energy is lost due to neutrino
radiation.  Behind the shock these neutrinos heat the matter, as
described above.  In the simulations, however, the heating is not
sufficient for reviving the shock.  Instead it falls back and
no explosion takes place.  Quite naturally, the main focus of model
builders is to find the missing input physics in order to obtain an
explosion.

In spite of the numerical failure of obtaining explosions, there are
good reasons to believe in the delayed-explosion scenario.
Observational hints suggest that the core collapse paradigm should, 
at least to some extent, be realized in nature.  We know that SNe
occur after stars have reached their final stage of nuclear burning.
For example, SN1987A was a blue supergiant before it exploded.  There
were neutrinos observed from SN1987A (Bionta et al.\ 1987, Hirata et
al.\ 1987) about three hours before the first photons were seen, which
is a clear evidence for a core collapse. Although there has not been a
neutron star identified at the site of SN1987A, for other SN remnants
neutron stars have been observed.  While the details of the
explosion mechanism are still under debate, basic features appear to
be robust.  If a core collapse occurs the rise of neutrino
luminosity and also the prompt neutrino burst are unavoidable as long
as the collapse does not continue to form a black hole immediately.
An ongoing neutrino emission for several seconds by the proto-neutron
star is also not doubted.  Throughout this thesis we will assume that
the delayed-explosion mechanism is realized in nature.

\begin{figure}[t]
\columnwidth=12.1cm
\plotone{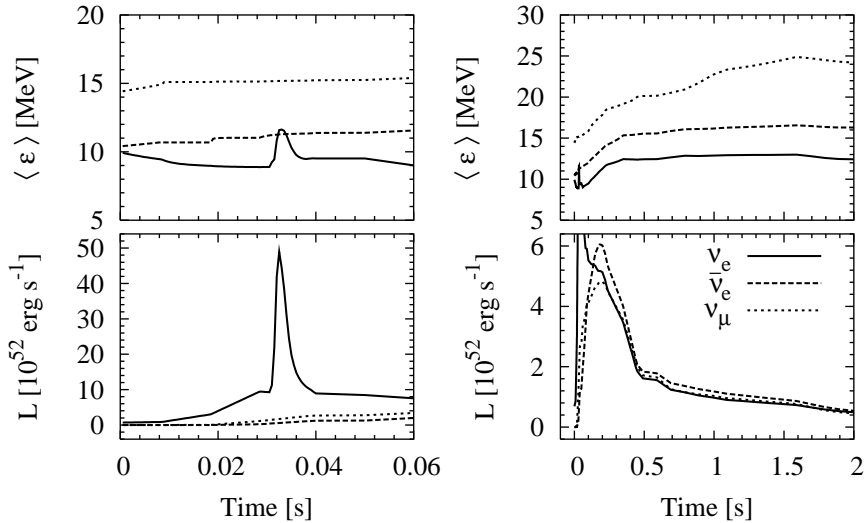}
\caption{\label{fig:lllLumi} Numerical results of the first 2 s
after core bounce taken from Totani et al.~(1998).  {\em Top left:}
Only the mean energy of $\nue$ is affected by the prompt burst.  {\em Top
right:} The evolution of mean energies shows the traditional hierarchy.
{\em Bottom left:} The prompt $\nue$ burst lasting for a few
milliseconds.  {\em Bottom right:} Time evolution of the neutrino
luminosities, scaled on both axes as compared to the left panel.}
\end{figure}

All numerical simulations agree on this qualitative behavior.  An
example that we took from Totani et al.~(1998) is given in
Figure~\ref{fig:lllLumi}.  Neutrino mean energies and luminosities are
shown as functions of time post bounce.  Solid lines represent $\nue$,
dashed $\bar\nue$, and dotted $\numu$.  In the bottom-left panel
we show a prominent feature, namely the prompt $\nue$ burst right
after the bounce.  The mean energy of $\nue$ in the top-left panel
shows a short rise during the prompt burst.  The left panels show the
mean energies and luminosities during the first 60~ms and the right
panels during the first 2~s after core bounce.  Right after the prompt
$\nue$ burst the accretion phase takes over, visible in the rise and
fall of luminosities.  After the end of the accretion phase
deleptonization and cooling of the proto-neutron star powers the
neutrino emission. The cooling phase starts at 0.5~s.  How the
flavor-dependent differences arise is the topic of the next section.

\section{Flavor-Dependent Neutrino Emission}
\label{sec:NuEmission}

As soon as the collapse starts the neutrino luminosity increases.  In
this early stage only electron neutrinos are produced in large amounts
due to electron captures on free protons.  The protons originate from
photo-disintegrated iron nuclei.  Since these processes happen most
frequently in the dense inner core the major part of the released
electron neutrinos is trapped.  This trapping is the reason for the
Fermi sea of $\nue$ to develop.  When the shock wave passes the region
where the star becomes transparent to neutrinos, i.e., the ``neutrino
sphere,'' the trapped $\nue$ can escape.  Additionally, the infalling
iron nuclei are disintegrated and the electron capture rate in the
vicinity of the neutrino sphere goes up.  Altogether, this results in
the prompt $\nue$ burst (left panels Figure~\ref{fig:lllLumi}).

\begin{figure}[t]
\columnwidth=8cm
\plotone{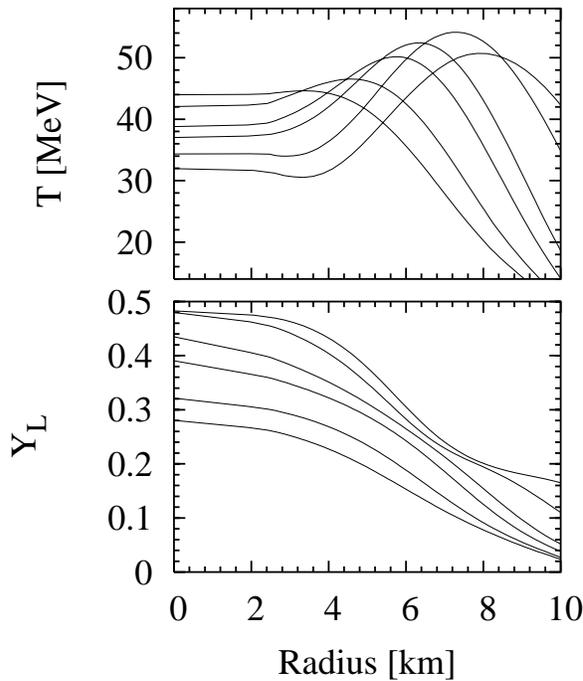}
\caption{\label{fig:lllYl} Deleptonization of a proto-neutron star as
obtained by the Livermore group (Totani et al.\ 1998).  {\em Top:} We
show the temperature as a function of radius at 0.5, 1, 3, 5, 10, and
15 s after bounce.  With increasing time the peak moves to lower
radii.  {\em Bottom:} For the same times we give the lepton number per
nucleon.  As time increases lepton number gets lower.} 
\end{figure}

Neutrinos below the neutrino sphere are still trapped.  During the
accretion phase, lasting for a few 100~ms, infalling matter
deposits energy on the surface of the proto-neutron star, increasing
the neutrino luminosities (bottom right panel Figure~\ref{fig:lllLumi}).
Afterwards, the proto-neutron star still gains energy by
contraction. The Fermi seas of electrons and $\nue$ have a huge
amount of energy stored.  The star releases this energy over a timescale of
several seconds.  For the calculation of the Livermore group
(Totani et al.~1998) we plotted the deleptonization in
Figure~\ref{fig:lllYl}.  The lower panel shows the number of leptons
per nucleon ($Y_{\rm L}$) as a function of radius. From top to bottom
the lines correspond to 0.5, 1, 3, 5, 10, and 15 s after bounce.
Accordingly, the top panel shows the temperature profile at these
times.  The maximum moves to lower radii as time increases.  At the
same time the innermost part heats up.

Even though neutrinos are trapped below their neutrino spheres, due to
the steep gradient in $Y_{\rm L}$ they diffuse outwards.  Since 
electrons are very degenerate in this region, even more degenerate
than $\nue$, neutrinos will scatter towards regions of lower electron
degeneracy.  Additionally, when scattering on electrons, neutrinos will
always lose energy, since the electron has to end up with an energy
above the surface of the electrons' Fermi sea.  By this
down scattering the electron-neutrino- and electron-degeneracy energy
is mainly released as heat.  Therefore the temperature maximum follows
the deleptonization (see e.g. Raffelt 1996).

The reason why neutrinos are trapped is their frequent interaction
with the surrounding matter via various interaction processes.  The
medium basically consists of neutrons, protons, electrons, positrons,
and of course neutrinos.  Due to the presence of electrons and
positrons, $\nue$ and $\bar\nue$ can undergo charged current reactions
that are absent for all other flavors.  Qualitatively the transport of
$\nue$ and $\bar\nue$ is therefore different from the transport of
$\numu$ and $\nu_\tau$.  To be exact, there is also a small population
of muons and anti-muons present.  In central regions, where the
temperature is maximal, the muon mass is twice the temperature.  Since
for a non-degenerate particle the mean energy is roughly three times
the temperature , $\langle \epsilon \rangle \approx 3\; T$, there is a
thermal distribution of muons present in the inner core.
However, in the vicinity of the neutrino sphere of $\numu$ and
$\bar\numu$, where muons would affect the emerging neutrino spectra,
the temperature is about 1/10 of the muon mass.  Thus muons are
strongly suppressed and do not affect the neutrino emission.
\begin{figure}[t]
\columnwidth=12cm
\plotone{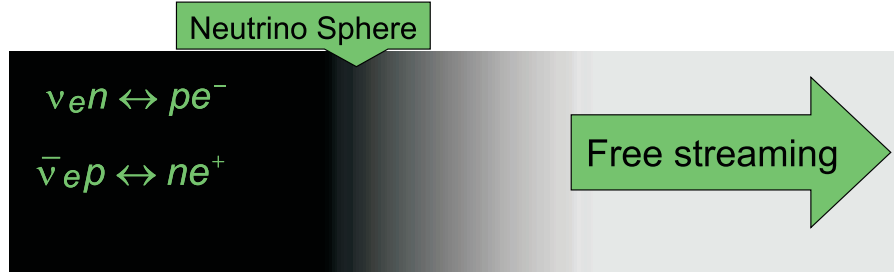}
\caption{\label{fig:nuespheres} Schematic picture of the $\nue$ and 
$\bar\nue$ spheres in a proto-neutron star.}
\end{figure}

Schematically, the emission of $\nue$ and $\bar\nue$ is displayed in
Figure~\ref{fig:nuespheres}.  In the central region of the
proto-neutron star, corresponding to the black shaded area on the left
hand side of Figure~\ref{fig:nuespheres}, $\beta$-processes  keep 
$\nue$ and $\bar\nue$ in LTE.  Due to the decreasing density this
reaction becomes inefficient at some radius, defining the neutrino
sphere.  From that radius neutrinos start streaming freely.  Since the
$\beta$-cross section depends on neutrino energy also the radial
position of the neutrino sphere is energy dependent.  Therefore it is
not possible to define a unique  neutrino-sphere radius.  Depending on
their energy neutrinos decouple from the proto-neutron star at
different radii and therefore different local temperatures.  For this
reason the spectrum of neutrinos leaving the star does not represent a
thermal distribution with a temperature corresponding to the medium
temperature at the neutrino sphere that is, anyway, not well defined.

The energy dependence of the interaction processes of $\nue$ and
$\bar\nue$  is exactly the same.  The proto-neutron star, however,
contains more neutrons than protons, which leads to a higher
absorption rate for $\nue$ than for $\bar\nue$.  Therefore, the $\nue$
sphere lies at a larger radius than the $\bar\nue$ sphere.  Since this
is true for all neutrino energies the mean energy of emitted $\nue$ is
always less than the mean energy of $\bar\nue$. 

For rough estimates, the neutrino sphere can be considered a
blackbody radiating neutrinos.  The surface area is thus proportional
to the number of emitted neutrinos.  Again, the energy dependence
of the neutrino-sphere radius makes a more reliable estimate very
difficult.  

\begin{figure}[t]
\columnwidth=12cm
\plotone{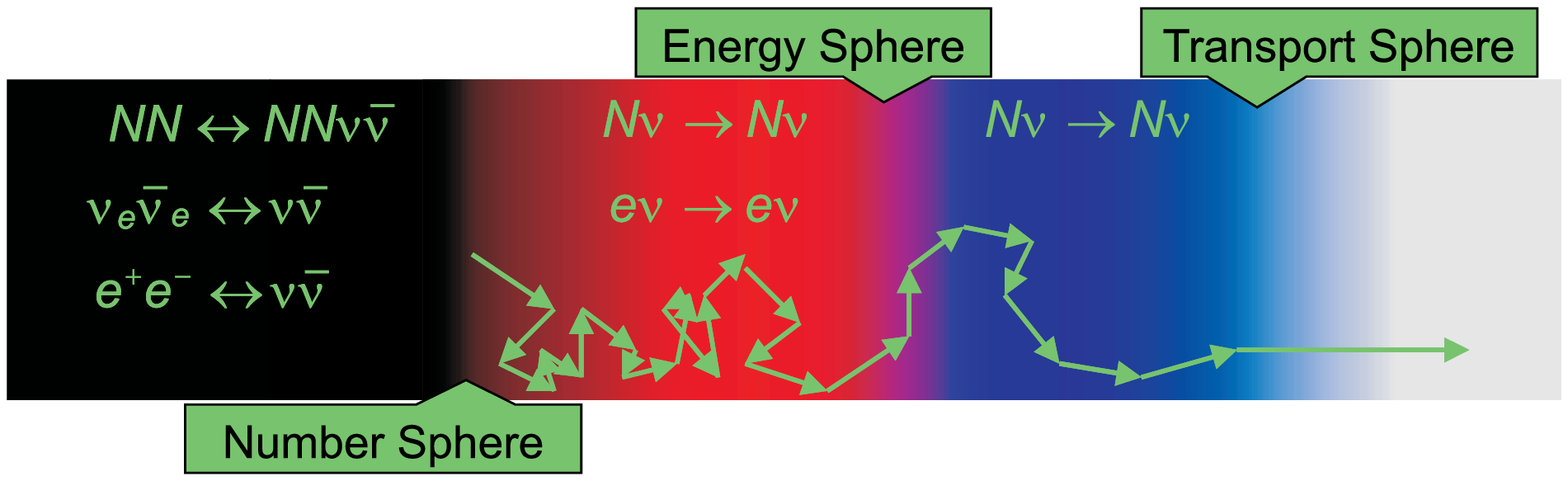}
\caption{\label{fig:numuspheres} Same as Figure~\ref{fig:nuespheres}
but for $\numu$.}
\end{figure}

In Figure~\ref{fig:numuspheres} we show a schematic view of the
$\numu$ transport.  Comparing Figures~\ref{fig:numuspheres} and
\ref{fig:nuespheres} tells us that the transport of $\numu$ is more
complicated.  There are no charged current interactions of $\numu$
with the medium.  Instead, there is a variety of neutral current
interactions that qualitatively differ from one another and, at the
same time, are of comparable importance for the transport process.
Naively, one would expect that $\numu$ decouple deeper inside the star
than $\bar\nue$, because the neutral-current interaction rates are
lower.  Then, the mean energy of emerging $\numu$ would be higher
compared to $\bar\nue$ and the flux lower.  In reality, the situation
is more subtle.  

In the inner-most region the $\numu$ are kept in LTE by
energy exchanging scattering processes and the following pair
processes:

\begin{itemize}
\item Bremsstrahlung $\rN\rN\leftrightarrow\rN\rN\numu\bar\numu$
\item Neutrino-pair annihilation $\nue\bar\nue\leftrightarrow\numu\bar\numu$
\item Electron-positron-pair annihilation
$\re^+\re^-\leftrightarrow\numu\bar\numu$ 
\end{itemize}

The radius where these reactions become inefficient defines the
number sphere.  Outside this sphere no particle creation or
annihilation takes place and $\numu$ can no longer be in LTE.  

Due to the scattering reactions 
\begin{itemize}
\item $\rN \numu \to \rN \numu$
\item $\elpo \numu \to \elpo \numu$
\end{itemize}
there is still energy exchange with the medium.  However, the two 
processes are qualitatively very different.  Scattering on $\elpo$
is less frequent since there are less $\elpo$ than nucleons.  On the 
other hand the amount of energy exchanged in each interaction with 
$\elpo$ is very large compared to the small recoil of nucleons.  At
the radius, where scattering on $\elpo$ freezes out lies the energy
sphere.  A diffusive regime starts, where neutrinos only scatter on
nucleons and therefore exchange little energy in each reaction.  This
regime is terminated by the transport sphere, defined by the radius at
which also scattering on nucleons becomes ineffective and the $\numu$
start streaming freely.

Even if we assume that nucleon scattering happens iso-energetically,
i.e., no energy can be exchanged outside the thermalization sphere, the
mean energy of neutrinos escaping the star is only about 50--60\% of
those close to the thermalization sphere.  Due to its dependence on
the square of neutrino energy the nucleon scattering cross section
has a filter effect, because it tends to scatter high energy neutrinos
more frequently (Raffelt 2001).  If one were to specify a neutrino
sphere according to the crude approximation of a blackbody radiating
neutrinos, it would coincide with the number sphere.  Neutrinos leave
LTE at the number sphere.  The position of the number sphere
determines the flux, because neutrino creation is not effective beyond
that radius.  The $\numu$ flux that passes the number sphere is
conserved.  On the other hand, the mean energy of $\numu$ in this area
is still significantly lowered due to scattering processes before the
$\numu$ leave the star. The mean energy of $\numu$ emerging the star
is usually found to be larger than that of $\bar\nue$.  Simulations
that include the $\numu$ interactions only very approximate obtain a
large hierarchy in the mean energies (Figure~\ref{fig:lllLumi}).
The systematic study of all interaction processes is a major part of
this work.  Our findings show that the transport of $\numu$ is more
subtle than  previously assumed.  Once all interaction
processes are taken into account, the mean energies of $\numu$ and
$\bar\nue$ are very similar and might even cross over.  The number
fluxes can differ by large amounts.

\cleardoublepage

\chapter{Neutrino Interactions}
\label{chap:rates}

We discuss neutrino interactions with the stellar medium that we
assume to consist of protons, neutrons, electrons, positrons, and
neutrinos of all flavors.  The most important reactions for electron
neutrinos and electron anti-neutrinos are charged current processes on
nucleons ($\beta$-processes).  All other flavors undergo
neutral-current interactions only. 

The neutral-current processes fall into three qualitatively different
groups.  True LTE is only obtained by the pair creation and
annihilation processes, i.e., bremsstrahlung, the $\nue\bar\nue$, 
and $\re^+\re^-$ pair processes.  Scattering on $\re^\pm$ and $\nue$,
$\bar\nue$ exchanges large amounts of energy in a single
reaction. Neutrino-nucleon scattering qualitatively differs from the
other scattering reactions in being more frequent but at the same time
exchanging only small amounts of energy in each reaction.  Going
beyond the traditional iso-energetic nucleon scattering we introduce
recoil and weak magnetism.

For the first time, we have investigated the $\nue\bar\nue$ pair
process in detail and were able to show that it is far more important
than the traditional $\re^+\re^-$ process.  By crossing symmetry the
$\nue\bar\nue$ pair process is related to $\numu\nue$ scattering. This
scattering process, however, is negligible compared to $\re^\pm$
scattering.  We have presented these results in a similar form in our
publication:\\
 R.~Buras, H.~T.~Janka, M.~T.~Keil, G.~G.~Raffelt, and M.~Rampp,
``Electron-neutrino pair annihilation: A new source for muon and tau  neutrinos in supernovae,''
Astrophys.\ J.\  {\bf 587} (2003) 320.

\section{Beta-Processes}

\begin{figure}[b]
\columnwidth=6.5cm
\plotone{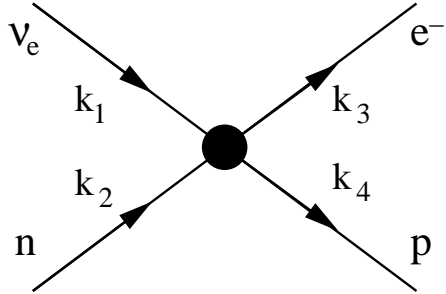}
\caption{\label{fig:beta} Feynman diagram for $\nue$ absorption.}
\end{figure}

For electron neutrinos, the most important neutrino-matter
interactions are the $\beta$-processes $\nue \rn \lra \re^- \rp$
and $\bar\nue \rp \lra \re^+ \rp$.  In principle, also $\numu$ can
undergo charged current reactions.  The central region of the SN core
certainly contains muons, but their presence is commonly neglected in
hydrodynamic simulations.  In the vicinity of the neutrino sphere the
temperature is too low for a thermal distribution of muons.  For our
simulations it is therefore justified to neglect the charged current
reactions of $\numu$.

The squared spin-summed matrix element for neutrino absorption by
neutrons is given by (Yueh \& Buchler 1976b)
\bea
\label{eq:BetaProcessM}
\sum_{\rm spins}|{\cal M}|^2 &=&
32\, \GF^2\,\Bigl [(\alpha + 1)^2\; (k_2\cdot k_1)(k_4\cdot k_3)\nonumber\\
&+&(\alpha - 1)^2  (k_2\cdot k_3)(k_4\cdot k_1)+(\alpha^2 -
1)\;m_\rn \;m_\rp (k_1\cdot k_3)\Bigr ]\,.\nonumber\\
\eea
The momenta $k_1$, $k_2$, $k_3$, and $k_4$ are assigned to the corresponding
particles as shown in Figure~\ref{fig:beta}; $\GF$ is the Fermi
coupling constant, $\alpha = C_A/C_V =1.26$, and $m_\rn$ and $m_\rp$
the masses of the neutron and proton, respectively.  Together with
phase-space blocking and the phase-space integrations,
Equation~(\ref{eq:BetaProcessM}) yields the absorption rate or
equivalently the inverse of the mean free path (mfp)
\bea
\lambda^{-1} &=& \frac{n_\rn}{(2\pi)^5 2\epsilon_1}\int\frac{d^3
k_2}{2\epsilon_2} \frac{d^3 k_4}{2\epsilon_4}\frac{d^3 k_3}{2\epsilon_3} \,\sum_{\rm spins}|{\cal
M}|^2 \nonumber\\
&& \times f_\rn(\epsilon_2)
\,(1-f_\rp(\epsilon_4))\,(1-f_\re(\epsilon_3))\,
\delta^4(k_1+k_2-k_3-k_4) \,, 
\eea
where $n_\rn$ is the neutron density, $\epsilon_j$ the energy of
particle $j$, and $f_\rn(\epsilon_2)$, $f_\rp(\epsilon_4)$, and
$f_\re(\epsilon_3)$ are the occupation numbers of neutrons, protons,
and electrons, respectively.

For anti-neutrinos we have to interchange neutrons and
protons and replace electrons by positrons.  For the reverse rate one
has to replace blocking factors and the occupation numbers
appropriately. 

Since there are more neutrons than protons present in the SN core, the
$\beta$-rates for $\nue$ are always higher than for $\bar\nue$.  All
interaction processes presented in the remaining sections of this
chapter are neutral-current interactions.  Due to the smaller coupling
constants they are sub-dominant for electron neutrinos.
The following sections are more relevant to $\numu$, because charged
current reactions do not contribute to the spectra formation of
$\numu$.  (If not stated differently, $\numu$ always refers to
$\nu_{\mu,\tau}$ and $\bar\nu_{\mu,\tau}$.)

\section{Neutrino-Nucleon Scattering}
\label{sec:nuNscat}

The most frequent neutral-current process is neutrino-nucleon
scattering and therefore the dominant opacity source for $\numu$.  For
a neutrino with initial energy $\epsilon_1$ and final energy
$\epsilon_2$, the differential cross section is given by
\be
\label{eq:nucscat} 
\frac{d\sigma}{d\epsilon_2\,d\cos\theta}
=\frac{C_A^2(3-\cos\theta)}{2\pi}\,\GF^2\,\epsilon_2^2\,
\frac{S(\omega,k)}{2\pi} \; ,
\ee 
with $\omega = \epsilon_1 - \epsilon_2$, $k$ the modulus of the
momentum transfer to the medium, and $\theta$ the scattering angle.
$S(\omega,k)$ represents the dynamical structure function that
parameterizes the response of the nuclear medium.  

\begin{figure}[b]
\columnwidth=10cm
\plotone{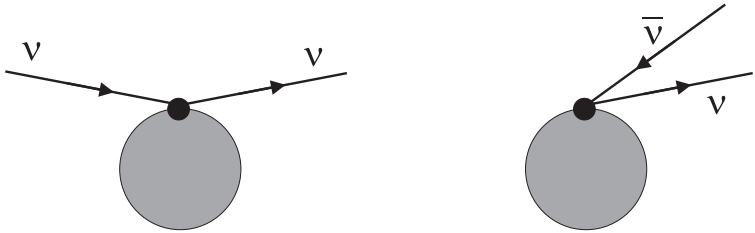}
\caption{\label{fig:FullS} {\em Left:\/} Diagram of neutrino nucleon
scattering. The grey shaded blob represents the nuclear medium. {\em
Right:\/} Bremsstrahlung is obtained by crossing one neutrino leg.}
\end{figure}

The dynamical structure function includes all effects that change the
behavior of the nucleon in the medium, like fluctuations,
correlations, and degeneracy effects.  The full dynamic structure
function covers the whole $(\omega,k)$ plane.  Therefore it does not
only describe neutrino scattering on nucleons but also neutrino
bremsstrahlung off nucleons.  Both processes are related by crossing
symmetry as shown in Figure~\ref{fig:FullS}, where the grey shaded
blob represents the nuclear medium.  However, the full structure
function is unknown.  Common approximations treat the space-like
($\omega^2\le k^2$) and time-like ($\omega^2\ge k^2$) domains on a
different footing.

The common simplification of the neutrino-nucleon scattering cross
section in traditional SN simulations was to assume that the reaction
is iso-energetic, i.e., the nucleons have an ``infinite mass'' and
therefore cannot absorb energy but arbitrary amounts of momentum.  In
this approximation the structure function is
\be 
\label{eq:ConservativeS}
S_{\rm no-recoil}(\omega,k)=2\pi\delta(\omega) \; .
\ee
The only multi-particle effect that is usually taken into account is
final-state blocking due to the nucleon degeneracy.

More elaborate recent approximations include correlated nucleons and a
finite nucleon mass.  (Burrows and Sawyer 1998, Reddy, Prakash, and
Lattimer 1998, Reddy et~al.\ 1999).  These results were calculated in
the random-phase approximation, that includes nucleon-spin
correlations but not nucleon-spin fluctuations.  Therefore,
bremsstrahlung is not allowed, because neutrinos couple to the nucleon
spin and the emission of a neutrino pair requires a spin flip.  In the
recent simulations by the Garching group (Rampp and Janka 2002), the
correlation effects were taken into account together with recoil.

\begin{figure}[b]
\columnwidth=12cm
\plotone{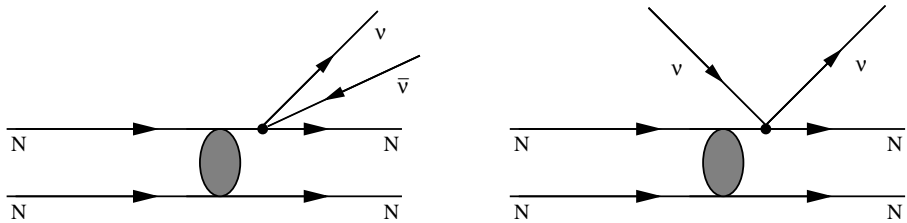}
\caption{\label{fig:brems} {\em Left:\/} Diagram of neutrino
bremsstrahlung off nucleons.  This process is
subdominant for energy exchange compared to recoil (Raffelt 2001). {\em
Right:\/} The ``crossed process'' corresponds to inelastic
neutrino-nucleon scattering.}
\end{figure}

Some different aspects of multi-particle effects are taken into
account by nucleon excitations.  The diagrammatic structure is given
in Figure~\ref{fig:brems}.  Raffelt (2001) developed a schematic
structure function in this spirit for bremsstrahlung.  By crossing a
neutrino leg one obtains inelastic scattering, that allows for an
energy transfer unrelated to nucleon recoil.  Even though, this energy
transfer is of the same size as nucleon recoil, the quadratic density
dependence of the inelastic scattering process yields lower rates in
more dilute regions, where recoil is still effective.  In a detailed
study Raffelt (2001) shows that once recoil is included in a SN
simulation the inelastic scattering contribution can be neglected.
For the implementation of $\numu$-nucleon scattering Raffelt (2001)
provides a structure function for recoiling nucleons that we adapted
to our code.

We do not distinguish between protons and neutrons.  Since for
non-relativistic nucleons the scattering cross section is proportional
to $C_V^2+3C_A^2$, the vector current ($C_V=-\frac{1}{2}$ for neutrons
and $\frac{1}{2}-2\sin^2\theta_{\rm W}$ for protons) is small compared
to the axial component, where we use $|C_A| = 1.26/2$.  Neglecting the
vector part simplifies the calculations significantly, since
otherwise, there are different structure functions for the axial
current, the vector current, and the mixed term.

Ignoring nucleon degeneracy effects, the structure function that
incorporates nucleon recoils is (Raffelt 2001)
\be 
\label{eq:s_recoil}
S_{\rm recoil}(\omega,k)=\sqrt{\frac{\pi}{\omega_k T}} \exp \left (
-\frac{\omega - \omega_k}{4 T \omega_k} \right )\,,
\ee
with $\omega_k = k^2/2m$, $m$ the nucleon mass, and $T$ the medium
temperature.

\begin{figure}[b]
\columnwidth=12.5cm
\plotone{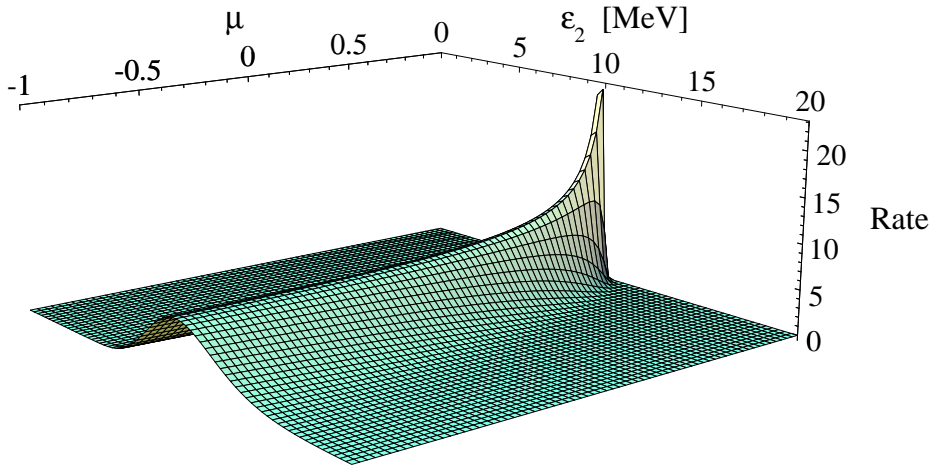}
\caption{\label{fig:recoilrate} The differential interaction rate for
$\nu\rN$ scattering according to Equations~(\ref{eq:nucscat}) and
(\ref{eq:s_recoil}) for initial energy $\epsilon_1=10$ MeV as a
function of the outgoing energy $\epsilon_2$ and scattering angle
$\mu=\cos\theta$. }
\end{figure}

Multiplying Equation~(\ref{eq:nucscat}) with the density of nucleons
and ignoring phase-space blocking of the essentially non-degenerate
nucleons yields the differential rates that can be integrated for
obtaining the required energy and angular differential rates.  In 
the case of recoil the numerical integrations are rather tricky
because Equation~(\ref{eq:nucscat}) is strongly forward peaked,
cf. Figure~\ref{fig:recoilrate}.  In our
code we employ the ``rejection method'' for obtaining the integrated
rates (Press et al.\ 1992).

In the case of $\nue$ and $\bar\nue$, scattering on nucleons is less
important than the $\beta$-processes.  Therefore we left the original
implementation of our code in the electron-neutrino sector unchanged.
The implementation follows Tubbs \& Schramm (1975) and can be found in
Janka (1987).  It basically corresponds to using the structure
function given in Equation~(\ref{eq:ConservativeS}).

\section{Weak Magnetism}
\label{sec:WeakMag} 

Weak magnetism arises as a correction to nucleon scattering due to
our schematic treatment of the dynamical structure function.  The
complete structure function would, of course, contain weak magnetism.
The effect is absent if we use the ``traditional''
approximation of infinite nucleon mass.

As mentioned earlier, there are essentially no muons or taus present
in the vicinity of the neutrino sphere.  Therefore, $\numu$ and
$\bar\numu$ are usually treated in the same way.  However, the
neutrino-nucleon cross section is different for neutrinos and
anti-neutrinos once weak magnetism is included.  The contributions to
the cross section from the anomalous magnetic moment of nucleons and
the interference term of the axial and vector current have different
signs for neutrinos and anti-neutrinos due to parity violation in the
standard model (Horowitz 2002).  The interaction rate is higher for
neutrinos than for anti-neutrinos.

Neutrinos in the relevant regime of the proto-neutron star have
energies on the order of several 10 MeV.  To first order in
$\epsilon/m$, where $\epsilon$ is the neutrino energy and $m$ the
nucleon mass, the correction factor to the neutrino-nucleon cross
section due to weak magnetism is (Horowitz 2002)
\be
\label{eq:weakmag_corr}
1 \pm \frac{4 C_A (C_V + F_2)}{C_A^2(3-\cos\theta)}
\frac{k}{m} \,,
\ee
where $k$ is the momentum transfer, $F_2$ a structure function
parameterizing the structure of nucleons. For neutrino-proton
scattering $F_2 = \frac{1}{2}  (\mu_\rp-\mu_\rn)-2
\sin^2(\theta_{\rm w})\mu_\rp\approx 1.019$, where  $\mu_{\rp,\rn}$ are the
magnetic moments of neutron and proton, respectively, and $\theta_{\rm w}$ is the
weak mixing angle. For neutrino-neutron scattering, the indices n and
p have to be interchanged yielding $F_2 = -0.963$.  The upper sign in
Equation~(\ref{eq:weakmag_corr}) is to be used for neutrinos and the
lower sign for anti-neutrinos.

In order to be consistent with Equation~(\ref{eq:nucscat}) we keep
terms $\propto C_A^2$ and neglect terms $\propto C_V^2$.  In addition,
we substituted $\epsilon_1 \cos\theta$, taken at the rest frame of the
nucleon, by our momentum transfer $k$.  This is correct for forward
and backward scattering but only an approximation for other angles.
It does not change the angular dependence by much, but in order to
keep our structure-function prescription of the cross section, we
cannot multiply factors that depend on $k$.

The first-order correction factor has the disadvantage that it becomes
negative for large momentum transfer.  In order to avoid numerical
problems we use
\be
\left ( 1 \pm \frac{4 C_A (C_V + F_2)}{C_A^2(3-\cos\theta)} \frac{k}{2
m}  \right )^2 
\ee
instead, which corresponds to Equation~(\ref{eq:weakmag_corr}) up to
first order in $k/m$.  For neutrino momenta in the region of the
neutrino sphere this approximation induces an error of less than 1\%.

\section{Bremsstrahlung}

Neutrino bremsstrahlung off nucleons inside the stellar medium is
related to neutrino nucleon scattering by crossing symmetry
(Section~\ref{sec:nuNscat}).  Raffelt (2001) showed that the emerging
neutrino spectra from a SN do not sensitively depend on the detailed
rate, a fact that we confirm in our simulations.  A schematic approach
is therefore well justified.  Since bremsstrahlung only plays a
crucial role for $\numu$ we do not consider $\nue$ and $\bar\nue$.  We
adopt the approach given in Raffelt (2001).  The rate for the
absorption of a $\nu_\mu$ by inverse bremsstrahlung
$\rN\rN\nu_\mu\bar\nu_\mu\to\rN\rN$ is given by 
\be
\label{eq:s_brems_mfp}
\lambda^{-1}=
\frac{C_A^2 \GF^2}{2}\, n_B\, \frac{1}{2\epsilon}
\int\frac{d^3\bar k}{2\bar\epsilon\,(2\pi)^3}\,
f(\bar\epsilon)\,24\,\epsilon\,\bar\epsilon\,
S(\omega)\,,
\ee 
where $\omega=\epsilon+\bar\epsilon$, $n_{\rm B}$ is the nucleon
density, and $\abs{C_A} = 1.26/2$ as in the scattering case.  The
over-barred quantities belong to the $\bar\nu_\mu$ that is absorbed
together with the primary $\nu_\mu$.  

The energy-differential rate for emission of bremsstrahlung is
obtained by adjusting the phase-space blocking:
\be
\label{eq:s_brems_rate}
\frac{{\rm d}\dot n}{{\rm d}\epsilon}=
\frac{C_A^2 \GF^2}{2}\, n_B\, \frac{\epsilon^2}{2\pi^2}\left [
1-f(\epsilon) \right ]
\int\frac{d^3\bar k}{2\bar\epsilon\,(2\pi)^3}\,
\left [1-f(\bar\epsilon)\right ]\,24\,\epsilon\,\bar\epsilon\,
S(-\omega)
\ee 

The heuristic ansatz for $S(\omega)$ given by Raffelt (2001) has the
form of a Lorentzian
\be
\label{eq:lorentzian}
S(\omega)=\frac{2\Gamma}{\omega^2+\Gamma^2}\;\frac{2}{1+\exp(-\omega/T)}\;,\quad
\Gamma=\frac{\sqrt\pi\,\alpha_\pi^2 n_B\,T}{m^2\sqrt{mT+m_\pi^2}}\,,
\ee
with the constant $\alpha_\pi\approx (2 m/ m_\pi)^2/4\pi$, where $m$ is
the nucleon mass and $m_\pi$ the pion mass.  For carrying out the
integration in Equations~(\ref{eq:s_brems_mfp}) and
(\ref{eq:s_brems_rate}) the $\Gamma^2$ in the denominator of $S$
can be neglected.  A very detailed analysis of this bremsstrahlung
rate can be found in Raffelt (2001).

\section{Pair Annihilation}
\label{sec:PairAnn}

We now turn to the leptonic source reactions for $\numu\bar\numu$
pairs, $\re^+ \re^- \to \nu_\mu \bar\nu_\mu$ and $\nue \bar\nue\to
\nu_\mu \bar\nu_\mu$.  Traditionally, only $\re^+ \re^-$ pair
annihilation was implemented in SN simulations.  As one part of this
dissertation we have for the first time investigated the importance of
$\nue \bar\nue$ pair annihilation that turns out to be a factor of
2--3 more important (Buras, Janka, Keil, Raffelt, \& Rampp~2003a).

\begin{figure}[b]
\columnwidth=6.5cm
\plotone{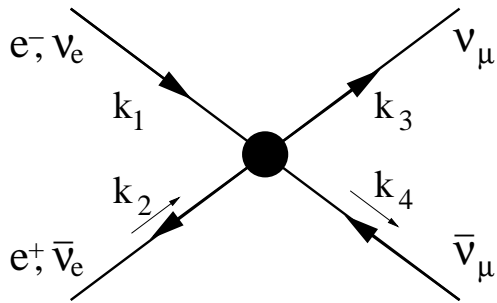}
\caption{\label{fig:feyn} Pair annihilation processes producing
$\nu_\mu\bar\nu_\mu$ pairs.}
\end{figure}

The matrix elements for both processes are identical up to coupling
constants while the phase-space integrations only differ by the
chemical potentials.  After summing over all spins and neglecting the
rest masses, the squared matrix element is
\be
\sum_{\rm spins}|{\cal M}|^2 =
8~\GF^2 \Bigl[(C_V+C_A)^2u^2 +(C_V-C_A)^2t^2\Bigr]
\label{eq:matrixel} 
\ee
with the Mandelstam variables $t=-2k_1\cdot k_3$ and $u=-2k_1\cdot
k_4$. The momenta are assigned to the particles as indicated in
Figure~\ref{fig:feyn}.  The weak interaction constants for  
$\re^+\re^-$ annihilation are
\be 
\label{eq:weakcoefel}
C_V = -\frac{1}{2}+2\sin^2\theta_{\rm W}\; ,\;\;\; C_A = -\frac{1}{2}
\ee
while for $\nue \bar\nue$ annihilation they are
\be 
\label{eq:weakcoefnu}
C_A = C_V = \frac{1}{2} \; .
\ee

For the interaction rates we have to perform the phase-space
integrations, using blocking factors for the final states and
occupation numbers for initial-state particles (Hannestad \& Madsen
1995, Yueh \& Buchler 1976a).  Three integrations remain that can not
be carried out analytically.

In order to perform the phase-space integrals we have to specify
chemical potentials for all interacting particles.  Mu- and
tau-leptons are almost absent in proto-neutron star atmospheres,
therefore, chemical potentials of the corresponding neutrinos can
only arise due to differences in the neutrino and anti-neutrino
interactions, i.e., weak magnetism.  This chemical potential builds up
dynamically, because anti-neutrinos escape more easily than neutrinos.
For the $\re^+\re^-$ reactions the local value of $\mu_\re$ can be
obtained from $\rho$, $T$, and $\Ye$ by inverting 
\be 
\frac{n_{\re^-}(\mu_\re)-n_{\re^+}(-\mu_\re)}{n_{\rm baryons}} 
= \Ye  \; ,
\ee
where $n_{\re^-}(\mu_\re)$ and $n_{\re^+}(-\mu_\re)$ are the number
densities of electrons and positrons, respectively.  The baryon number
density is 
\be
n_{\rm baryons} = \frac{\rho}{m_\rn (1-\Ye)+m_\rp\Ye}\,.
\ee 
For $\nue$ and $\bar\nue$ the chemical potential is obtained by the
relation 
\be 
\label{eq:EleChemPo}
\mu_{\nue} = \mu_\re + \mu_\rp - \mu_\rn\,,
\ee
with the chemical potentials $\mu_\rp$ and $\mu_\rn$ of protons and
neutrons, respectively.

The reason why the $\nue\bar\nue$ pair process is an important source
for $\numu\bar\numu$ pairs is that below the $\numu$ number sphere,
$\nue$ and $\bar\nue$ are basically a part of the stellar medium like
$\re^\pm$.  The energy sphere of $\nu_{\mu}$ lies always deeper inside
the star than the $\nue$ and $\bar\nue$ spheres
(Section~\ref{sec:ThermaDepthApp}).  Thus $\nue$ and $\bar\nue$ are in
LTE and are part of the medium as far as the transport of $\nu_\mu$ is
concerned. 

\begin{figure}[t]
\columnwidth=7.5cm
\plotone{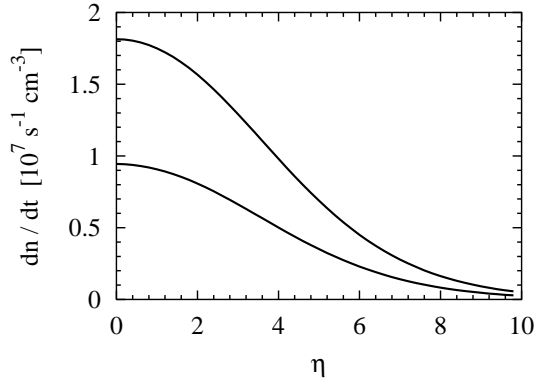}
\caption{\label{fig:easyexpl} Pair production rates by the process
$\nu_{\rm e}\bar\nu_{\rm e}\to\nu_{\mu}\bar\nu_{\mu}$ as a function of
$\eta_{\nu_{\rm e}}$ (upper line) and ${\rm e}^+{\rm
e}^-\to\nu_{\mu}\bar\nu_{\mu}$ as a function of $\eta_{\rm e}$ (lower
line).  We used $T=12$~MeV and $\eta_{\nu_\mu}=0$.}
\end{figure}

Until very recently $\re^+\re^-$ annihilation was considered to be the
most important leptonic source for $\numu$.  In most numerical
simulations it is still the only included creation mechanism for the
$\numu$ flavors.  Comparing the rates for SN conditions it is obvious
that $\nue\bar\nue$ pairs are more important.  In
Figure~\ref{fig:easyexpl} we show both leptonic pair rates as a
function of their respective degeneracy parameters.  The chemical
potential of $\numu$ was set to zero and $\nue$, $\bar\nue$, and
$\re^\pm$ are assumed to be in LTE.  In the region below the neutrino
spheres the degeneracy parameters obey $\eta_{\nue}=\eta_\re+\eta_\rp
- \eta_\rn < \eta_\re$ and therefore the neutrino pair process is
always more important than $\re^+\re^-$ annihilation.

\begin{figure}[t]
\columnwidth=9cm
\plotone{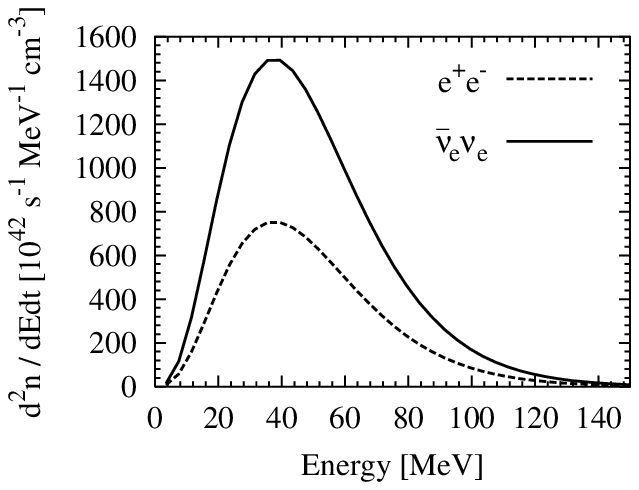}
\caption{\label{fig:energy-rates-eta0} Differential $\nu_\mu$ and
$\bar\nu_\mu$ production rates ${\rm d}^2n/{\rm d}\epsilon{\rm d}t$
vs.\ neutrino energy for $\eta_\re=\eta_{\nue}=0$ and $T=12$ MeV.}
\end{figure}

\begin{figure}[b]
\columnwidth=7.7cm
\plotone{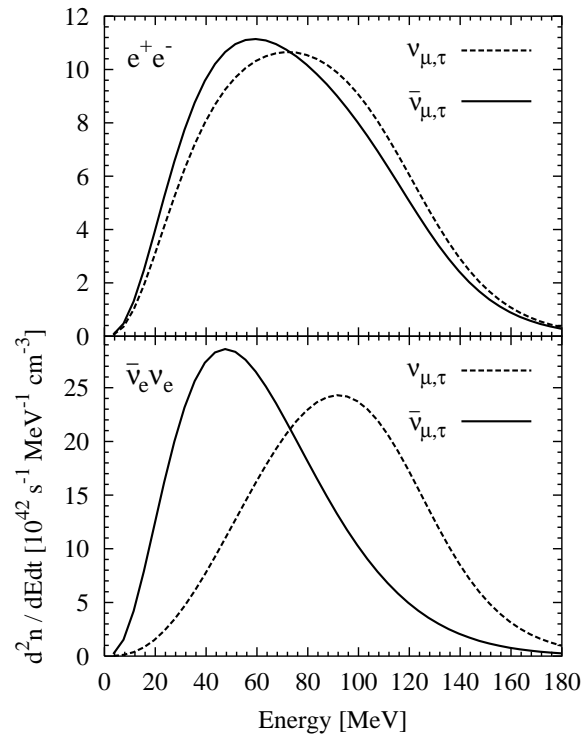}
\caption{\label{fig:energy-rates-eta10} Differential $\nu_\mu$ and
$\bar\nu_\mu$ production rates ${\rm d}^2n/{\rm d}\epsilon{\rm d}t$ 
for $\eta_{\re}=\eta_{\nue}=10$
and $T=12$ MeV.  Upper panel for $\re^+\rme^-\to\numu\bar\numu$, lower
panel for $\nu_{\re}\bar\nue\to\numu\bar\numu$.} 
\end{figure}

Another qualitative difference of both rates becomes apparent by
looking at the energy-differential production rate.  In the case of
vanishing degeneracy of $\re^\pm$ and $\nue$, $\bar\nue$ the
differential production rate ${\rm d}^2 n/{\rm d}\epsilon{\rm d}t$ is
equal for both $\numu$ and $\bar\numu$, as shown in
Figure~\ref{fig:energy-rates-eta0}.  In general, however, electrons
and electron neutrinos will be significantly degenerate.  For the
extreme case $\eta_{\rm e}=\eta_{\nu_{\rm e}}=10$ we show the
differential production rates in Figure~\ref{fig:energy-rates-eta10}.
In the upper panel both $\numu$ and $\bar\numu$ creation rates are
very similar for $\re^+\re^-$ annihilation.  This represents the fact
that the prefactors $u^2$ and $t^2$ in Equation~(\ref{eq:matrixel}) are
similar, namely $(C_{\rm V}+C_{\rm A})^2\approx 0.54^2$ and $(C_{\rm
V}-C_{\rm A})^2\approx 0.46^2$.  Comparing the rates for $\numu$ and
$\bar\numu$, which corresponds to exchanging $u$ and $t$, then has no
big effect.

For the $\nue\bar\nue$ process the situation is different.  In this
case $(C_{\rm V}+C_{\rm A})^2=1$ and $(C_{\rm V}-C_{\rm A})^2=0$ and
only $u^2$ contributes.  Interchanging $\bar\numu$ with $\numu$
corresponds to replacing $u^2$ by $t^2$.  Therefore, the kinematics for
particles and anti-particles are different.  Since in the case of
$\eta_{\nue}\neq 0$ also the distribution functions of particles and
anti-particles are different this leads to a difference in the energy
dependence of the $\numu$ and $\bar\numu$ production rate.  The lower
panel of Figure~\ref{fig:energy-rates-eta10} displays both rates.  The
numbers of produced $\numu$ and $\bar\numu$ are equal, of course.

The reason for the average energy of $\numu$ to be larger than of
$\bar\numu$ can be understood by looking at the reaction
$\nue\bar\nue\to\numu\bar\numu$ in the center of momentum (CM) frame.
The differential cross section is \be \frac{{\rm
d}\sigma}{{\rm d} \cos \theta} = \frac{\GF^2}{4 \pi}\,\epsilon^2
(1+\cos\theta)^2 \,, \ee where $\theta$ is the angle between the
ingoing $\nue$ and the outgoing $\nu_\mu$, or equivalently, between
the ingoing $\bar\nue$ and the outgoing $\bar\nu_\mu$.  Put another
way, forward scattering is favored and backward scattering
forbidden.  This is due to angular momentum conservation.  The ingoing
$\nu_{\rm e}$ and $\bar\nu_{\rm e}$ have opposite helicities and, in
the CM frame, opposite momenta, so that their combined spins add up to
1.  The same is true for the outgoing particles so that backward
scattering would violate angular momentum conservation.  In the rest
frame of the medium the ingoing $\nue$ tends to have energies of the
order of its Fermi energy, while the ingoing $\bar\nue$ tends to have
energies of order $T$.  Because forward scattering is favored, the
outgoing $\nu_\mu$ tends to inherit the larger energy of the ingoing
$\nu_{\rm e}$.

The differences of the source spectra, however, do not translate into
significant spectral differences of the $\nu_\mu$ and $\bar\nu_\mu$
fluxes emitted from the SN core. While pair annihilations and nucleon
bremsstrahlung are responsible for producing or absorbing neutrino
pairs and thus their equilibration with the stellar medium below the
``neutrino-energy sphere,'' other processes, notably $\nu_\mu {\rm
e}^\pm$ scattering and nucleon recoils, are more efficient for the
exchange of energy between neutrinos and the medium between the
equilibration and transport spheres.  In our numerical runs we will
find in fact that adding the new process to a SN simulation primarily
modifies the flux with only minor modifications of the spectrum.

The numerical implementation is vastly simplified by exploiting the
fact that $\nue$ and $\bar\nue$ are in LTE in the region where the
pair processes are effective.  Instead of using the actual neutrino
distributions in the phase-space integration we can use the
equilibrium distribution with the local medium temperature.  This
approximation breaks down at larger radii where the $\nue\bar\nue$
process is unimportant anyway.

In order to further reduce computation time one of the remaining three
phase-space integrations can be approximated by the analytic
expressions given in Takahashi, El~Eid, \& Hillebrandt (1978).  This
also requires simplifying the blocking factors.  With $\mu_\re =
-\mu_{\re^+} \ge 0$ we can approximate the positron occupation number
by a Maxwell-Boltzmann distribution.  For $\mu_\re/T \gsim 2$ this
holds to very good accuracy.  The greatest deviation is at $\mu_\re/T
=0$ and yields blocking factors too low by about 10\%.  However,
$\re^-$ and $\nue$ are always degenerate in the relevant regions.

\section{Scattering on Electrons and Electron Neutrinos}
\label{sec:LeptoScatt}

\begin{figure}[t]
\columnwidth=6.5cm
\plotone{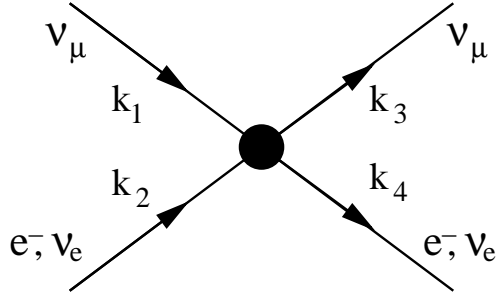}
\caption{\label{fig:feynsc} Leptonic scattering processes.}
\end{figure}

Even though the scattering reactions on $\re^\pm$ and on
$\nue$, $\bar\nue$ are closely related to the pair annihilations
discussed in
the previous section, scattering on $\re^\pm$ is more important than on
$\nue$ or $\bar\nue$.  The matrix elements for these reactions are
just the crossed versions of the leptonic pair processes,
\be 
\label{eq:matrixelsc}
\sum_{\rm spins}|{\cal M}|^2 =
8~\GF^2 \Bigl[(C_V+C_A)^2s^2 +(C_V-C_A)^2u^2\Bigr]
\ee
with the weak interaction coefficients of
Equations~(\ref{eq:weakcoefel}) or 
(\ref{eq:weakcoefnu}) for scattering on $\re^-$ or on $\nue$,
respectively.  For $s=2k_1\cdot k_2$ and $u=-2k_1\cdot k_4$ the
momenta are assigned to the particles according to
Figure~\ref{fig:feynsc}.  Crossing the matrix element
Equation~(\ref{eq:matrixelsc}) again by interchanging $u \lra t$, we obtain
scattering on $\re^+$ or $\bar\nue$. This is also true for scattering
of $\bar \nu_\mu$ on $\re^-$ or $\nue$; scattering of $\bar\nu_\mu$ on
$\re^+$ or $\bar\nue$ brings us back to Equation~(\ref{eq:matrixelsc}).
The numerical procedure for calculating the rates is the same as for
the pair-annihilation processes, given in the previous section.

\begin{figure}[b]
\columnwidth=7.5cm
\plotone{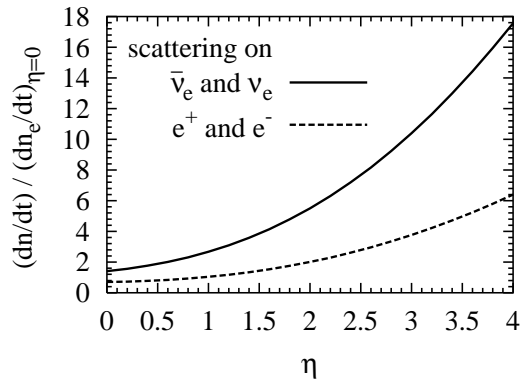}
\caption{\label{fig:easyexpl_scat} Thermally averaged scattering rate
for $\numu$ on $\re^\pm$ as a function of $\eta_{\re}$ (lower
line) and for $\numu$ on $\nue$ and $\bar\nue$ (upper line) as a
function of $\eta_{\nue}$. The rates are normalized to the scattering
rate on $\re^\pm$ at $\eta_{\re}=0$.  We used $T=12$~MeV and
$\eta_{\numu}=0$.}
\end{figure} 

In contrast to the pair rates of the previous section, the
differential scattering rates are monotonically rising functions of
the degeneracy.  In Figure~\ref{fig:easyexpl_scat} we show the rates
for $\numu$ scattering on $\re^\pm$ (lower line) and the rates for
$\numu$ scattering on $\nue$ or $\bar\nue$, normalized to the latter
rate.  Scattering on $\nue$ and $\bar\nue$ is always more frequent
than scattering on $\re^\pm$ in case of $\eta_{\nue}=\eta_{\rm
e}$. However, for realistic situations with $\eta_{\nue}<\eta_{\rm e}$
we expect that scattering on ${\rm e}^\pm$ has 1--2 times the rate of
scattering on $\nue$ and $\bar\nue$. Therefore, neutrino-neutrino
scattering is expected to be a relatively minor correction.  In our
numerical studies we will indeed find that this process has only a
small effect on the neutrino spectra and fluxes.

\section{Other Reactions}

Interactions that also take place inside a SN, but are unimportant for
our purpose, are absorption and emission of $\nue$ and $\bar\nue$ by
nuclei as well as scattering of neutrinos on nuclei.  In the vicinity
of the neutrino sphere all nuclei are disintegrated.  Reactions on
nuclei take place at larger radii and are therefore important for the
heating process behind the stalled shock wave.

Throughout the past literature, oftentimes the plasmon process was
taken into account as a source for neutrino pairs.  The plasmon is
an excitation of the medium that can emit a $\nu\bar\nu$ pair once it
decays.  In the relevant area close to the neutrino sphere this
process is several orders of magnitude less frequent than the other
pair processes.

\cleardoublepage

\chapter{Setting the Stage for Neutrino Transport}
\label{chap:SetStage}
Proto-neutron star atmospheres are characterized by their radial
profiles of temperature, nucleon density, and the number of electrons
per nucleon.  In order to probe a variety of possible atmospheric
structures we use two self-consistent profiles obtained by
hydrodynamic simulations and, in addition, power-law parameterizations
with constant electron fractions.  After defining the optical depth
and an effective mean free path for thermalizing processes we
calculate the thermalization depths, i.e., the last points of
interaction for all the thermalizing processes in each of the
background models. This simple approach is a powerful test of the
relative importance of various neutrino-matter interactions and their
dependence on stellar parameters.

\section{Characterizing Proto-Neutron Stars}
\label{sec:PNSChar}

All the neutrino interaction rates with the background medium depend
on the density of scatterers and their temperature.  Both density
$\rho$ and temperature $T$ are functions of the radius.  In the region
relevant for neutrino spectra formation we assume that the
proto-neutron star consist only of
\begin{itemize}
\item protons 
\item neutrons
\item electrons
\item positrons
\item neutrinos of all flavors.  
\end{itemize}
In order to calculate every interaction rate for all radii,
we need the number density and the chemical potential for each of
these matter constituents as a function of radius in addition to the
temperature.  As mentioned earlier no muons and taus are present.
Therefore, usually we set the chemical potential of $\numu$ to
zero.  Only in the case of weak magnetism a small $\numu$ degeneracy
develops and we obtain a chemical potential for $\numu$.

Our neutron-star atmospheres are characterized by $T(r)$, $\rho(r)$,
and the electron fraction per baryon $\Ye(r)$.  Since $m_\re \ll
m_{\rn,\rp}$, the density is given by the number densities of neutrons
and protons times their masses
\be
\rho(r)=n_\rn \, m_\rn + n_\rp \, m_\rp \;.
\ee
In addition we use
\be
\Ye(r) = \frac{n_\rp}{n_\rn + n_\rp} \; 
\ee
to obtain the densities of neutrons and protons. By solving 
\be
n_{\rn,\rp}(r) = \int {\rm d} \epsilon \,
f_{\rn,\rp}(\epsilon,\mu_{\rn,\rp}) =\int {\rm d} \epsilon \,
\frac{\epsilon^2}{1+\exp(\frac{\epsilon-\mu_{\rn,\rp}}{T})}\;, 
\ee
we obtain the chemical potentials of protons and neutrons $\mu_\rn$
and $\mu_\rp$.

Similarly, we determine $\mu_\re$ by
\be
\label{eq:ElChemPo}
n_{\rm B} \Ye = n_{\re^-} - n_{\re^+} = \int {\rm d} \epsilon \left(
f(\epsilon,\mu_\re) - f(\epsilon,-\mu_\re) \right)\,.
\ee
A vanishing electron fraction does not mean that there are no
electrons present, but rather that there is an equal thermal
distribution of electrons and positrons.

After all other chemical potentials are known, $\mu_{\nue}$ is given by
Equation~(\ref{eq:EleChemPo})
\bea 
\mu_{\nue} = \mu_\re + \mu_\rp - \mu_\rn\;.\nonumber
\eea
This completes the set of thermodynamic quantities needed for the
transport of neutrinos in a SN core.  

\begin{figure}[b]
\columnwidth=8cm
\plotone{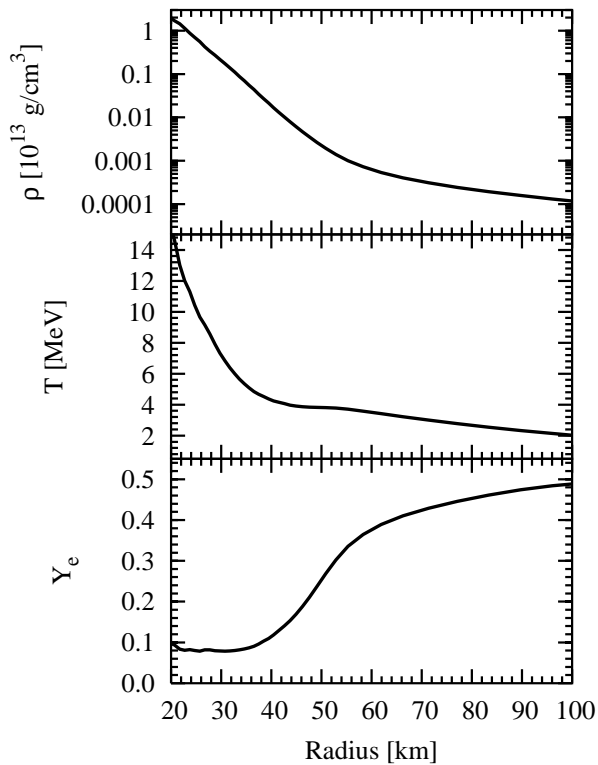}
\caption{\label{fig:Messermodel} Accretion-Phase Model~I, a SN model
324~ms after bounce from a Newtonian calculation (O.E.B.~Messer,
personal communication).}
\end{figure}

\begin{figure}[b]
\columnwidth=8cm
\plotone{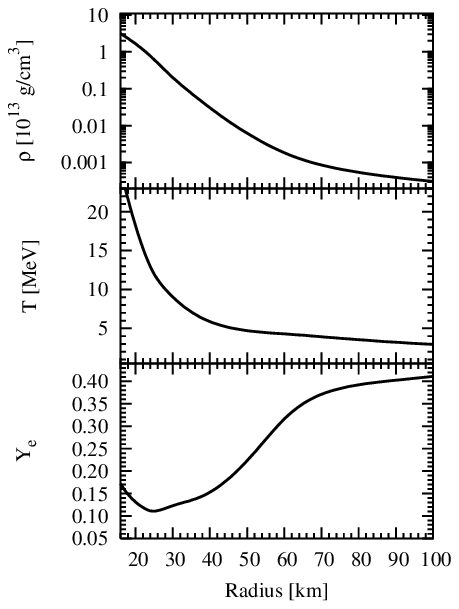}
\caption{\label{fig:Ramppmodel} Accretion-Phase Model~II, a SN core at
150~ms post bounce from a general-relativistic simulation.  (M.~Rampp,
personal communication).}
\end{figure}

\section{Proto-Neutron-Star Profiles}
\label{sec:PNSProfiles}

One possibility for our study of the influence of neutrino
interactions and the stellar background model on the emerging neutrino
spectra is to use proto-neutron star atmospheres that were obtained by
self-consistent hydrodynamic simulations.  As a first example we use a
model representative for the accretion phase; henceforth we will refer
to it as the ``Accretion-Phase Model~I'' (Fig.~\ref{fig:Messermodel}).
It was provided to us by O.~E.~B.~Messer and was already used in
Raffelt (2001) for a more schematic study.  Based on the Woosley \&
Weaver $15\,M_\odot$ progenitor model labeled s15s7b, the Newtonian
collapse simulation was performed with the SN code developed by
Mezzacappa et~al.\ (2001).  The snapshot is taken at 324~ms after
bounce when the shock is at about 120~km, i.e., the proto-neutron star
still accretes matter.  In this simulation the traditional
microphysics for $\nu_\mu$ transport was included, i.e., iso-energetic
scattering on nucleons, $\re^+\re^-$ annihilation and $\nu_\mu\re^-$
scattering.

As another self-consistent example (Accretion-Phase Model~II,
Fig.~\ref{fig:Ramppmodel}) we obtained a 150~ms post bounce model from
M.~Rampp (personal communication) that uses a very similar progenitor
(s15s7b2).  The simulation includes an approximate general
relativistic treatment in spherical symmetry as described by Rampp \&
Janka (2002).  The three neutrino flavors are transported with all
relevant interactions except $\nue\bar\nue \to \nu_{\mu}\bar\nu_{\mu}$
(Section~\ref{sec:ourflavorcomp} and Rampp et al.\ 2002).

When we use these self-consistent models with the same microphysics
that was used in the original models by Messer and Rampp we find good
agreement with their results.  Changing the microphysical input then
enables us to study the influence of the various interaction
processes (Chapter~\ref{chap:rates}) on the emerging neutrino spectra
in a systematic way, which has never been done before.  Since we are
simulating a static proto-neutron star and also vary the input
physics, our simulations are not self consistent.  However, once our
findings are implemented in the self-consistent hydrodynamic
simulations they qualitatively find the same results as we do (Buras
et al.~2003b).

The self-consistently obtained stellar background models are in the
relevant range to a good approximation just power laws for the radial
dependence of temperature and density.  As a different approach we can
thus parameterize the profiles in a simple way and thereby study the
influence of the stellar profiles.  Especially steep profiles that are
characteristic for the late phase of the proto-neutron star are not
available to us from self-consistent simulations.

We use two power-law profiles of the form
\begin{equation}
\label{eq:powerlaws}
\rho=\rho_0\left(\frac{r_0}{r}\right)^p,\quad 
T=T_0\left(\frac{r_0}{r}\right)^q,
\end{equation}
with a constant electron fraction per baryon $\Ye$.  We adjust
parameters such that $\langle\epsilon\rangle\approx 20$--25 MeV for
the emerging neutrinos to obtain model atmospheres in the ball park of
results from proto-neutron star evolution calculations.  We define a
``steep'' power-law model, corresponding to the one used by Raffelt
(2001), and a ``shallow'' one; the characteristics are given in
Table~4.1.  The shallow model could be
characteristic of a SN core during the accretion phase while the steep
model is more characteristic for the neutron-star cooling phase. The
constant electron fraction $\Ye$ is another parameter that allows us
to investigate the relative importance of the leptonic processes as a
function of the assumed $\Ye$.

\begin{center}
Table 4.1. \quad Characteristics of power-law models.

\vspace{.5cm}

\begin{tabular}{lll}

\hline
\hline

&\rule[-.35cm]{0cm}{.9cm}Steep&Shallow\\

\hline
$p$ \rule[0cm]{0cm}{.6cm}     & 10    & 5   \\
$q$                              & 2.5   & 1   \\
$q/p$                            & 0.25  & 0.2 \\
$\rho_0~\rm [10^{14}~g\,cm^{-3}]$ & 2.0   & 0.2 \\
$T_0~\rm [MeV]$                  & 31.66 & 20.0\\
$r_0~\rm [km]$\rule[-.2cm]{0cm}{.1cm}   & 10    & 10  \\

\hline

\end{tabular}
\end{center}

\vspace{.5cm}

\section{Optical Depth vs. Thermalization Depth}
\label{sec:odeVStherm}

By comparing the optical depths due to different neutrino interactions
we can assess the relative importance of the neutrino processes in
the stellar background model.  In principle the optical depth $\tau$
is defined as the integral of the inverse mfp $\lambda^{-1}$ (see
e.g.\ Shapiro \& Teukolsky 1983, Suzuki 1989):
\be
\tau_i(r,\epsilon)
=\int_r^\infty \!\! dr' \frac{1}{\lambda_i(r',\epsilon)}
=\int_r^\infty \!\! dr' 
\sigma_i(\epsilon) \,
n_i(r')\,,
\ee
where $i$ stands for any interaction channel, $\sigma_i$ is the
corresponding cross section, and $n_i$ the number density.  The
optical depth specifies how frequently a specific process occurs on a
neutrino's way out of the SN core.

Since we want to know how efficient a certain process is at exchanging
energy we have to take into account the qualitatively different nature
of the neutrino-matter interactions.  The relevant interaction
processes fall into three classes:
\begin{enumerate}
\item {\bf Creation processes}, that are able to keep neutrinos in
LTE.  They can create and absorb neutrinos and therefore also
contribute to the energy exchange between neutrinos and the background
medium.  
\item {\bf Scattering with large energy transfer}.  In this case the
neutrinos scatter on particles with a low mass compared to the
neutrino's momentum (i.e., $\re^\pm$ and neutrinos).
\item {\bf Scattering on nucleons}.  Here the energy transfer is very small
in a single scattering, The fact that allows for the effective mfp
description discussed in the previous section.
\end{enumerate}
All three classes contribute to the opacity, but only the first two
classes of reactions can exchange large amounts of energy in single
interactions.

Making use of this qualitative difference between nucleon scattering
and the other interactions we can define the location (thermalization
depth) where a neutrino of given energy last exchanged energy
efficiently with the medium by a reaction such as
$\nu_\mu\re^-\to\re^-\nu_\mu$.  Looking at the example of $\numu
\re^-$ scattering we have two different mfps, namely one for nucleon
scattering $\lambda_T$ and one for the energy exchanging $\numu \re^-$
process  $\lambda_E$.  The total mfp is then given by $\lambda_{\rm
tot}=1/(\lambda_T^{-1}+ \lambda_E^{-1})$.  Between two energy
exchanging scatterings the neutrino undergoes many isoenergetic
scatterings.  This yields an effective mfp for the energy exchange
reaction that is larger than $\lambda_E$.  

The way of a neutrino through the medium can be considered a random
walk. Therefore the traveled distance scales as the square root of
time that is proportional to $\sqrt{N}$, where $N$ represents the
number of interactions.  For each reaction the probability of energy
exchange is $\lambda_{\rm tot} / \lambda_E$, so the inverse is the
number of reactions until the next energy exchange.  Altogether the
traveled distance from one energy-exchange reaction to the other
corresponds to the effective mfp of the energy exchanging process:
\be
\lambda_E^{\rm eff}=\lambda_{\rm tot} \sqrt{\frac{\lambda_E}{\lambda_{\rm
tot}}} = \sqrt{\frac{\lambda_E}{\lambda_T^{-1}+ \lambda_E^{-1}}} 
\ee
With the effective mfp we can then calculate the optical depth for
thermalization
\be
\label{eq:definetautherm}
\tau_{\rm therm}(r,\epsilon)
=\int_r^\infty \!\! dr'\sqrt{\frac{1}{\lambda_E(r',\epsilon)}
\left[\frac{1}{\lambda_T(r',\epsilon)}+\frac{1}{\lambda_E(r',\epsilon)}\right]}\,.
\ee
When the optical depth becomes less than of order unity, the
corresponding process is no longer relevant.  The corresponding radius
is referred to as the thermalization depth:
\be
\label{eq:defineRtherm}
\tau_{\rm therm}(R_{\rm therm})=\frac{2}{3}\,,
\ee
where $R_{\rm therm}$ depends on the neutrino energy $\epsilon$.

With this concept we can determine the energy-dependent sphere
where a certain class of reactions becomes inefficient.  For a
specific neutrino energy the corresponding radius is given by the
largest $R_{\rm therm}$ of any reaction in the class.  For the first
class this sphere corresponds to the number sphere, where particle
creation freezes out, as defined in Section~\ref{sec:NuEmission}.  The
second class of reactions freezes out at the radius that determines
the energy sphere.

For the third class, i.e., neutrino-nucleon scattering, we are not
able to obtain the thermalization depth.  In each reaction the energy
transfer is very small, thus many reactions are necessary to
thermalize the neutrino.  We can therefore not determine the location
of the transport sphere, i.e., the radius at that neutrino-nucleon
scattering becomes inefficient.

As we show in the next section, due to the energy dependence it is not
possible to define a unique neutrino sphere. Even for $\nue$ and
$\bar\nue$, where the $\beta$-processes dominate and therefore the
only relevant sphere is their number sphere, its location is strongly
energy dependent.  The other flavors are emitted at their number
sphere, lose energy in elastic scattering up to their transport
sphere, and then diffuse before they start streaming freely from their
transport sphere.  Compared to $\nue$ and $\bar\nue$ one would refer
to the number sphere as neutrino sphere of $\numu$.  The spectrum
emitted at this neutrino sphere, however, is very different from the
emitted spectrum.

In order to illustrate the $\numu$ transport let us consider
bremsstrahlung and iso-energetic scattering on nucleons only.  For
nucleon bremsstrahlung the energy dependence is rather weak and at the
same time the density dependence is very strong.  Therefore, to a good
approximation one may picture the bremsstrahlung-number sphere as a
blackbody surface that injects neutrinos into the scattering 
atmosphere and absorbs those scattered back (Raffelt 2001).  With
iso-energetic scattering on nucleons, the neutrino flux and spectrum
emerging from the transport sphere is then easily understood in terms
of the energy-dependent transmission probability of the blackbody
spectrum launched at the number sphere, that is in this simple picture
equal to the energy sphere.  The transport cross section scales as
$\epsilon^2$, implying that the transmitted flux spectrum is shifted
to lower energies relative to the temperature at the energy
sphere.  This simple ``filter effect'' accounts surprisingly well for
the emerging flux spectrum (Raffelt 2001).  For typical conditions the
mean flux energies are 50--60\% of those corresponding to the
blackbody conditions at the energy sphere.

Moreover, it is straightforward to understand that the effective
temperature of the emerging flux spectrum is not overly sensitive to
the exact location of the number sphere.  If the pair processes are
somewhat more effective, the number sphere is at a larger radius with
a lower medium temperature.  However, the scattering atmosphere has a
smaller optical depth so that the higher-energy neutrinos are less
suppressed by the filter effect, partly compensating the smaller
energy-sphere temperature.  For typical situations Raffelt (2001)
found that changing the bremsstrahlung rate by a factor of 3 would
change the emerging neutrino energies only by some 10\%.  This finding
suggests that the emitted average neutrino energy is not overly
sensitive to the details of the pair processes.  

With the complete set of interactions neutrinos undergo energy
exchanging scattering reactions below the energy sphere before they
enter the diffusive regime.  Therefore the mean energy of the emerging
flux is even less sensitive to the exact location of the number sphere
than found by Raffelt (2001).  In Section~\ref{sec:SteepPowerLaw} we
show that changing the bremsstrahlung rate by a factor of three up or
down has almost no effect on the mean energy of the emerging neutrino
flux.


\section{Thermalization Depths in Our Stellar Models}
\label{sec:ThermaDepthApp}

\begin{figure}[b]
\columnwidth=8cm
\plotone{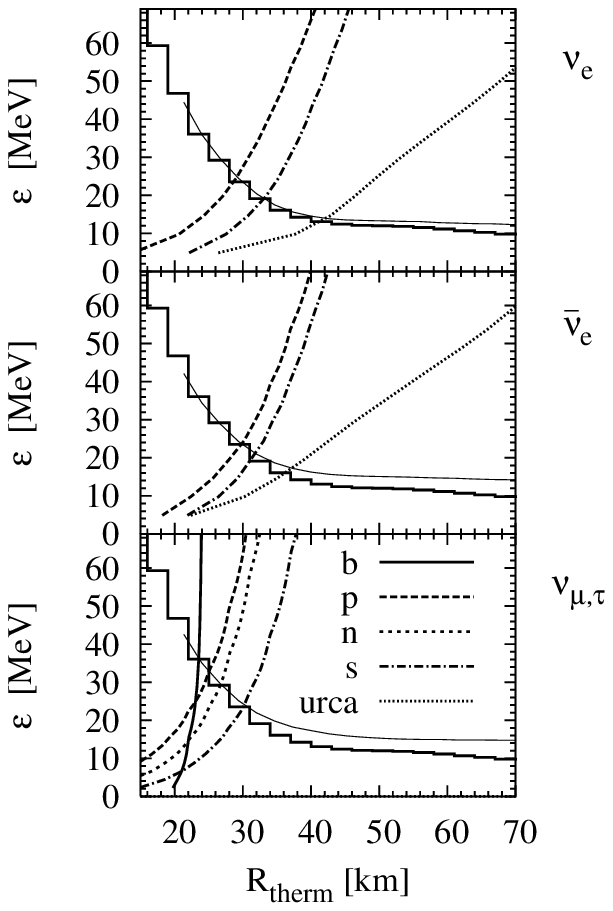}
\caption{\label{fig:rthermrealistic} $R_{\rm therm}$ as a function of
neutrino energy $\epsilon$ for our Accretion-Phase Model~I.  From top
to bottom the panels show the results for $\nue$, $\bar\nue$, and
$\nu_\mu$.  Energy exchanging processes: bremsstrahlung (solid line),
$\re^+\re^-$ annihilation (dashed), $\nue \bar\nue$ annihilation
(dotted), and scattering on $\re^\pm$ (dash-dotted). ``Urca'' denotes
the charged-current reaction of $\nue$ and $\bar\nue$ on nucleons.
The steps represent $\langle\epsilon\rangle = 3.15\,T$; the thin line
close to $3.15\,T$ is the mean energy of the flux
$\langle\epsilon\rangle_{\rm flux}$, cf. Chapter~\ref{chap:specchar}.}
\end{figure}

\begin{figure}[b]
\columnwidth=8cm
\plotone{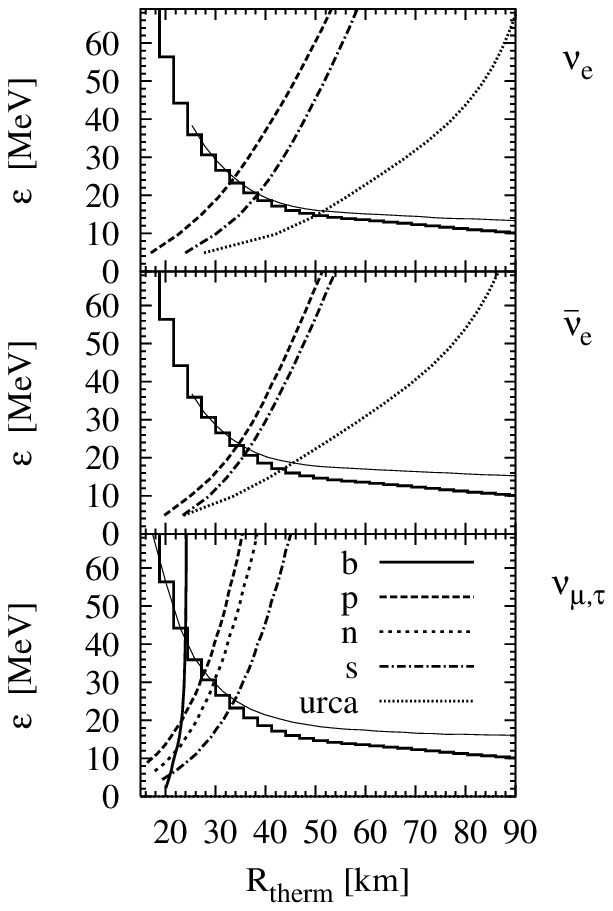}
\caption{\label{fig:rthermrealistic1} Same as
Figure~\ref{fig:rthermrealistic} for the Accretion-Phase Model~II.}
\end{figure}

We now turn to applying the concept of thermalization depths to the
neutron-star atmospheres described above
(Section~\ref{sec:PNSProfiles}).  We consider the neutrino mfp for
nucleon bremsstrahlung $\rN\rN\to\rN\rN\nu_\mu\bar\nu_\mu$, pair
annihilation $\re^+\re^-\to \nu_\mu\bar\nu_\mu$ and $\nue\bar\nue \to
\nu_\mu\bar\nu_\mu$, and scattering on charged leptons $\nu_\mu
\re^\pm\to\re^\pm\nu_\mu$.  The reaction rates used were described in
Chapter~\ref{chap:rates}.

In Figures~\ref{fig:rthermrealistic} and \ref{fig:rthermrealistic1} we
give the thermalization depth $R_{\rm therm}$ as a function of
neutrino energy $\epsilon$ for the two hydrodynamically
self-consistent accretion-phase models.  From top to bottom the panels
show the results for $\nue$, $\bar\nue$, and $\nu_\mu$, respectively.
The step-like curves represent the temperature profiles in terms of
the mean neutrino energy, $\langle\epsilon_{\nu}\rangle=3.15\,T$ for
non-degenerate neutrinos at the local medium temperature; the steps
correspond to the radial zones of our numerical simulation.
Slightly above these steps the thin line represents the mean energy of
neutrinos flowing in radial direction obtained by our numerical
simulation.  The other curves represent $R_{\rm therm}$ for
bremsstrahlung (b), $\re^+ \re^-$ annihilation (p), $\nue\bar\nue$
annihilation (n), and scattering on $\re^\pm$ (s).  In the case of
$\nue$ and $\bar\nue$ we do not include bremsstrahlung and
$\nue\bar\nue$ annihilation. Particle creation is dominated by the
charged current reactions on nucleons (urca).

\begin{figure}[b]
\columnwidth=8cm
\plotone{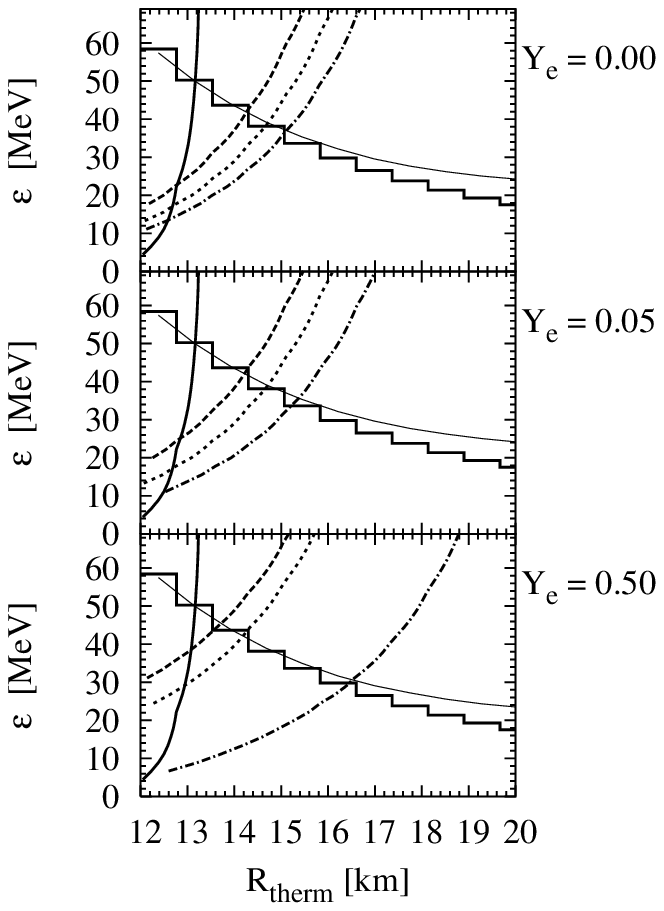}
\caption{\label{fig:rthermsteep} $R_{\rm therm}$ for $\nu_\mu$ in the
steep power-law model with the indicated values of $\Ye$. This figure
corresponds to the bottom panel of Figure~\ref{fig:rthermrealistic}.}
\end{figure}

\begin{figure}[b]
\columnwidth=8cm
\plotone{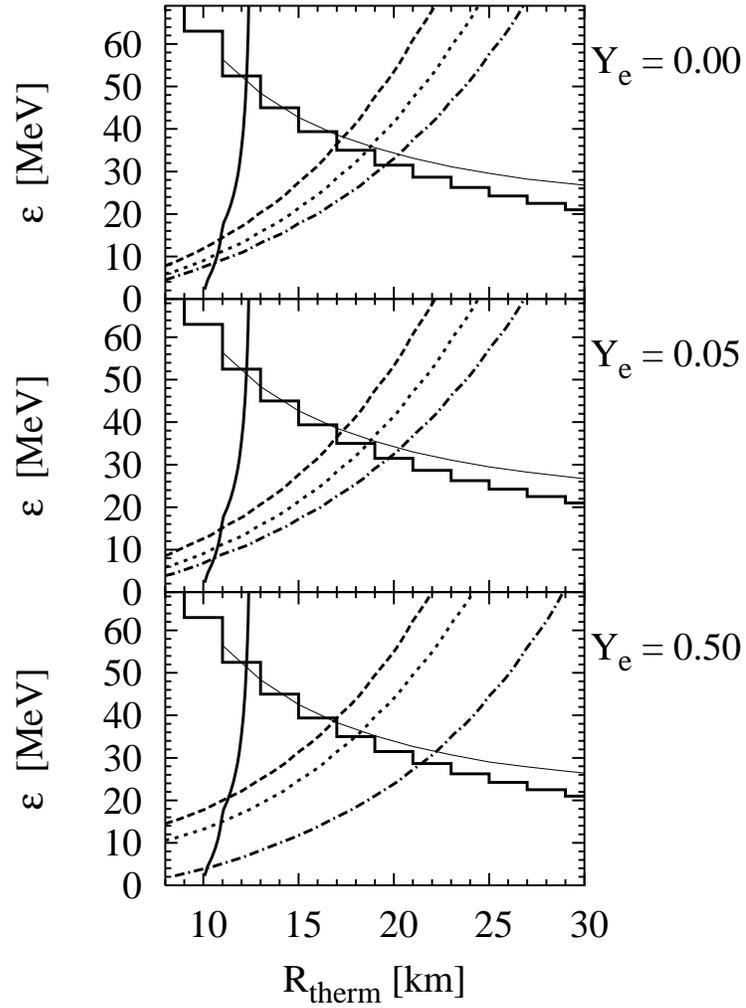}
\caption{\label{fig:rthermshallow}
 Same as Fig.~\ref{fig:rthermsteep} for the shallow 
power-law model.}
\end{figure}

For the power-law models we show $R_{\rm therm}$ for $\nu_\mu$ in
Figures~\ref{fig:rthermsteep} and~\ref{fig:rthermshallow}.  The
different panels correspond to the indicated values of the electron
fraction $\Ye$.  Note that $\Ye$ represents the net electron density
per baryon, i.e., the $\re^-$ density minus that of $\re^+$ so that
$\Ye=0$ implies that there is an equal thermal population of $\re^-$
and $\re^+$.

The $\nu_\mu$ absorption rate for the bremsstrahlung process varies
approximately as $\epsilon^{-1}$, the $\nu_\mu\rN$ transport cross
section as $\epsilon^2$ so that the inverse mfp for thermalization
varies only as $\epsilon^{1/2}$.  This explains why $R_{\rm therm}$
for bremsstrahlung is indeed quite independent of $\epsilon$.
Therefore, bremsstrahlung alone allows one to specify a rather
well-defined number sphere.  The other processes depend much more
sensitively on $\epsilon$ so that a mean energy sphere is much
less well defined.

Both electron scattering and the leptonic pair processes are so
ineffective at low energies that true LTE can not be established even
for astonishingly deep locations.  Bremsstrahlung easily ``plugs''
this low-energy hole so that one can indeed expect LTE for all
relevant neutrino energies below a certain radius.  For higher
energies, the leptonic processes dominate and shift the energy sphere
to larger radii than bremsstrahlung alone.  The relative importance of
the various processes depends on the density and temperature profiles
as well as $\Ye$.

To assess the role of the various processes for the overall spectra
formation one needs to specify some typical neutrino energy.  One
possibility would be $\langle\epsilon\rangle$ for neutrinos in LTE.
Another possibility is the mean energy of the neutrino flux, in
particular the mean energy of those neutrinos which actually leave the
star.  In Figures~\ref{fig:rthermrealistic}--\ref{fig:rthermshallow}
the thin line represents this flux mean energy.  In the case of
$\numu$ it is clearly visible that even after the last energy
exchanging process, i.e., $\re^\pm$ scattering, became ineffective the
mean energy of the flux still drops.  This drop is due to scattering
on nucleons.  Among the pair processes it is always the $\nue\bar\nue$
annihilation process that has the largest $R_{\rm therm}$ in the range
of flux energy.  However, especially in the steep power-law profile
the intersection of the bremsstrahlung's and $\nue\bar\nue$
annihilation's thermalization depths are rather close at the mean flux
energy.  The $\re^+\re^-$ process is always less important than
$\nue\bar\nue$, as was already discussed in
Section~\ref{sec:PairAnn}.  In our numerical simulations we find the
same results (Chapter~\ref{chap:importance}).

The final shape of the $\numu$ spectra and therefore also their mean
energy cannot depend much on any of the $\numu$-creation processes.
The energy sphere represented by $R_{\rm therm}$ of $\re^\pm$
scattering stretches in all profiles far beyond the other
thermalization depths.  Additionally, also the scattering on nucleons
has a significant impact on the mean energies.  Its importance
compared to electron scattering is very difficult to guess.  Recall,
it is not possible to define a thermalization depth for nucleon
recoils.  Both processes are qualitatively very different in that
neutrinos transfer only a small fraction of their energy per nucleon 
scattering.  Numerically we have compared the importance of all
processes and present our results in Chapter~\ref{chap:importance}.

\cleardoublepage

\chapter{Characterizing Non-Thermal Neutrino Spectra}
\label{chap:specchar}

The neutrino fluxes emerging from a SN core are not thermal.  There
are numerous ways to describe spectral characteristics with a few
parameters, for example energy moments.  For many applications it is
useful to have an analytic description of the spectra.  Throughout the
literature it is common to use an nominal Fermi-Dirac distribution.
We present a simpler two-parameter fit that describes our Monte Carlo
data better, and compare both descriptions.

\section{Global Parameters}

While a SN core can be crudely understood as a blackbody source for
neutrinos of all flavors, the escaping fluxes are not strictly thermal.
A number of observables have been defined in the literature in order
to describe the properties of these spectra.

Let us assume $f(\epsilon,\mu)$ is the distribution of neutrinos as a
function of energy $\epsilon$ and $\mu=\cos\theta$, where $\theta$ is
the angle between the radial direction and the direction of motion.
Far away from the SN core neutrinos stream freely in radial
direction, therefore $\mu=1$.

As the most intuitive quantity characterizing the spectrum the mean
energy is commonly used,
\be
\label{eq:meanenergy}
\langle\epsilon\rangle=
\frac{\int_0^\infty d\epsilon\,\epsilon\,\int_{-1}^{+1}d\mu\,
f(\epsilon,\mu)}
{\int_0^\infty d\epsilon\,\int_{-1}^{+1}d\mu\,f(\epsilon,\mu)}\,.
\ee 
In this ratio the numerator is the energy density and the denominator
is the number density.

Within the number sphere neutrinos are in LTE, but in the outer
regions some net flux of neutrinos develops.  Therefore, it is often
useful to extract spectral characteristics for those neutrinos which
are actually flowing by removing the isotropic part of the
distribution.  Specifically, we define the mean flux energy by
\be
\label{eq:fluxenergy}
\langle\epsilon\rangle_{\rm flux}=
\frac{\int_0^\infty d\epsilon\,\epsilon\,\int_{-1}^{+1}d\mu\,\mu\,
f(\epsilon,\mu)}
{\int_0^\infty d\epsilon\,\int_{-1}^{+1}d\mu\,\mu\,f(\epsilon,\mu)}\,.
\ee 
Outside the transport sphere all neutrinos will flow essentially in
the radial direction, implying that the angular distribution becomes a
delta-function in the forward direction so that
$\langle\epsilon\rangle_{\rm flux}=\langle\epsilon\rangle$. In the
trapping regions the two averages are very different because the
distribution function is dominated by its isotropic term.  However,
for most discussions only the emerging spectra are relevant and both
definitions of the mean energy, Equations~(\ref{eq:meanenergy}) and
(\ref{eq:fluxenergy}) are equivalent.  The right hand side of
Equation~(\ref{eq:fluxenergy}) then corresponds to the ratio of
luminosity to number flux.

Deep inside the core, below the number sphere neutrinos are in LTE.
With a vanishing chemical potential their distribution function is
given by
\be
\label{eq:eqdist}
f(\epsilon,\mu) = \frac{\epsilon^2}{1+\exp(\epsilon/T)}
\ee
and therefore
\be\label{eq:avenergytot}
\langle\epsilon\rangle=\frac{7\pi^4}{180\,\zeta_3}\,T
\approx3.1514\,T\,.
\ee
We use this relation to compare the numerically obtained mean energy
to the medium temperature, cf.\  e.g.\ 
Figure~\ref{fig:rthermrealistic}.

Since the emerging spectra are not thermal, additional characteristics
are useful.  A rather general way to characterize the
spectrum beyond mean values is by its energy moments (Janka \&
Hillebrandt 1989a)
\be
\label{eq:energymoments}
\langle\epsilon^n\rangle=
\frac{\int_0^\infty d\epsilon\,\epsilon^n\,\int_{-1}^{+1}d\mu\,
f(\epsilon,\mu)}
{\int_0^\infty d\epsilon\,\int_{-1}^{+1}d\mu\,f(\epsilon,\mu)}\,.
\ee 
Usually, values are given only up to n=2 in the literature, i.e., the
mean energy and the second moment.  

With the first two moments an easy way to see by how much a spectrum
deviates from being thermal is the pinching parameter, defined by
Raffelt (2001).  It is basically the ratio of the first two moments
\be
\label{eq:pinch}
p\equiv \frac{1}{a}\,
\frac{\langle\epsilon^2\rangle}{\langle\epsilon\rangle^2}\,,
\ee
normalized such that a Fermi-Dirac distribution with vanishing
chemical potential yields $p=1$.
Therefore, 
\be
\label{eq:pinchingnorm}
a\equiv\frac{\langle\epsilon^2\rangle_{\rm
FD}}{\langle\epsilon\rangle_{\rm FD}^2} 
=\frac{486000\,\zeta_3\zeta_5}{49\,\pi^8}\approx1.3029\,.
\ee
For a Maxwell-Boltzmann distribution this quantity would be 4/3. 
A spectrum that is thermal up to its second moment has $p=1$, while
$p<1$ signifies a pinched spectrum (high-energy tail suppressed),
$p>1$ an anti-pinched spectrum (high-energy tail enhanced).

In some publications the root-mean-square energy
$\langle\epsilon\rangle_{\rm rms}$ is given instead of the average
energy.  The rms-energy involves the ratio of third and first energy
moment 
\be
\label{eq:rmsdef}
\langle\epsilon\rangle_{\rm rms}=\sqrt{
{\int_0^\infty\!d\epsilon\int_{-1}^{+1}\!d\mu\,
\epsilon^3f(\epsilon,\mu) \over
\int_0^\infty\!d\epsilon\int_{-1}^{+1}\!d\mu\,
\epsilon\,f(\epsilon,\mu)} } =
\sqrt{\frac{\langle\epsilon^3\rangle}{\langle\epsilon\rangle}}\,.
\ee
This characteristic spectral energy is useful for estimating 
the energy transfer from neutrinos to the stellar medium in
reactions with cross sections proportional to $\epsilon^2$.  However,
comparing the mean energy defined in Equation~(\ref{eq:meanenergy})
and the rms-energy is difficult, because the relation of both
quantities depends on the spectral shape.

\section{Analytic Fits}

For applications like neutrino oscillations it is useful to have an
analytic fit to the SN neutrino spectra.  We present a two-parameter
fit that reproduces the first five or more moments of our numerically
obtained spectra to a very good accuracy
(Section~\ref{sec:High-Statistics}).  In addition to the overall
normalization we introduce the parameters $\alpha$ and $\bar\epsilon$
by defining
\be
\label{eq:powerlawfit}
f_\alpha(\epsilon)=\frac{1}{c}\left(\frac{\epsilon}{\bar\epsilon}\right)^\alpha 
{\rm e}^{-(\alpha+1)\, \epsilon/\bar\epsilon}\,,
\ee 
with the normalization
\be
\label{eq:alphanorm}
c=\int_0^\infty {\rm d}\epsilon \; f_\alpha(\epsilon)=
(\alpha+1)^{-(\alpha+1)} \;\Gamma(\alpha+1)\;\bar\epsilon \,.
\ee
The average energy is always $\langle\epsilon\rangle=\bar\epsilon$,
while $\alpha$ controls the pinching of the function.  Just like the
mean energy all observables discussed in the previous section can be
expressed in terms of $\alpha$ and $\bar\epsilon$ by simple analytic
relations.

The pinching parameter $p$ is closely related to the width
\be
\label{eq:alphawidth}
w=\sqrt{\langle\epsilon^2\rangle-\langle\epsilon\rangle^2}
=\sqrt{a\;p-1}\;\langle\epsilon\rangle
=\sqrt{\frac{1}{1+\alpha}}\;\bar\epsilon\,,
\ee
with $a\approx 1.3029$ as defined in Equation~(\ref{eq:pinchingnorm}).
The relation for the pinching parameter can be read off
Equation~(\ref{eq:alphawidth}),
\be
a\;p=
\label{eq:pinchingalpharelation}
\frac{\langle\epsilon^2\rangle}{\langle\epsilon\rangle^2}
=\frac{2+\alpha}{1+\alpha}\,.
\ee

For the more general case of the ratios of arbitrary neighboring
moments we find
\be
\label{eq:momentsalpharelation}
\frac{\langle\epsilon^n\rangle}{\langle\epsilon^{n-1}\rangle}
= \frac{n+\alpha}{1+\alpha} \;\bar\epsilon \,,
\ee
that can be easily generalized to arbitrary ratios of moments with
Equation~(\ref{eq:alphanorm}).  

\begin{figure}[t]
\columnwidth=12cm
\plotone{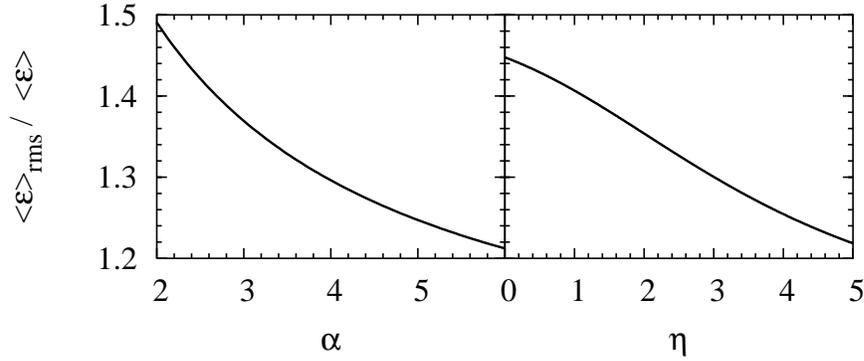}
\caption{\label{fig:ErmsAlphaEta} The ratio of rms energy and mean
energy.  The left panel shows the dependence on $\alpha$ as given in
Equation~(\ref{eq:ErmsAlpha}); in the right panel we plot the dependence on
$\eta$.}
\end{figure}

Taking the ratio of third and first energy moment yields the rms-energy
\be
\label{eq:ErmsAlpha}
\langle\epsilon\rangle_{\rm rms}=\sqrt{
\frac{(2 + \alpha)(3 + \alpha)}{(1 + \alpha)^2}}\;\bar\epsilon\,.
\ee
As shown in in the left panel of Figure~\ref{fig:ErmsAlphaEta},
$\langle\epsilon\rangle_{\rm rms}$ is always some 25--45\% larger than
$\langle\epsilon\rangle$ in the range of $\alpha=2$--5, that turns
out to be the parameter range obtained by numerical simulations.
Therefore, the two different energy averages contain information on
the spectral shape.

From any three moments it is possible
to obtain the normalization, $\alpha$, and $\bar\epsilon$.  From all
our studies we found that the spectra are very well described by these
three parameters.

In the literature one frequently encounters a nominal Fermi-Dirac
distribution characterized by a temperature $T$ and a degeneracy
parameter $\eta$ according to 
\be\label{eq:FermiDiracfit}
f_\eta(\epsilon)=
\frac{\epsilon^2}{1+\exp\left(\frac{\epsilon}{T}-\eta\right)}
\ee
as a representation of the SN neutrino spectrum (Janka \& Hillebrandt
1989a).  The functional form is motivated by the equilibrium
distribution of neutrinos inside the star.  The values of the
parameters $T$ and $\eta$, however, are not easy to interpret, because
the emerging spectra are not thermal.

\begin{figure}[b]
\columnwidth=7.5cm
\plotone{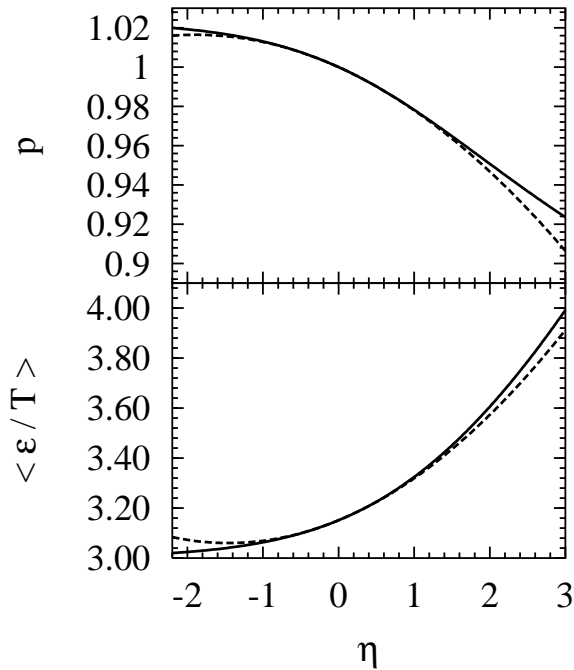}
\caption{\label{fig:fermidirac} Mean energy and pinching parameter
as a function of the degeneracy parameter for a Fermi-Dirac
distribution.  As dashed lines we show the expansions given in
Equation~(\ref{eq:fdexpansion}).}
\end{figure}

The functional form of $f_\eta$ is more complicated than
$f_\alpha$.  Most calculations involving $f_\eta$ have to be carried
out numerically and/or approximately.  In Figure~\ref{fig:fermidirac}
we show $\langle\epsilon/T\rangle$ and $p$ as a function of $\eta$. Up
to second order, expansions are
\bea
\label{eq:fdexpansion} 
\langle\epsilon/T\rangle&\approx& 3.1514+0.1250\,\eta+0.0429\,\eta^2
\nonumber\\
p&\approx& 1 - 0.0174\,\eta -0.0046\,\eta^2\,.
\eea
These expansions are shown in Figure~\ref{fig:fermidirac} as dashed
lines.

There are no analytic expressions for the energy moments and therefore
also the rms-energy can only be obtained numerically.  For the special
case of $\eta=0$ the expression is 
\be
\langle\epsilon\rangle_{\rm rms}=\sqrt{\frac{930}{441}}\,\pi\,T
\approx 4.5622\,T\,.
\ee
With Equation~(\ref{eq:avenergytot}) this corresponds to
$\langle\epsilon\rangle \approx 0.691\langle\epsilon\rangle_{\rm
rms}$, but recall that SN neutrino spectra are not thermal spectra and
$\eta\neq 0$.  The numerically obtained dependence of the ratio of rms
and mean energy is displayed in the right panel of
Figure~\ref{fig:ErmsAlphaEta}.  It is rather similar to the case of
the $\alpha$ fit.

In view of analytic simplicity the $\alpha$ fit is certainly
superior, but the shapes of both functions are rather similar.  The
numerically determined spectra are similarly well described by both
$f_\eta$ and $f_\alpha$.  As we show in the following for a reasonable
range for the parameters both functions differ by no more than about
10\%.

\begin{figure}[h!]
\columnwidth=7.3cm
\plotone{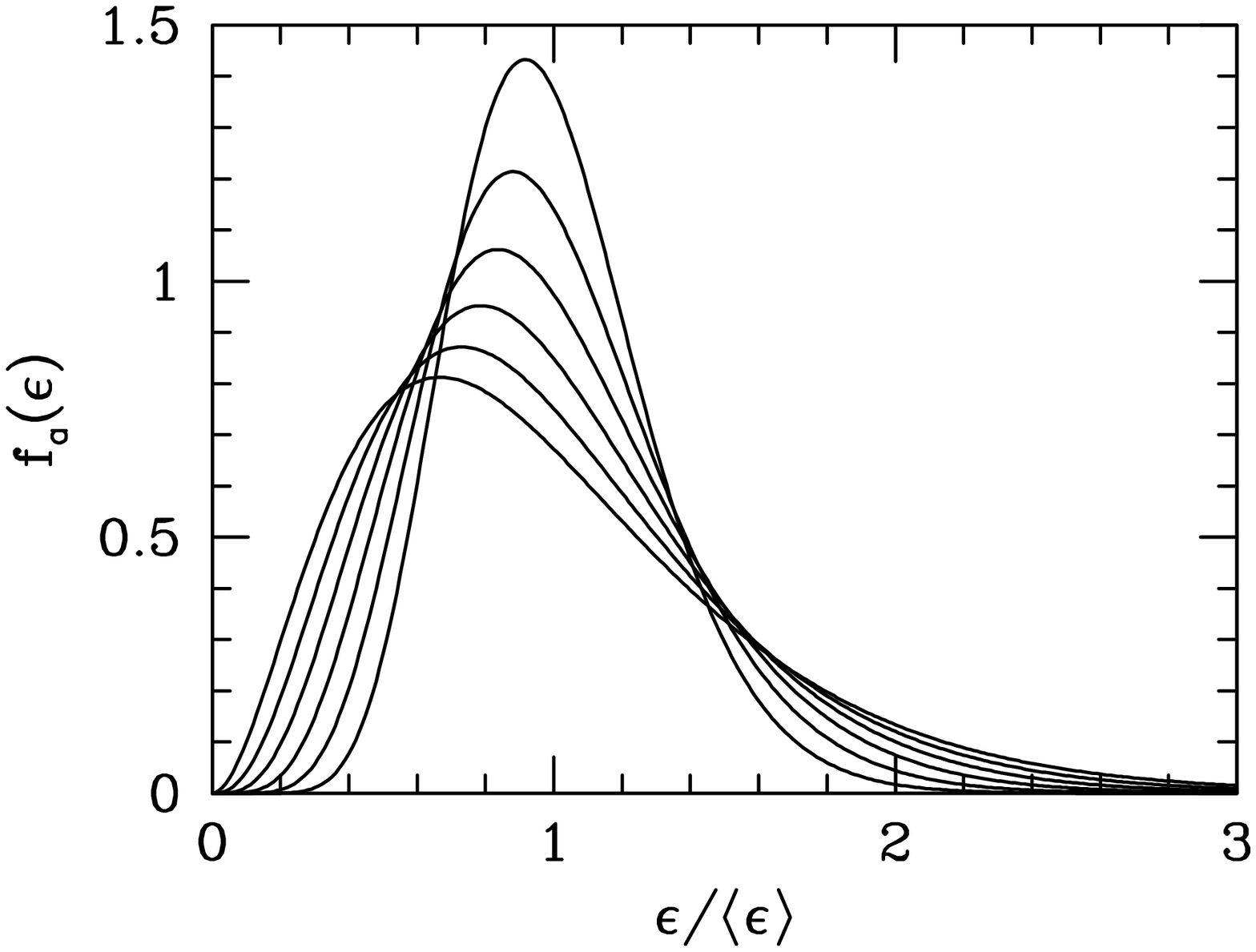}

\plotone{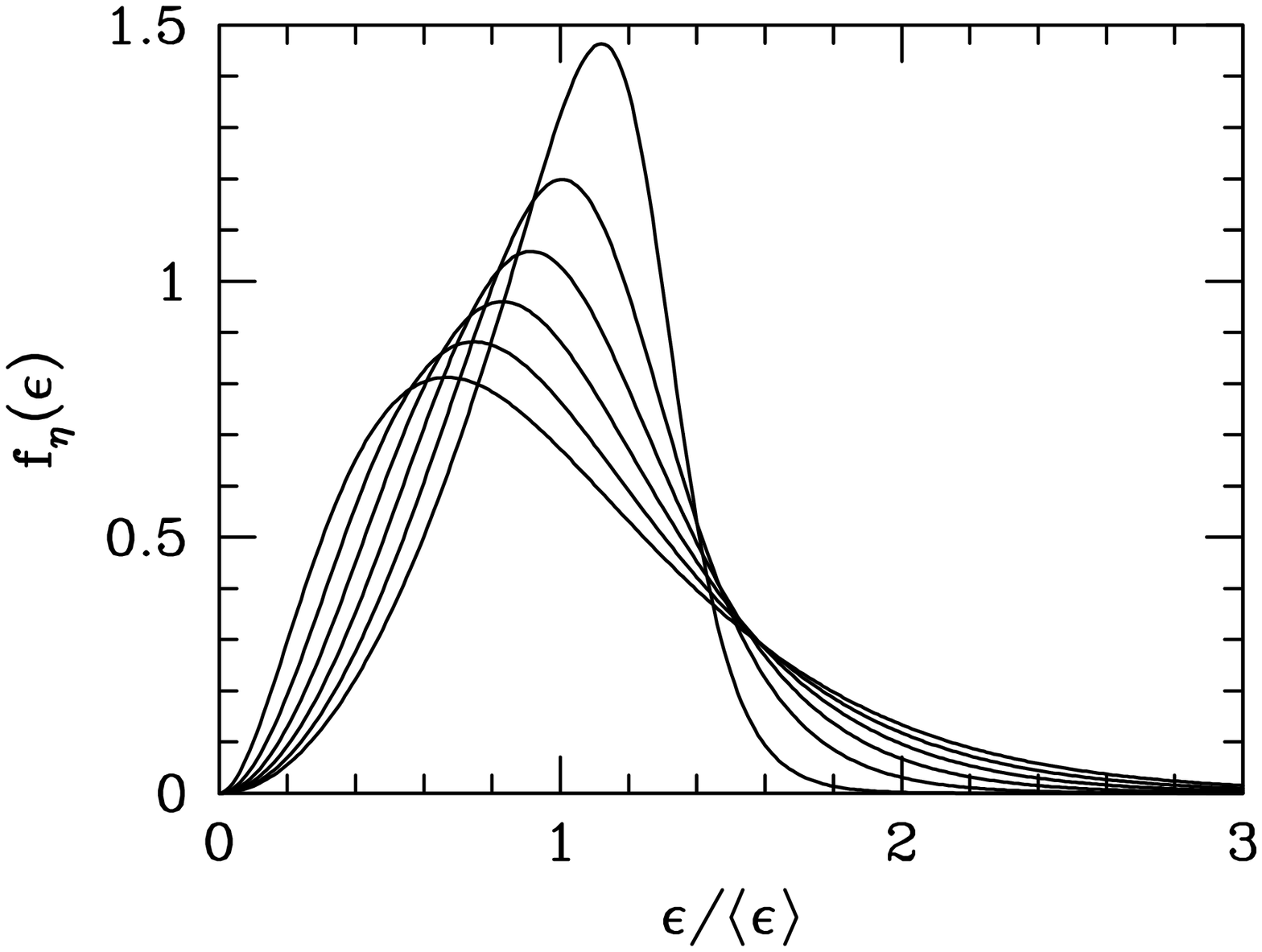}

\plotone{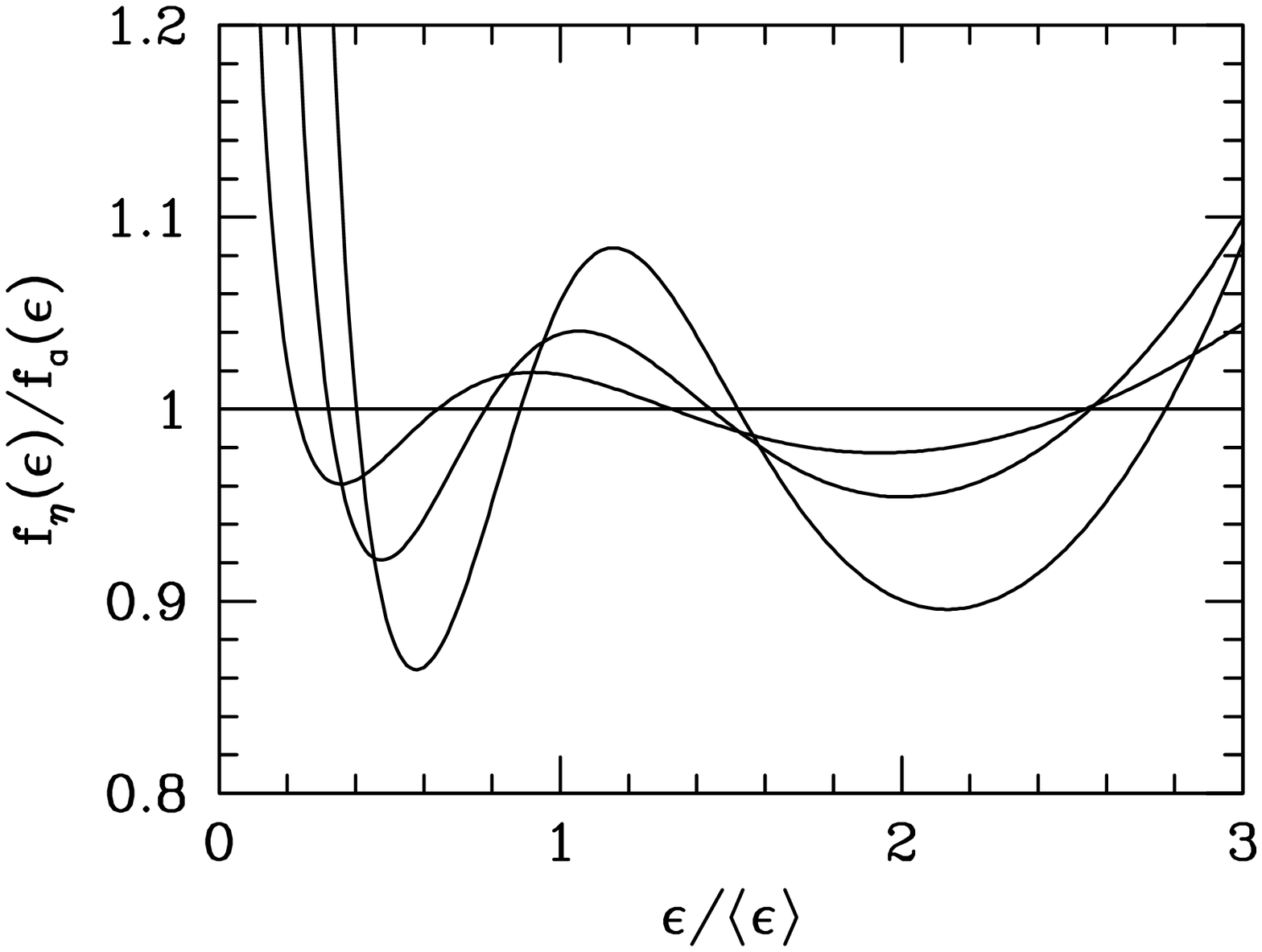}
\caption{\label{fig:fitfunctions} Normalized fit functions.  {\it
Upper panel:} $\alpha$-fit according to Equation~(\ref{eq:powerlawfit}).
{\it Middle panel:} Fermi-Dirac fit according to
Equation~(\ref{eq:FermiDiracfit}).  In both panels the broadest curve
corresponds to $f(\epsilon)=\epsilon^2 \exp(-3\epsilon/\bar\epsilon)$,
i.e., to $\alpha=2$ and $\eta=-\infty$, respectively.  For the other
curves the width was decreased in decrements of 10\%, see
Table~\ref{tab:fitfunctions}.  {\it Bottom panel:} Ratio of the fits
$f_\alpha/f_\eta$ for the first four cases.}
\end{figure}

\begin{deluxetable}{lll}
\tablecaption{\label{tab:fitfunctions} 
Parameters for fit-functions of Figure~\ref{fig:fitfunctions}.}
\tablewidth{0pt}
\tablehead{Width&$\alpha$&$\eta$}
\startdata
$w_0=\langle\epsilon\rangle/\sqrt{3}$&2.    &$-\infty$\\
$0.9\,w_0$                           &2.7037&1.1340\\
$0.8\,w_0$                           &3.6875&2.7054\\
$0.7\,w_0$                           &5.1225&4.4014\\
$0.6\,w_0$                           &7.3333&6.9691\\
$0.5\,w_0$                           &11.   &13.892\\
\enddata
\end{deluxetable}

In the upper panel of Figure~\ref{fig:fitfunctions} we show
$f_\alpha(\epsilon)$, the integral normalized to unity, for several
values of $\alpha$. The broadest curve is for $\alpha=2$ while for the
narrower ones the width given in Equation~(\ref{eq:alphawidth}) was
decreased in 10\% decrements as shown in
Table~\ref{tab:fitfunctions}. The middle panel of
Figure~\ref{fig:fitfunctions} shows the corresponding curves
$f_\eta(\epsilon)$ with the $\eta$-values given in
Table~\ref{tab:fitfunctions}. The broadest curves in each panel are
identical and correspond to $\epsilon^2\exp(-3\epsilon/\bar\epsilon)$
with a width $w_0=\langle\epsilon\rangle/\sqrt{3}$. 

The limiting behavior of $f_\alpha(\epsilon)$ for large $\alpha$ is
$\delta(\epsilon-\bar\epsilon)$ while for $f_\eta(\epsilon)$ the
limiting width is $w_0/\sqrt{5}\approx0.44721\,w_0$. Evidently the
curves $f_\alpha(\epsilon)$ can accommodate a much broader range of
widths than the curves $f_\eta(\epsilon)$.

We find that the neutrino spectra are always fit with parameters in
the range $2\lsim\alpha\lsim5$ or $0\lsim\eta\lsim4$
(Section~\ref{sec:ourflavorcomp}), i.e.\ with a width above about
$0.75\,w_0$. In the bottom panel of Figure~\ref{fig:fitfunctions} we
show the ratios of the fit functions for the widths down to
$0.7\,w_0$. From this plot we find that except for the lowest energies
and very high energies the two fit functions are equivalent to better
than 10\%. Therefore, the two types of fits are largely equivalent for
most practical purposes.

On the basis of a few high-statistics Monte Carlo runs we show in
Section~\ref{sec:High-Statistics} that the numerical spectra are
actually better approximated over a broader range of energies by the
``power-law'' fit functions $f_\alpha(\epsilon)$.  In addition, these
functions are more flexible at representing the high-energy tail of
the spectrum that is most relevant for studying the Earth effect in
neutrino oscillations.

\cleardoublepage

\chapter{Relative Importance of Neutral-Current Reactions}
\label{chap:importance}
We briefly introduce our Monte Carlo code, that was developed to study
neutrino transport in proto-neutron-star atmospheres.  For the
proto-neutron-star atmospheres presented in
Section~\ref{sec:PNSProfiles} we show our results that were obtained
by switching on and off various neutral-current interaction
processes.

The traditional set of processes included in $\numu$ transport turns
out to be a rather poor approximation.  Contrary to what had been
assumed in the past, nucleon-nucleon bremsstrahlung and the
$\nue\bar\nue$ pair process are the main sources for $\numu\bar\numu$
production.  Above the number sphere the spectra 
are shaped by electron scattering and scattering on recoiling
nucleons.  The traditional $\re^+\re^-\to \nue\bar\nue$ process and
scattering on $\nue$/$\bar\nue$ turn out to be rather negligible.

These results and those of Chapter~\ref{chap:AllFlavComp}
represent a major part of this dissertation and were already published
in a similar form as: \\
M.~T.~Keil, G.~G.~Raffelt, and H.~T.~Janka, ``Monte Carlo study of
supernova neutrino spectra formation,'' Astrophys.\ J.\  in press,
astro-ph/0208035.

\section{Monte Carlo Setup}
\label{sec:MonteCarlo}

For the stellar background models introduced in
Section~\ref{sec:PNSProfiles} we have run our Monte Carlo code using
various sets of neutrino-matter interactions.  In
Appendix~\ref{app:MonteCarlo} we give a detailed description of the
code.  Our main interest is to assess the impact of the various
neutrino interactions on the flux and spectrum formation. Since
neutrinos are in LTE below the number sphere, it is sufficient to
simulate the neutrino transport starting slightly below that radius
where we have to specify a boundary condition.

At that lower boundary of our proto-neutron-star atmosphere we always
impose a blackbody boundary condition.  We assume neutrinos to be
in LTE, i.e., at the local temperature and the appropriate chemical
potential.   For numerical runs discussed in this chapter the latter
is taken to vanish, because only $\numu$ are involved and weak
magnetism is discussed in the next chapter.  As a consequence of this
inner boundary condition, the luminosity emerging at the surface is
generated within the computational domain and calculated by our
Monte Carlo transport.  A small flux across the inner boundary
develops because of the negative gradients of temperature and density
in the atmosphere.  Its magnitude depends on the radial resolution of
the proto-neutron-star atmosphere and is therefore an artifact of the
numerical implementation.  However, for our setups this flux is small
compared to the luminosity at the surface and therefore the emerging
neutrino spectra do not depend on the lower boundary condition.

The shallow energy dependence of the thermalization depth of the
nucleon bremsstrahlung implies that whenever we include this process
it is not difficult to choose a reasonable location for the lower
boundary.  Taking the latter too deep in the star is very
CPU-expensive as one spends most of the simulation for calculating
frequent scatterings of neutrinos that are essentially in LTE.

\begin{deluxetable}{ll}
\tablecaption{\label{tab:interacs}
Neutral-current processes included in our $\numu$ tranport.}
\tablewidth{0pt}
\tablehead{Process & Label}
\startdata
bremsstrahlung                      & b \\
nucleon recoil                      & r \\ 
scattering on $\re^+$ and $\re^-$   & s \\
$\re^+ \re^-$ pair annihilation     & p \\
$\nue \bar\nue$ annihilation        & n \\  
scattering on $\nue$ and $\bar\nue$ & sn \\  
\enddata
\end{deluxetable}

The outer boundary is determined by the requirement that neutrinos
stream freely.  In the outer regions the transport is not
CPU-expensive.  Therefore, the exact location of the outer boundary is
not very crucial.  However, we always use a grid of 30 equally spaced
radial zones and can thus obtain better spatial resolution by choosing
the outer boundary fairly low.

We always include $\nu_\mu\rN$ scattering as the main opacity source.
For energy exchange, we switch on or off the processes given in
Table~\ref{tab:interacs}.  We never include inelastic nucleon
scattering $\nu_\mu\rN\rN\to\rN\rN\nu_\mu$ as this process is never
important relative to recoil (Raffelt 2001).  For confirming that
scattering on $\nue$ and $\bar\nue$ is really unimportant if 
$\nu_\mu\re^\pm$ is included (Section~\ref{sec:LeptoScatt}) we
used the Accretion-Phase Model~I.  In all other models we ignore this
process.  We also neglect $\nu_\mu\nu_\mu$ or $\nu_\mu\bar\nu_\mu$
scattering even though such processes may have a larger rate than some
of the included leptonic processes.  Processes of this type do not
exchange energy between the neutrinos and the background medium.  They
are therefore not expected to affect the emerging fluxes and should
also have a minor effect on the emitted spectra.  We study weak
magnetism in Chapter~\ref{chap:AllFlavComp}.

\section{Accretion-Phase Model~I}
\label{sec:messer}

Our first goal is to assess the relative importance of different
energy-exchange processes for the $\nu_{\mu}$ transport.  As a first
example we begin with our Accretion-Phase Model~I. The results from
our numerical runs are summarized in
Table~\ref{tab:NumericalResultsMesser}, where for each run we give
$\langle\epsilon\rangle_{\rm flux}$, $\langle\epsilon^2\rangle_{\rm
flux}$, our fit parameter $\alpha$ determined by
Equation~(\ref{eq:pinchingalpharelation}), and the pinching parameter
$p_{\rm flux}$ for the emerging flux spectrum, defined in
Equation~(\ref{eq:pinch}).  We fitted the temperature and degeneracy
parameter of an effective Fermi-Dirac spectrum producing the same
first two energy moments, and give the luminosity.  Only
$\langle\epsilon\rangle_{\rm flux}$, $\langle\epsilon^2\rangle_{\rm 
flux}$, and the luminosity were obtained from the numerical
simulations.  All other characteristics were calculated from these
first two energy moments.

The first row contains the muon neutrino flux characteristics of the
original Boltzmann transport calculation by Messer.  To make a
connection to these results we ran our code with the same input
physics, i.e., $\nu_\mu\re^\pm$ scattering (s) and $\re^+\re^-$
annihilation (p).  There remain small differences between the original
spectral characteristics and ours. These can be caused by differences
in the implementation of the neutrino processes, by the limited number
of energy and angular bins in the Boltzmann solver, the coarser
resolution of the radial grid in our Monte Carlo runs, and by our
simple blackbody lower boundary condition. We interpret the first two
rows of Table~\ref{tab:NumericalResultsMesser} as agreeing
sufficiently well with each other that a detailed understanding of the
differences is not warranted. Henceforth we will only discuss
differential effects within our own implementation.

In the next row (bsp) we include nucleon bremsstrahlung which has the
effect of increasing the luminosity by a sizable amount without
affecting much the spectral shape. This suggests that bremsstrahlung
is important as a source for $\nu_\mu\bar\nu_\mu$ pairs, but that the
spectrum is then shaped by the energy-exchange in scattering with
$\re^\pm$. In the next row (bp) we switch off $\re^\pm$ scattering so
that no energy is exchanged except by pair-producing processes.  The
spectral energy indeed increases significantly.  However, the biggest
energy-exchange effect in the scattering regime is nucleon recoil.  In
the next two rows we include recoil (brp) and then additionally
$\re^\pm$ scattering (brsp), both lowering the spectral energies and
also the luminosities.

\begin{deluxetable}{lllllrrcrcrc}
\tablecaption{\label{tab:NumericalResultsMesser}
Monte Carlo results for Accretion-Phase Model~I.}
\tablewidth{0pt}
\tablehead{\multicolumn{5}{l}{Energy exchange}
&$\langle\epsilon\rangle_{\rm flux}$&$\langle\epsilon^2\rangle_{\rm
flux}$&$\alpha$&$p_{\rm flux}$&$T$&$\eta$&$L_\nu$}
\startdata
\multicolumn{5}{l}{original run}
              &17.5& 388.& 2.7& 0.97&  5.2&  1.1& 14.4\\  
--&--&s &p &--&16.6& 362.& 2.2& 1.01&  5.3& $-$0.3& 15.8\\
b &--&s &p &--&16.3& 351.& 2.1& 1.02&  5.4& $-$2.2& 19.1\\
b &--&--&p &--&17.8& 419.& 2.1& 1.02&  5.9& $-$1.9& 20.1\\
b &r &--&p &--&15.1& 285.& 3.0& 0.96&  4.3&  1.6& 18.6\\  
b &r &s &p &--&14.2& 255.& 2.8& 0.98&  4.2&  1.1& 14.8\\  
b &r &s &p &n &14.4& 264.& 2.7& 0.97&  4.3&  1.2& 17.6\\  
b &--&s &p &n &16.6& 358.& 2.3& 1.00&  5.2&  0.2& 21.7\\  
--&--&s &p &n &16.9& 369.& 2.4& 0.99&  5.3&  0.4& 20.2\\  
b &r &s &--&--&14.0& 251.& 2.6& 0.99&  4.3&  0.6& 13.1\\  
b &r &s &--&n &14.4& 263.& 2.7& 0.97&  4.3&  1.2& 17.0\\  
--&r &s &p &--&14.5& 265.& 2.8& 0.97&  4.3&  1.2& 13.0\\  
--&r &s &p &n &14.7& 269.& 3.1& 0.96&  4.2&  1.7& 16.8\\  
b &r &sn&p &n &14.3& 260.& 2.7& 0.97&  4.3&  1.2& 17.9\\  
\enddata
\tablecomments{The Energy exchange is specified in
Table~\ref{tab:interacs}.  We give $\langle\epsilon\rangle_{\rm flux}$
and $T$ in MeV, $\langle\epsilon^2\rangle_{\rm flux}$ in MeV$^2$, and
$L_\nu$ in $10^{51}~\rm erg~s^{-1}$.}
\end{deluxetable}

The picture of all relevant processes is completed by adding
$\nue\bar\nue$ pair annihilation (brspn), which is similar to
$\re^+\re^-$ pair annihilation, but a factor of 2--3 more important
(Section~\ref{sec:PairAnn}).  The luminosity is again increased, an
effect which is understood in terms of our blackbody picture for the
number and energy spheres.  In the lower panel of
Figure~\ref{fig:rthermrealistic} we see that $R_{\rm therm}$ moves to
larger radii once ``n'' is switched on, the radiating surface of the
``blackbody'' increases and more pairs are emitted.  For both ``p''
and ``n'' $R_{\rm therm}$ is strongly energy dependent and therefore
it is impossible to define a sharp thermalization radius.  Switching
off ``r'' again (bspn) shows that also with ``n'' included, ``r''
really dominates the mean energy and shaping of the spectrum.

To study the relative importance of the different pair processes, we
switch off the leptonic ones (row ``brs'') and compare this to only
the leptonic processes (row ``rspn'').  In this stellar model both
types contribute significantly.  Comparing then ``brsp'' with ``brsn''
shows that among the leptonic processes ``n'' is clearly more
important than ``p''.

The last row (brsnpn) includes in addition to all other processes
scattering on $\nue$ and $\bar\nue$.  As already shown in
Section~\ref{sec:PairAnn}, this process is about half as important as
scattering on $e^\pm$ and its influence on the neutrino flux and
spectra is negligible.  We show this case for completeness but do not
include scattering on $\nue$ and $\bar\nue$ for any of our further
models.

\begin{figure}[b]
\columnwidth=9.5cm
\vspace{-.2cm}
\plotone{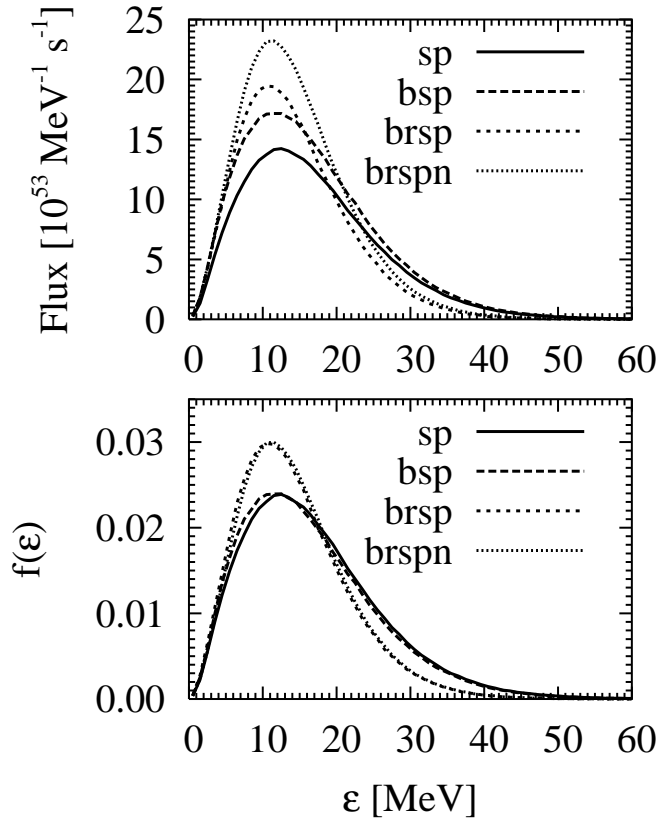}
\vspace{-.2cm}
\caption{\label{fig:ap1spectra} High-statistics spectra for
Accretion-Phase Model~I with different input physics as in
Table~\ref{tab:NumericalResultsMesser}.  {\it Upper
panel:}~Differential particle fluxes.  {\it Lower panel:}~Spectra
normalized to equal particle fluxes.}
\end{figure}

In order to illustrate some of the cases of
Table~\ref{tab:NumericalResultsMesser} we show in the upper panel of
Figure~\ref{fig:ap1spectra} several flux spectra from high-statistics
Monte Carlo runs.  Starting again with the input physics of the
original hydrodynamic simulation (sp) we add bremsstrahlung (b),
recoil (r), and finally $\nue\bar\nue$ pair annihilation (n).  Each of
these processes has a significant and clearly visible influence on the
curves. The pair-creation processes (``b'' and ``n'') hardly change
the spectral shape but increase the number flux, whereas recoil (r)
strongly modifies the spectral shape.  In the lower panel of
Figure~\ref{fig:ap1spectra} we show the same curves, normalized to equal
particle fluxes.  In this representation it is particularly obvious
that the pair processes do not affect the spectral shape.

The very different impact of pair processes and nucleon recoils has a
simple explanation. The thermalization depth for the pair processes is
deeper than that of the energy-exchanging reactions, i.e., the
number sphere is below the energy sphere. Therefore, the particle flux
is fixed more deeply in the star while the spectra are still modified
by energy-exchanging reactions in the scattering atmosphere.

\section{Steep Power Law}
\label{sec:SteepPowerLaw}

As another example we study the steep power-law model defined in
Equation~(\ref{eq:powerlaws}) and Table~4.1.  This
model is supposed to represent the outer layers of a late-time
proto-neutron star but without being hydrostatically self-consistent.
It connects directly with Raffelt (2001), where the same profile was
used in a plane parallel setup, studying bremsstrahlung and nucleon
recoil.  The results of our runs are displayed in
Table~\ref{tab:NumericalResultsSteep} and agree very nicely with those
obtained by Raffelt (2001), corresponding to our cases ``b'' and~``br''.

For investigating the importance of leptonic processes, we run our
code with a variety of neutrino interactions and in addition assume a
constant electron fraction $\Ye$ throughout the whole stellar 
atmosphere.  This
assumption is somewhat artificial, but gives us the opportunity to
study extreme cases in a controlled way.  In the relevant region $\Ye
= 0.5$ yields the highest possible electron density. In addition we
study the electron fraction being one order of magnitude smaller, $\Ye
= 0.05$, and finally the extreme case with an equal number of
electrons and positrons, $\Ye = 0$.

The first leptonic process we consider is $\re^+\re^-$ pair
annihilation.  Comparing the rows ``bp'' with the row ``b'' shows a
negligible effect on the spectrum, but a rise in luminosity.
Increasing $\Ye$ brings the luminosity almost back to the ``b'' case,
because the electron degeneracy rises and the positron density 
decreases so that the pair process becomes less important. 

Adding scattering on $\re^\pm$ forces the transported neutrinos to
stay closer to the medium temperature, i.e., reduces their mean energy.
Of course, the scattering rate increases with the number density of
electrons and positrons, i.e., for higher $\Ye$ we get lower spectral
energies.  For the luminosity the situation is more complicated.
Since the neutrino flux energies decrease when we switch on $\re^\pm$
scattering we would expect a lower luminosity.  However, the opacity
of the medium to neutrinos is strongly energy dependent and low energy
neutrinos can escape more easily than high-energy ones, increasing the
number flux.  On balance, the ``bsp'' luminosities are larger compared
to the ``bp'' ones.

\begin{deluxetable}{llllllrrcrcrc}
\tablecaption{\label{tab:NumericalResultsSteep}
Monte Carlo results for the steep power-law model.}
\tablewidth{0pt}
\tablehead{\multicolumn{5}{l}{Energy exchange}&$\Ye$
&$\langle\epsilon\rangle_{\rm flux}$&$\langle\epsilon^2\rangle_{\rm
flux}$&$\alpha$&$p_{\rm flux}$&$T$&$\eta$&$L_\nu$}
\startdata
        b &--&--&--&--&--- & 25.8& 962. & 1.2 & 1.11&---&---&  21.0\\     
        b &r &--&--&--&--- & 19.5& 487. & 2.6 & 0.98&6.0&0.7&  14.5\\ 
        b &--&--&p &--&0   & 25.4& 890. & 1.6 & 1.06&---&---&  23.8\\     
        b &--&--&p &--&0.05& 25.6& 908. & 1.6 & 1.06&---&---&  23.2\\     
        b &--&--&p &--&0.5 & 25.5& 917. & 1.4 & 1.08&---&---&  21.6\\     
        b &--&s &p &--&0   & 24.2& 787. & 1.9 & 1.03&---&---&  24.5\\     
        b &--&s &p &--&0.05& 23.8& 753. & 2.0 & 1.02&---&---&  24.5\\     
        b &--&s &p &--&0.5 & 21.3& 591. & 2.3 & 1.00&  6.8&$-$0.3&  23.1\\
        b &r &--&p &--&0.5 & 20.0& 507. & 2.7 & 0.98&  6.0&  1.0&  16.8\\ 
        b &r &s &p &--&0   & 20.3& 518. & 2.9 & 0.97&  5.9&  1.4&  19.7\\ 
        b &r &s &p &--&0.05& 20.3& 518. & 2.9 & 0.97&  5.9&  1.4&  19.5\\ 
        b &r &s &p &--&0.5 & 19.6& 488. & 2.7 & 0.98&  5.9&  1.1&  18.7\\ 
        b &r &s &p & n&0   & 20.7& 535. & 3.0 & 0.96&  5.8&  1.8&  23.9\\ 
b${\times}3$&r &s &p&n&0.05& 20.3& 522. & 2.7 & 0.97&  6.0&  1.3&  24.2\\ 
        b &r &s &p & n&0.05& 20.6& 530. & 3.0 & 0.96&  5.9&  1.7&  23.8\\ 
b${\times}0.3$&r&s&p&n&0.05& 20.7& 534. & 3.1 & 0.96&  5.8&  1.8&  23.4\\ 
        b &r &s &p & n&0.5 & 19.8& 499. & 2.7 & 0.97&  5.9&  1.2&  21.4\\ 
\enddata
\end{deluxetable}

To compare the scattering on $\re^\pm$ with that on nucleons, we
turn off ``s'' again and instead switch on recoil (r).  Qualitatively,
the energy exchange is very different from the earlier case.  In the
scattering on $\re^\pm$ a neutrino can exchange a large amount of
energy, while for scattering on nucleons the energy exchange is
small. But since neutrino-nucleon scattering is the dominant source of
opacity that keeps the neutrinos inside the star, the scatterings are
very frequent. This leads to a stronger suppression in the high-energy
tail of the neutrino spectrum and therefore to a visibly smaller mean
flux energy and lower effective spectral temperature, but higher
$\alpha$ and higher effective degeneracy. Many nucleon scatterings,
however, are needed to downgrade the high-energy neutrinos (different
from e$^\pm$ scattering). Therefore neutrinos stay longer at high
energies and experience a larger opacity and a larger amount of
backscattering. This suppresses the neutrino flux significantly.
 
In the runs including both scattering reactions (brsp), we find a
mixture of the effects of e$^\pm$ and nucleon scatterings and an
enhanced reduction of the mean flux energy.

Finally, adding the neutrino pair process yields almost no change in
energy and pinching, but an increased luminosity as expected from the
analogous case in Section~\ref{sec:messer}.  Although this profile is
rather steep, leptonic pair processes are still important
(Figure~\ref{fig:rthermsteep}).
 
In order to estimate the sensitivity to the exact treatment of nucleon
bremsstrahlung we have performed one run with the bremsstrahlung rate
artificially enhanced by a factor of 3, and one where it was decreased
by a factor 0.3. All other processes were included.  The emerging
fluxes and spectra indeed do not depend sensitively on the exact
strength of bremsstrahlung as argued in Section~\ref{sec:odeVStherm}.

\section{Shallow Power Law}

For the shallow power law (Table~\ref{tab:NumericalResultsShallow})
almost the same discussion as for the steep case applies.  As we can
already infer from Figure~\ref{fig:rthermshallow}, leptonic processes
are more important. This leads to a much higher increase of the
neutrino flux once ``p'' or ``n'' are included, and to stronger
spectral pinching when e$^+$e$^-$ annihilation is switched
on. Scattering on $\re^\pm$ downgrades the transported neutrino flux
by a larger amount.

\section{Summary}

We find that the $\nu_\mu$ spectra are reasonably well described by
the simple picture of a blackbody sphere determined by the
thermalization depth of the nucleonic bremsstrahlung process, the
``filter effect'' of the scattering atmosphere, and energy transfers
by nucleon recoils. This is also true for the $\nu_\mu$ flux in case
of steep neutron-star atmospheres. For more shallow atmospheres pair
annihilation (e$^+$e$^-$ and $\nu_{\rm e}\bar\nu_{\rm e}$), however,
yields a large contribution to the emitted $\nu_{\mu}$ flux and
e$^\pm$ scattering reduces the mean flux energy significantly. It is
therefore important for state-of-the art transport calculations to
include these leptonic processes. The traditional process
$\re^+\re^-\to\nu_\mu\bar\nu_\mu$ is always sub-dominant compared to
$\nue\bar\nue\to\nu_\mu\bar\nu_\mu$.  The relative importance of the
various reactions depends on the stellar profile.

Neutrinos emitted from a blackbody surface and filtered by a
scattering atmosphere without recoils and leptonic processes have an
anti-pinched spectrum (Raffelt 2001). However, after all
energy-exchanging reactions have been included we find that the
spectra are always pinched. When described by our $\alpha$-fit,
$\alpha$ ranges from about 2.5 to 3.5.  For an effective Fermi-Dirac
distribution, the nominal degeneracy parameter $\eta$ is typically in
the range 1--2, depending on the profile and electron concentration.
 
The traditional set of neutrino interactions that is commonly included
in hydrodynamic SN simulations is not sufficient to account for
$\numu$ spectra formation.  Nucleon recoils significantly lower the
mean energy, and bremsstrahlung as well as the $\nu_{\rm
e}\bar\nu_{\rm e}$ pair process cause higher luminosities compared to
results obtained with the traditional interactions.

\begin{deluxetable}{llllllrrcrcrc}
\tablecaption{\label{tab:NumericalResultsShallow}
Monte Carlo results for the shallow power-law model.}
\tablewidth{0pt}
\tablehead{\multicolumn{5}{l}{Energy exchange}&$\Ye$
&$\langle\epsilon\rangle_{\rm flux}$&$\langle\epsilon^2\rangle_{\rm
flux}$&$\alpha$&$p_{\rm flux}$&$T$&$\eta$&$L_\nu$}
\startdata
b &--&--&--&--&--- & 27.7& 1120.& 1.2 & 1.12&---&---&  20.3\\     
b &r &--&--&--&--- & 20.1&  521.& 2.5 & 0.99&  6.3&  0.4&  13.4\\ 
b &--&--&p &--&0   & 27.7&  974.& 2.7 & 0.98&  8.3&  1.0&  43.1\\ 
b &--&--&p &--&0.05& 27.9&  990.& 2.7 & 0.98&  8.3&  1.1&  43.3\\ 
b &--&--&p &--&0.5 & 28.3& 1019.& 2.7 & 0.98&  8.5&  1.0&  38.3\\ 
b &--&s &p &--&0   & 25.5&  830.& 2.6 & 0.98&  7.6&  1.1&  46.2\\ 
b &--&s &p &--&0.05& 25.4&  815.& 2.8 & 0.97&  7.5&  1.2&  46.3\\ 
b &--&s &p &--&0.5 & 23.5&  706.& 2.6 & 0.98&  7.1&  1.0&  44.8\\ 
b &r &--&p &--&0.5 & 22.5&  624.& 3.3 & 0.95&  6.1&  2.2&  33.1\\ 
b &r &s &p &--&0   & 22.3&  612.& 3.3 & 0.95&  6.1&  2.1&  39.6\\ 
b &r &s &p &--&0.05& 22.2&  609.& 3.2 & 0.95&  6.1&  2.1&  39.1\\ 
b &r &s &p &--&0.5 & 21.7&  585.& 3.1 & 0.95&  6.1&  1.9&  39.2\\ 
b &r &s &p & n&0   & 22.2&  608.& 3.3 & 0.94&  6.0&  2.2&  54.7\\ 
b &r &s &p & n&0.05& 22.4&  615.& 3.4 & 0.94&  6.1&  2.3&  54.9\\ 
b &r &s &p & n&0.5 & 21.8&  587.& 3.3 & 0.95&  6.1&  1.9&  51.3\\ 
\enddata
\end{deluxetable}

\cleardoublepage


\chapter{Comparison of All Flavors}
\label{chap:AllFlavComp}

The new energy-exchange channels studied in the previous section lower
the average $\nu_{\mu}$ energies.  In order to compare the $\nu_\mu$
fluxes and spectra with those of $\nue$ and $\bar\nue$ we perform a
new series of runs where we include the full set of relevant
microphysics for $\nu_\mu$ and also simulate the transport of $\nue$
and $\bar\nue$.

Our findings for $\nue$ and $\bar\nue$ agree with the previous
literature, whereas our results for $\numu$ show lower mean energies.
The mean energies of $\bar\nue$ and $\numu$ are almost equal.  The
$\numu$ luminosity can differ from that of $\nue$ and $\bar\nue$  by
up to a factor of 2.  Weak magnetism causes a difference between the
mean energies of $\numu$ and $\bar\numu$ of a few per cent.

We present an overview of the results obtained by other groups and
find that the more recent simulations agree rather well on the
differential effects in energies and luminosities of the different
flavors.  These findings as well as our results suggest that
assumptions commonly made for calculating oscillation effects, i.e.,
equal luminosities and a difference in the mean energies of $\bar\nue$
and $\bar\numu$ of up to a factor of 2, need to be changed.

We show results of the detailed spectral shapes obtained in our
simulations.  Our $\alpha$ fit describes the spectra slightly better
than a nominal Fermi-Dirac distribution.

Parts of this chapter were already published in a similar form
as:\\
M.~T.~Keil, G.~G.~Raffelt, and H.~T.~Janka, ``Monte Carlo study of
supernova neutrino spectra formation,'' Astrophys.\ J.\  in press,
astro-ph/0208035.

\section{Results from Our Monte Carlo Study}
\label{sec:ourflavorcomp}

For our $\nue$ and $\bar\nue$ transport we employ the same
microphysics as in Janka \& Hillebrandt (1989a,b), i.e.,
charged-current reactions of e$^\pm$ with nucleons, iso-energetic
scattering on nucleons, scattering on $\re^\pm$, and $\re^+\re^-$ pair
annihilation.  In principle one should also include nucleon
bremsstrahlung and the effect of nucleon recoils for the transport of
$\nue$ and $\bar\nue$, but their effects will be minimal.  Therefore,
we preferred to leave the original working code unmodified for these
flavors.

\begin{deluxetable}{llrccrcrr}
\tablecaption{\label{tab:MultiFlavorAccretion}
Comparing Monte Carlo results for different flavors.
Accretion-Phase Models}
\tablewidth{0pt}
\tablehead{Flavor
&$\langle\epsilon\rangle_{\rm
flux}$&$\langle\epsilon^2\rangle_{\rm flux}$&$\displaystyle
\frac{\langle\epsilon\rangle_{\rm
flux}}{\langle\epsilon_{\bar\nue}\rangle_{\rm flux}}$&  
$\alpha$&$p_{\rm flux}$
&$T$&$\eta$&$L_\nu$}
\startdata
\hline
\multicolumn{9}{c}{\bf Accretion-Phase Model~I}\\
\hline
\multicolumn{8}{l}{Original}\\
\quad$\nu_\mu$, $\bar\nu_\mu$ 
                 &  17.5 &  388. & 1.20 &  2.7& 0.97&  5.2&  1.1&  14.4 \\ 
\quad$\bar\nue$ &  14.6 &  253. & 1    &  4.4& 0.91&  3.5&  3.4&  29.2 \\ 
\quad$\nue$     &  12.5 &  190. & 0.86 &  3.6& 0.93&  3.2&  2.8&  30.8 \\ 
\multicolumn{8}{l}{Our runs}\\                                    
\quad$\nu_\mu$, $\bar\nu_\mu$ (sp)                                 
                 &  16.6 &  362. & 1.19 &  2.2& 1.01&  5.3& $-$0.3&15.8 \\ 
\quad$\nu_\mu$, $\bar\nu_\mu$ (brspn)                              
                 &  14.3 &  260. & 1.02 &  2.7& 0.97&  4.3&  1.2&  17.9 \\ 
\quad$\nu_\mu$ (wm)                                   
                 &  14.2 &  254. & 1.01 &  2.9& 0.96&  4.1&  1.6&  17.4 \\ 
\quad$\bar\nu_\mu$ (wm)                               
                 &  14.9 &  281. & 1.06 &  2.8& 0.97&  4.3&  1.5&  18.3 \\ 
\quad$\bar\nue$ &  14.0 &  237. & 1    &  3.8& 0.93&  3.6&  2.7&  31.7 \\ 
\quad$\nue$     &  11.8 &  175. & 0.84 &  2.9& 0.97&  3.4&  1.4&  31.9 \\ 
\hline
\multicolumn{9}{c}{\bf Accretion-Phase Model~II}\\
\hline
\multicolumn{8}{l}{Original}\\                                    
\quad$\nu_\mu$, $\bar\nu_\mu$                                          
                 &  17.2 &  380. & 1.09 &  2.5& 0.98&  5.2&  0.8&  32.4 \\ 
\quad$\bar\nue$ &  15.8 &  300. & 1    &  4.0& 0.92&  4.0&  3.0&  68.1 \\ 
\quad$\nue$     &  12.9 &  207. & 0.82 &  3.1& 0.96&  3.7&  1.7&  65.6 \\ 
\multicolumn{8}{l}{Our runs}\\                                    
\quad$\nu_\mu$, $\bar\nu_\mu$                                          
                 &  15.7 &  317. & 1.02 &  2.5& 0.98&  4.8&  0.8&  27.8 \\ 
\quad$\bar\nue$ &  15.4 &  283. & 1    &  4.2& 0.92&  3.8&  3.2&  73.5 \\ 
\quad$\nue$     &  13.0 &  207. & 0.84 &  3.4& 0.95&  3.6&  2.1&  73.9 \\ 
\enddata
\tablecomments{We give $\langle\epsilon\rangle_{\rm flux}$ and $T$ in
MeV, $\langle\epsilon^2\rangle_{\rm flux}$ in MeV$^2$, $L_\nu$ in
$10^{51}~\rm erg~s^{-1}$.}  
\end{deluxetable}

In the first three rows of Table~\ref{tab:MultiFlavorAccretion} we
give the spectral characteristics for the Accretion-Phase Model~I from
the original simulation of Messer.  The usual hierarchy of
average neutrino energies is found, i.e.,
$\langle\epsilon_{\nue}\rangle:\langle\epsilon_{\bar\nue}\rangle
:\langle\epsilon_{\nu_\mu}\rangle=0.86 : 1 : 1.20$. The luminosities
are essentially equal between $\bar\nue$ and $\nue$ while $\nu_\mu$,
$\bar\nu_\mu$, $\nu_\tau$, and $\bar\nu_\tau$ each provide about half
of the $\bar\nue$ luminosity.

Our Monte Carlo runs of this profile establish the same picture for
the same input physics (label ``sp''). Although our mean energies are
slightly offset 
to lower values for all flavors relative to the original run, our
energies relative to each other are
$\langle\epsilon_{\nue}\rangle:\langle\epsilon_{\bar\nue}\rangle
:\langle\epsilon_{\nu_\mu}\rangle=0.84:1:1.19$ and thus very
similar. However, once we include all energy exchanging processes
(brspn) we find $0.84:1:1.02$ instead.  Therefore,
$\langle\epsilon_{\nu_\mu}\rangle$ no longer exceeds
$\langle\epsilon_{\bar\nue}\rangle$ by much.  The luminosity of
$\nu_\mu$ is about half that of $\nue$ or $\bar\nue$ which are
approximately equal, in rough agreement with the original results.
Even though the additional processes lower the mean energy of
$\nu_\mu$ they yield a more than 10\% higher $\nu_{\mu}$ luminosity, 
mainly due to $\nue\bar\nue$ annihilation.

Different rates for neutrinos and anti-neutrinos arise due to the weak
magnetism correction of the neutrino-nucleon scattering cross
section (Section~\ref{sec:WeakMag}).  The next two lines (wm) show
results from runs where we included weak magnetism in our $\numu$
transport.  Weak magnetism causes a significant correction to the
neutrino-nucleon cross section that mainly arises due to the large
anomalous magnetic moments of protons and neutrons (Vogel \& Beacom
1999, Horowitz 2002).  It increases the neutrino interaction rate but
lowers the rate for anti-neutrinos.  It is expected to be a small
correction in the SN context, but has never been implemented for
$\numu$ and $\bar\numu$ so far.  Following Horowitz (2002) we add weak
magnetism to our nucleon-recoil rate as given in
Section~\ref{sec:WeakMag}.

Our Monte Carlo code transports only one species of neutrinos at a
time. In order to test the impact of weak magnetism we assumed that a
chemical potential for $\nu_\mu$ would build up, and assumed a fixed
value for the $\nu_\mu$ degeneracy parameter throughout our stellar
model. We then iterated several runs for $\nu_\mu$ and $\bar\nu_\mu$
with different degeneracy parameters until their particle fluxes were
equal because in a stationary state there will be no net flux of
$\mu$-lepton number.

We performed this procedure for our Accretion-Phase Model I.  The mean
energy of $\nu_\mu$ goes down by $1\%$ and goes up for $\bar\nu_\mu$
by $4\%$.  The mean luminosities change in opposite directions by
about 2--3\%.  Since we adjusted the particle fluxes by hand in order
to fix the chemical potential, we cannot claim an effect on the
luminosities.

\begin{figure}[t]
\columnwidth=7cm
\plotone{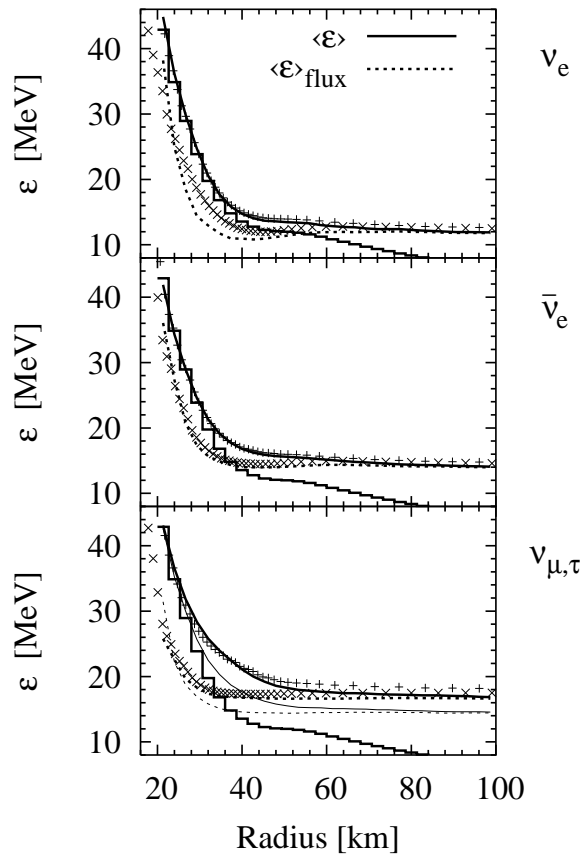}
\caption{\label{fig:messercomp} Comparison of the Accretion-Phase
Model~I calculations. Continuous lines show our Monte Carlo runs
while crosses represent the original simulation by Messer. The steps
correspond to $\langle\epsilon\rangle = 3.15\,T$.  Thin lines in the
lower panel represent our findings after including the full set of
relevant interaction processes.}
\end{figure}

In Figure~\ref{fig:messercomp} we compare our calculations for the
Accretion-Phase Model~I with those of the original simulation.  The
step-like curve represents the mean energy of neutrinos in LTE for
zero chemical potential.  The smooth solid line is the mean energy 
$\langle\epsilon\rangle$ from our runs, the dotted line gives
$\langle\epsilon\rangle_{\rm flux}$.  The crosses are the
corresponding results from the original runs.  In the case of $\numu$
(lower panel) we show the results of our ``sp'' run as thick lines and
the results of our ``brspn'' run, i.e., the complete set of
interactions included, as thin lines.  The boundaries of our
simulation domain were the same for all flavors, namely $R_{\rm in} =
20$~km and $R_{\rm out} = 100$~km.

As another example of an accreting proto-neutron star we use the
Accretion-Phase Model~II.  The neutrino interactions included in this
model were nucleon bremsstrahlung, scattering on $\re^\pm$, and $\rm
e^+ \rm e^-$ annihilation.  Nucleon correlations, effective mass, and
recoil were taken into account, following Burrows \& Sawyer (1998,
1999), as well as weak magnetism effects (Horowitz 2002) and quenching
of $g_A$ at high densities (Carter \& Prakash 2002).  All these
improvements to the traditional microphysics affect mainly $\nu_{\mu}$
and to some degree also $\bar\nu_{\rm e}$.  Weak magnetism terms
decrease the nucleon scattering cross sections for $\bar\nu_{\mu}$
more strongly than they modify $\nu_{\mu}$ scatterings.  In this
hydrodynamic calculation, however, $\nu_{\mu}$ and $\bar\nu_{\mu}$
were treated identically by using the average of the corresponding
reaction cross sections. The effects of weak magnetism on the
transport of $\nu_{\mu}$ and $\bar\nu_{\mu}$ are therefore not
included to very high accuracy. Note, moreover, that the original data
come from a general relativistic hydrodynamic simulation with the
solution of the Boltzmann equation for neutrino transport calculated
in the comoving frame of the stellar fluid. Therefore the neutrino
results are affected by gravitational redshift and, depending on where
they are measured, may also be blueshifted by Doppler effects due to
the accretion flow to the nascent neutron star.

Our Monte Carlo simulation in contrast was performed on a static
background without general relativistic corrections. It includes
bremsstrahlung, recoil, $\rm e^+ \rm e^-$ pair annihilation,
scattering on $\rm e^\pm$, and $\nu_{\rm e}\bar\nu_{\rm e}$
annihilation, i.e., our microphysics is similar but not identical with
that used in the original run.  As an outer radius we took 100~km; all
flux parameters are measured at this radius because farther out
Doppler effects of the original model would make it difficult to
compare the results.  Keeping in mind that we use very different
numerical approaches and somewhat different input physics, the
agreement in particular for $\nu_{\rm e}$ and $\bar\nu_{\rm e}$ is
remarkably good.  This agreement shows once more that our Monte Carlo
approach likely captures at least the differential effects of the new
microphysics in a satisfactory manner.

\begin{figure}[t]
\columnwidth=7cm
\plotone{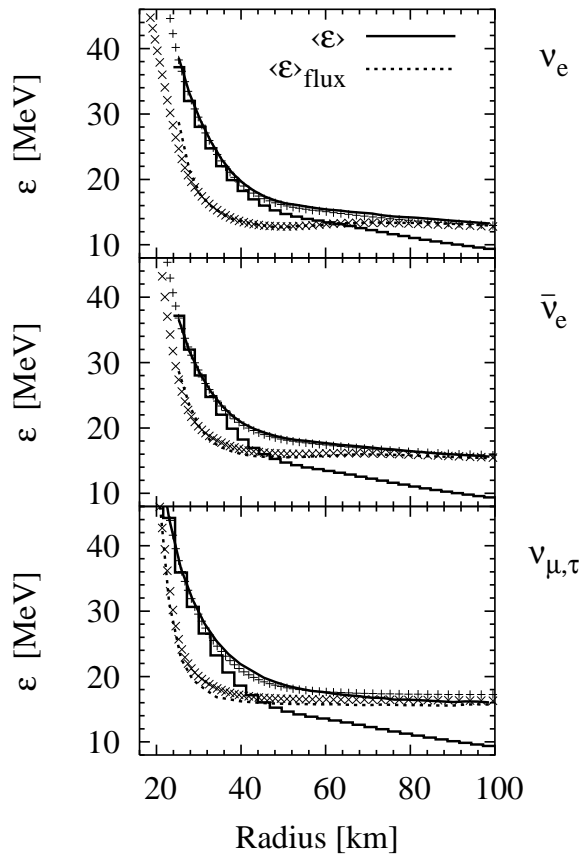}
\caption{\label{fig:ramppcomp} Same as Figure~\ref{fig:messercomp} for
the Accretion-Phase Model~II.}
\end{figure}

For the Accretion-Phase Model~II we show the comparison with our
runs in Figure~\ref{fig:ramppcomp}. In this case, for the transport of
$\nu_\mu$ our inner boundary is $R_{\rm in} = 16$~km, while for $\nue$
and $\bar\nue$ we use $R_{\rm in} = 24$~km.  For $\nue$ and $\bar\nue$
the charged-current processes (urca) keep these neutrinos in LTE up to
larger radii than pair processes in the case of $\nu_\mu$.  With our
choice of $R_{\rm in}$ the neutrinos are in LTE within the innermost
radial zones.  In this profile our choice of boundaries reduces the
CPU time needed and does not affect the results.  As stated before
outer boundaries were chosen to be at $R_{\rm out} = 100$~km.

The results are similar to the Accretion-Phase Model~I.  The
luminosities are not equipartitioned but instead follow roughly
$L_{\nue} \approx L_{\bar\nue}\approx 2 \, L_{\nu_\mu}$.  The ratios
of mean energies are
$\langle\epsilon_{\nue}\rangle:\langle\epsilon_{\bar\nue}\rangle
:\langle\epsilon_{\nu_\mu}\rangle=0.82 : 1 : 1.09$ in the original run
and $0.84 : 1 : 1.02$ in our run.

In summary, both accretion-phase models agree reasonably well in the
$\langle\epsilon_{\nue}\rangle:\langle\epsilon_{\bar\nue}\rangle$
ratio for all runs. Moreover, using traditional input physics one
finds something like $\langle\epsilon_{\bar\nue}\rangle
:\langle\epsilon_{\nu_\mu}\rangle=1 : 1.20$. Depending on the
implementation of the new input physics and depending on the model one
finds results between $\langle\epsilon_{\bar\nue}\rangle
:\langle\epsilon_{\nu_\mu}\rangle=1 : 1.02$ and $1:1.09$.

\begin{deluxetable}{llrrccrcrr}
\tablecaption{\label{tab:MultiFlavorPower}
Comparing Monte Carlo results for different flavors.
Power-Law~Models}
\tablewidth{0pt}
\tablehead{Flavor
&$\Ye$&$\langle\epsilon\rangle_{\rm
flux}$&$\langle\epsilon^2\rangle_{\rm flux}$&$\displaystyle
\frac{\langle\epsilon\rangle_{\rm
flux}}{\langle\epsilon_{\bar\nue}\rangle_{\rm flux}}$&  
$\alpha$&$p_{\rm flux}$
&$T$&$\eta$&$L_\nu$}
\startdata
\hline
\multicolumn{9}{c}{\bf Steep Power Law $\mathbold{p=10}$}\\ 
\hline
$q=2.5$\\                                                         
\quad$\nu_\mu$, $\bar\nu_\mu$                                          
                &0.15 & 20.4 &  525. & 1.10 &  2.8& 0.96&  5.9&  1.5&  23.5 \\ 
\quad$\bar\nue$&0.15 & 18.5 &  413. & 1    &  3.8& 0.92&  4.6&  3.0&  23.5 \\ 
\quad$\nue$    &0.15 & 12.7 &  198. & 0.69 &  3.4& 0.94&  3.4&  2.4&  12.8 \\ 
\quad$\nu_\mu$, $\bar\nu_\mu$                                          
                &0.2  & 20.4 &  521. & 1.14 &  3.0& 0.97&  5.9&  1.5&  23.3 \\ 
\quad$\bar\nue$&0.2  & 17.9 &  383. & 1    &  4.1& 0.92&  4.4&  3.1&  11.7 \\ 
\quad$\nue$    &0.2  & 13.4 &  218. & 0.75 &  3.7& 0.93&  3.4&  2.9&  24.4 \\ 
$q=3.0$\\                                                          
\quad$\nu_\mu$, $\bar\nu_\mu$                                          
                &0.1  & 17.7 &  393. & 1.14 &  2.9& 0.96&  5.0&  1.8&  12.7 \\ 
\quad$\bar\nue$&0.1  & 15.5 &  289. & 1    &  3.9& 0.93&  4.0&  2.8&   8.8 \\ 
\quad$\nue$    &0.1  & 10.5 &  132. & 0.68 &  4.1& 0.92&  2.6&  3.0&   6.6 \\ 
$q=3.5$\\                                                          
\quad$\nu_\mu$, $\bar\nu_\mu$                                          
                &0.07 & 15.8 &  310. & 1.22 &  3.1& 0.95&  4.4&  2.1&   7.9 \\ 
\quad$\nu_\mu$ (wm)                                   
                &0.07 & 15.5 &  296. & 1.19 &  3.3& 0.94&  4.2&  2.3&   7.7 \\ 
\quad$\bar\nu_\mu$ (wm)                               
                &0.07 & 16.5 &  337. & 1.27 &  3.2& 0.95&  4.5&  2.1&   8.3 \\ 
\quad$\bar\nue$&0.07 & 13.0 &  207. & 1    &  3.4& 0.94&  3.5&  2.3&   4.3 \\ 
\quad$\nue$    &0.07 &  9.4 &  103. & 0.72 &  5.0& 0.90&  2.1&  3.9&   4.1 \\ 
\hline
\multicolumn{9}{c}{\bf Shallow Power Law $\mathbold{p=5}$, $\mathbold{q=1}$}\\            
\hline
\quad$\nu_\mu$, $\bar\nu_\mu$                                          
                &0.3  & 22.0 &  596. & 1.14 &  3.3& 0.94&  6.0&  2.2&  53.9 \\ 
\quad$\bar\nue$&0.3  & 19.3 &  440. & 1    &  4.5& 0.91&  4.5&  3.7&  85.7 \\ 
\quad$\nue    $&0.3  & 14.7 &  262. & 0.76 &  3.7& 0.93&  3.8&  2.7&  56.5 \\ 
\enddata
\end{deluxetable}

In order to estimate the corresponding results for later stages of the
proto-neutron star evolution we employ our steep power-law model.  We
vary the power $q$ of the temperature profile within a reasonable
range so that $q/p = 0.25$--0.35, with $q$ and $p$ defined in
Equation~(\ref{eq:powerlaws}). $\Ye$ is fixed by demanding roughly equal
number fluxes for $\nue$ and $\bar\nue$ because a few seconds after
bounce deleptonization should be essentially complete.  The fluxes of
these neutrinos depend very sensitively on $\Ye$ so that this
constraint is only reached to within about 30\% without tuning $\Ye$
to three decimal places.  However, the mean energies are rather
insensitive to the exact value of $\Ye$.  This is illustrated in the
first section of Table~\ref{tab:MultiFlavorPower} by the steep
power-law model with $q = 2.5$ where we show results for $\Ye=0.15$
and 0.20.  The number fluxes of $\nue$ and $\bar\nue$ differ by less
than 30\% for $\Ye=0.15$, but differ by a factor of 3 for $\Ye=0.2$.
At the same time, the average spectral energies barely change.

The ratios of mean energies are not very different from those of the
accretion-phase models.  Of course, the absolute flux energies have no
physical meaning because we adjusted the stellar profile in order to
obtain realistic values.  For the luminosities we find $L_{\nue}<
L_{\nu_\mu}$, different from the accretion phase. The steep power law
implies that the radiating surfaces are similar for all flavors so
that it is not surprising that the flavor with the largest energies
also produces the largest luminosity.

For our steepest profile we also performed runs including weak
magnetism, because this should probe how much spectra are affected in
late-time profiles.  As already found for the Accretion-Phase Model~I,
the mean energy of $\bar\numu$ goes up by $4\%$, whereas in the case
of $\numu$ it decreases by 1\%.  The mean luminosities are almost
unaffected.  We conclude that weak-magnetism corrections are
small. Transporting $\nu_\mu$ and $\bar\nu_\mu$ separately in a
self-consistent hydrodynamic simulation is probably not worth the cost
in computer time.

We find that $\langle\epsilon_{\nu_\mu}\rangle$ always exceeds
$\langle\epsilon_{\bar\nue}\rangle$ by a small amount, the exact value
depending on the stellar model.  During the accretion phase the
energies seem to be almost identical, later they may differ by up to
20\%. We have not found a model where the energies differ by the large
amounts which are sometimes assumed in the literature.  At late times
when $\Ye$ is small the microphysics governing $\bar\nue$ transport is
closer to that for $\nu_\mu$ than at early times.  Therefore, one
expects that at late times the behavior of $\bar\nue$ is more similar
to $\nu_\mu$ than at early times. We do not see any argument for
expecting an extreme hierarchy of energies at late times for
self-consistent stellar models.

We never find exact equipartition of the flavor-dependent
luminosities.  Depending on the stellar profile the fluxes can
mutually differ by up to a factor of 2 in either direction.

\section{Previous Literature}
\label{sec:PreviousLiterature}

Studies of oscillation effects in SN neutrino spectra usually assume
exactly equal neutrino luminosities for all flavors and a strong
hierarchy for the mean energies.  Often times they assume the
difference between the mean energies of $\bar\nue$ and $\bar\numu$ to
be about a factor of 2.  We find an almost orthogonal picture in our
simulations.  Where does this come from?

To the best of our knowledge, except in the very recent models by
Buras et al.~(2003b), the microphysics employed for $\nu_\mu$
transport was roughly the same in all published simulations. It
included iso-energetic scattering on nucleons, $\re^+\re^-$
annihilation and $\nu_\mu\re^\pm$ scattering.  Of course, the
transport method and the numerical implementation of the neutrino
processes differ in the codes of different groups.  The new results of
self-consistent calculations point in the same direction as the
findings from our Monte Carlo studies presented in the previous
chapter. Therefore these findings are almost orthogonal to the
previous assumptions oscillation studies are based on.  Our
representative sample of pertinent results is summarized in
Table~\ref{tab:Literature}.   We have also included our
Accretion-Phase Models~I and II and a very recent model by the
Garching group including all relevant processes.  Note that the
simulations discussed below did not in all cases use the same stellar
models and equations of state for the dense matter in the SN core.  

We begin with the simulations of the Livermore group who find robust
explosions by virtue of the neutron-finger convection
phenomenon. Neutrino transport is treated in the hydrodynamic models
with a multigroup flux-limited diffusion scheme.  Mayle, Wilson, \&
Schramm (1987) gave detailed results for their SN simulation of a
$25\,M_\odot$ star.  For half a second after bounce they obtained a
somewhat oscillatory behavior of the neutrino luminosities.  After the
prompt peak of the electron neutrino luminosity, they got
$L_{\nue}\approx L_{\bar\nue}\approx 2\, L_{\nu_\mu}\approx 50$--$130
\times 10^{51}~{\rm erg~s^{-1}}$.  After about one second the values
stabilize.  This calculation did not produce the ``standard''
hierarchy of energies. However, there is clearly a tendency that
$\bar\nue$ behave more similar to $\nu_\mu$ at late times.

\begin{deluxetable}{llrrrccrrrrr}
\tablecaption{\label{tab:Literature}
Flavor dependent flux characteristics from the literature.}
\tablewidth{0pt}
\rotate
\tablehead{& tpb & $\langle\epsilon_{\nue}\rangle$ &
$\langle\epsilon_{\bar\nue}\rangle$ &
$\langle\epsilon_{\nu_\mu}\rangle$ & $\displaystyle
\frac{\langle\epsilon_\nu\rangle}{\langle\epsilon_{\bar\nue}\rangle}$ &
$L_{\nue}$ & $L_{\bar\nue}$ & $L_{\nu_\mu} $}
\startdata
Mayle et~al.\ (1987) &1.0 & 12 & 24 & 22 &  $0.50:1:0.92$ & 20
& 20 & 20\\ 
Totani et~al.\ (1998)&0.3 & 12 & 15 & 19 &  $0.80:1:1.26$ & 20
& 20 & 20\\ 
                     &10  & 11 & 20 & 25 &  $0.55:1:1.25$ & 0.5
& 0.5 & 1\\ 
Bruenn (1987)        &0.5 & 10 & 12 & 25 &  $0.83:1:2.08$ & 3
& 5 & 16\\ 
Myra \& Burrows (1990)&0.13&11 & 13 & 24 &  $0.85:1:1.85$ & 30
& 30 & 16\\ 
Janka \& Hillebrandt (1989b)
                     &0.3 &  8 & 14 & 16 &  $0.57:1:1.14$ & 30
& 220 & 65\\ 
Suzuki (1989)        & 1  & 9.5& 13 & 15 &  $0.73:1:1.15$ & 4
& 4 & 3\\ 
                     &20  & 8  & 10 &  9 &  $0.80:1:0.90$ & 0.3
& 0.3 & 0.07\\ 
Suzuki (1991)        & 1  & 9.5& 13 & 15 &  $0.73:1:1.15$ & 3
& 3 & 3\\ 
                     &15  & 8  &  9 & 9.5&  $0.89:1:1.06$ & 0.4
& 0.4 & 0.3\\ 
Suzuki (1993)        & 1  & 9  & 12 & 13 &  $0.75:1:1.08$ & 3
& 3 & 3\\ 
                     &15  & 7  &  8 &  8 &  $0.88:1:1.00$ & 0.3
& 0.3 & 0.3\\ 
\\
\\
\\
Accretion-Phase Model~I (original) 
&0.32&13 & 15 & 18 &  $0.86:1:1.20$ & 31 & 29 & 14\\ 
Accretion-Phase Model~I (our run) 
&0.32&12 & 14 & 14 &  $0.84:1:1.02$ & 32 & 32 & 18\\ 
Accretion-Phase Model~II (original) 
&0.15&13 & 16 & 17 &  $0.82:1:1.09$ & 66 & 68 & 32\\ 
Accretion-Phase Model~II (our run) 
&0.15&13 & 15 & 16 &  $0.84:1:1.02$ & 74 & 74 & 28\\ 
Buras et al.\ (personal comm.) 
&0.25&14.1 & 16.5 & 16.8 &  $0.85:1:1.02$ & 43 & 44 & 32\\ 
\\
\multicolumn{9}{l}{\bf The following lines show
\boldmath{$\langle\epsilon\rangle_{\rm rms}$} instead of
\boldmath{$\langle\epsilon\rangle$}}\\ 
Mezzacappa et~al.\ (2001)&0.5 & 16 & 19 & 24 &  $0.84:1:1.26$ & 25
& 25 & 8\\ 
Liebend\"orfer et~al.\ (2001)&0.5 & 19 & 21 & 24 & $0.90:1:1.14$ & 30
& 30 & 10\\ 
\enddata
\tablecomments{We give the time post bounce (tpb) in s,
$\langle\epsilon\rangle$ in MeV, and $L_\nu$ in $10^{51}~\rm
erg~s^{-1}$.}
\end{deluxetable}

The most recent published Livermore simulation is a $20\,M_\odot$ star
(Totani et~al.\ 1998).  It shows an astonishing degree of luminosity
equipartition from the accretion phase throughout the early
Kelvin-Helmholtz cooling phase.  About two seconds after bounce the
$\nu_\mu$ flux falls off more slowly than the other flavors.  In
Table~\ref{tab:Literature} we show representative results for an early
and a late time. The mean energies and their ratios are consistent
with what we would have expected on the basis of our study.

With a different numerical code, Bruenn (1987) found for a
$25\,M_\odot$ progenitor qualitatively different results for
luminosities and energies.  At about 0.5~s after bounce the
luminosities and energies became stable at the values given in
Table~\ref{tab:Literature}.  This simulation is an example for an
extreme hierarchy of mean energies.

In Burrows (1988) all luminosities are said to be equal.  In addition
it is stated that for the first 5 seconds
$\langle\epsilon_{\nu_\mu}\rangle \approx 24$ MeV and the relation to
the other flavors is
$\langle\epsilon_{\nue}\rangle:\langle\epsilon_{\bar\nue}\rangle
:\langle\epsilon_{\nu_\mu}\rangle = 0.9:1:1.8$.  Detailed results are
only given for $\bar\nue$, so we are not able to add this reference to
our table.  The large variety of models investigated by Burrows (1988)
and the detailed results for $\bar\nue$ go beyond the scope of our
brief description. In a later paper Myra \& Burrows (1990) studied a
$13\,M_\odot$ progenitor model and found the extreme hierarchy of
energies shown in our table.

With the original version of our code Janka \& Hillebrandt (1989b)
performed their analyses for a $20\,M_\odot$ progenitor from a
core-collapse calculation by Hillebrandt (1987).  Of course, like our
present study, these were Monte Carlo simulations on a fixed
background model, not self-consistent simulations.  Taking into
account the different microphysics the mean energies are consistent
with our present work.  The mean energies of $\nu_{\rm e}$ were
somewhat on the low side relative to $\bar\nu_{\rm e}$ and the
$\bar\nu_{\rm e}$ luminosity was overestimated. Both can be understood
by the fact that the stellar background contained an overly large
abundance of neutrons, because the model resulted from a post-bounce
calculation which only included electron neutrino transport.

Suzuki (1989) studied models with initial temperature and density
profiles typical of proto-neutron stars at the beginning of the
Kelvin-Helmholtz cooling phase about half a second after bounce.  He
used the relatively stiff nuclear equation of state developed by
Hillebrandt \& Wolff (1985). In our table we show the results of the
model C12.  From Suzuki (1991) we took the model labeled C20 which
includes bremsstrahlung.  The model C48 from Suzuki (1993) includes
multiple-scattering suppression of bremsstrahlung. Suzuki's models are
the only ones from the previous literature which go beyond the
traditional microphysics for $\nu_\mu$ transport. It is reassuring
that his ratios of mean energies come closest to the ones we find.

Over the past few years, first results from Boltzmann solvers coupled
with hydrodynamic simulations have become available, for example the
unpublished ones that we used as our Accretion-Phase Models I and~II.
For convenience we include them in Table~\ref{tab:Literature}.
Further, we include a very recent accretion-phase model of
the Garching group (Buras et al., personal communication) that
includes the full set of microphysical input.
Finally, we include two simulations similar to the Accretion-Phase
Model~I, one by Mezzacappa et~al.\ (2001) and the other by
Liebend\"orfer et~al.\ (2001).  These latter papers show rms energies
instead of mean energies. Recalling that the former tend to be about
45\% larger than the latter these results are entirely consistent with
our accretion-phase models.  Moreover, the ratios of
$\langle\epsilon\rangle_{\rm rms}$ tend to exaggerate the spread
between the flavor-dependent mean energies because of different
amounts of spectral pinching, i.e., different effective degeneracy
parameters.  To illustrate this point we take the first two rows from
Table~\ref{tab:MultiFlavorAccretion} as an example.  The ratio of the
mean energies $\langle \epsilon_1 \rangle = 17.5$ MeV and $\langle
\epsilon_2 \rangle = 14.6$ MeV  is $\langle \epsilon_1 \rangle /
\langle \epsilon_2 \rangle = 1.19$.  Using Equation~\ref{eq:ErmsAlpha}
for $\alpha_1=2.7$ and $\alpha_2=4.4$ the ratio of rms energies
equals~1.31.

To summarize, the frequently assumed exact equipartition of the
emitted energy among all flavors appears only in some simulations of
the Livermore group and some by Suzuki.  We note that the
flavor-dependent luminosities tend to be quite sensitive to the
detailed atmospheric structure and chemical composition.  On the other
hand, the often-assumed extreme hierarchy of mean energies was only
found in the early simulations of Bruenn (1987) and of Myra \& Burrows
(1990), possibly a consequence of the neutron-star equation of state
used in these calculations.

If we ignore results which appear to be ``outliers'', the picture
emerging from Table~\ref{tab:Literature} is quite consistent with our
own findings.  For the luminosities, typically $L_{\nue} \approx
L_{\bar\nue}$ and a factor of \hbox{2--3} between this and
$L_{\nu_\mu}$ in either direction, depending on the evolutionary
phase.  For the mean energies we read typical ratios in the range of
$\langle\epsilon_{\nue}\rangle:\langle\epsilon_{\bar\nue}\rangle
:\langle\epsilon_{\nu_\mu}\rangle=0.8$--$0.9 : 1 : 1.0$--1.3.  The
more recent simulations involving a Boltzmann solvers show a
consistent behavior and will in future provide reliable information
about neutrino fluxes and spectra.  There is a clear tendency that the
inclusion of all relevant microphysics decreases the hierarchy of mean
energies.

Analyzing detected neutrino spectra from a future galactic SN will
demand very accurate predictions from self-consistent simulations.
This is necessary for both, a possible measurement of
neutrino oscillation parameters and also for testing the current SN
paradigm.

\section{Spectral Shape}
\label{sec:High-Statistics}

Thus far we have characterized the neutrino spectra by a few simple
parameters.  However, it is extremely useful to have a simple analytic
fit to the overall spectrum that can be used, for example, to simulate
the response of a neutrino detector to a SN signal.  To study the
quality of different fit functions we have performed a few
high-statistics Monte Carlo runs for the Accretion-Phase Model~I and
two of the steep power-law models ($p=10$, $q=2.5$ and 3.5) including
all interaction processes. Moreover, we have performed these runs for
the flavors $\nu_\re$, $\bar\nu_\re$, and $\nu_\mu$.

In order to get smooth spectral curves we have averaged the output of
70,000 time steps.  In addition we have refined the energy grid of the
neutrino interaction rates.  Both measures leave the previous results
unaffected but increase computing time and demand for memory
significantly.
  
\begin{figure}
\columnwidth=12cm
\plotone{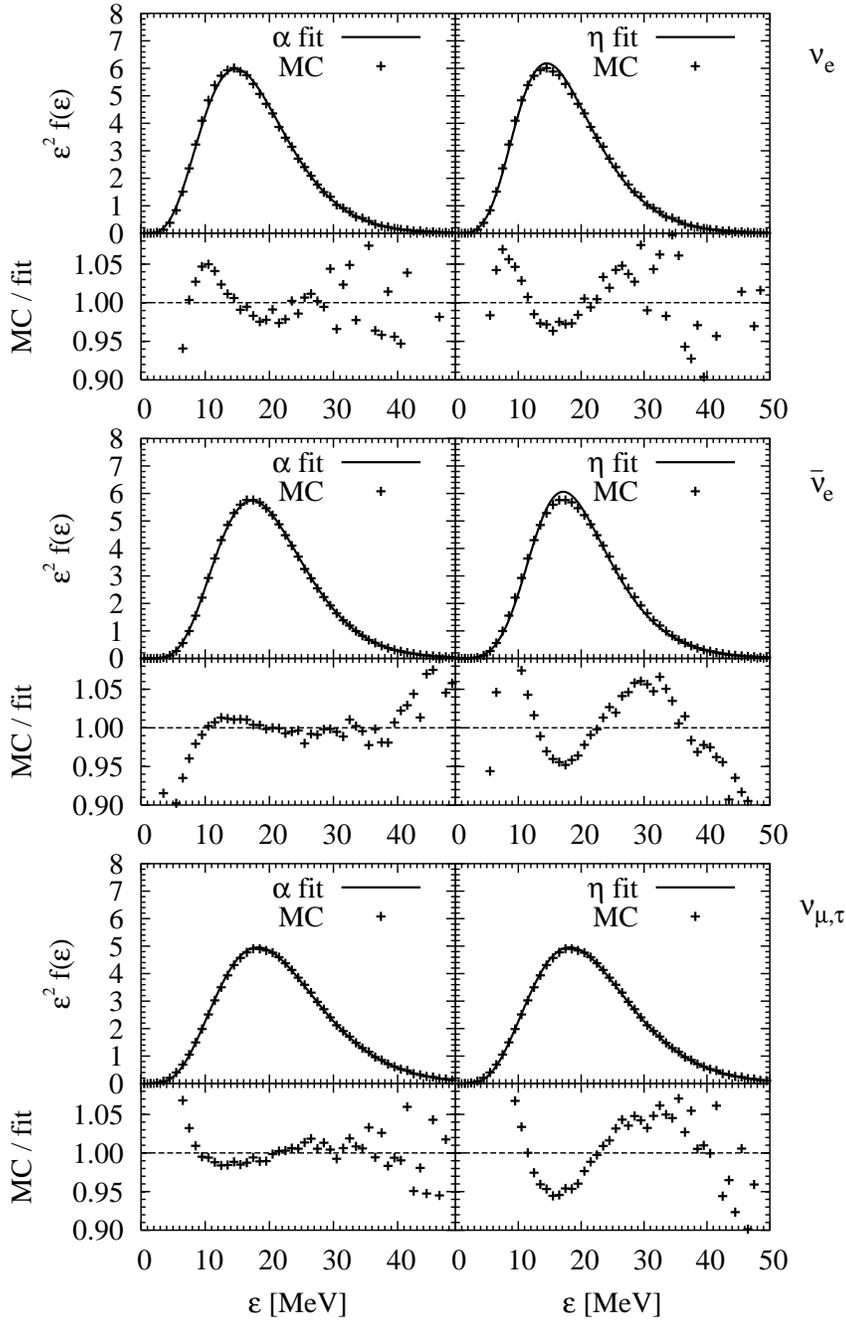}
\caption{\label{fig:highstatspecs} High-statistics spectra for
Accretion-Phase Model~I including all interaction processes.  The
Monte Carlo (MC) results are shown as crosses, the analytic fit
functions as smooth lines.  The left-hand panels use as fits
$f_\alpha(\epsilon)$ according to Equation~(\ref{eq:powerlawfit}), the
right-hand panels $f_\eta(\epsilon)$ according to
Equation~(\ref{eq:FermiDiracfit}).  Below the spectra we show the ratio
between Monte Carlo and fit.}
\end{figure}

In Figure~\ref{fig:highstatspecs} we show our high-statistics
Monte Carlo spectra (MC) for the Accretion-Phase Model~I together with
the $\alpha$-fit function $f_\alpha(\epsilon)$ defined in
Equation~(\ref{eq:powerlawfit}) and the $\eta$-fit function
$f_\eta(\epsilon)$ of Equation~(\ref{eq:FermiDiracfit}).  The analytic
functions can only fit the spectrum well over a certain range of
energies. We have chosen to optimize the fit for the event spectrum in
a detector, assuming the cross section scales with
$\epsilon^2$. Therefore, we actually show the neutrino flux spectra
multiplied with $\epsilon^2$.  Accordingly, the parameters $\alpha$
and $\bar\epsilon$, as well as $\eta$ and $T$ and the normalizations
are determined such that the energy moments
$\langle\epsilon^2\rangle$, $\langle\epsilon^3\rangle$, and
$\langle\epsilon^4\rangle$ are reproduced by the fits.

Below each spectrum we show the ratio of our Monte Carlo results with
the fit functions.  In the energy range where the statistics in a
detector would be reasonable for a galactic SN, say from 5--10~MeV up
to around 40~MeV, both types of fits represent the Monte Carlo results
nicely.  However, in all cases the $\alpha$-fit works somewhat better
than the $\eta$-fit.

\begin{figure}
\columnwidth=12cm
\plotone{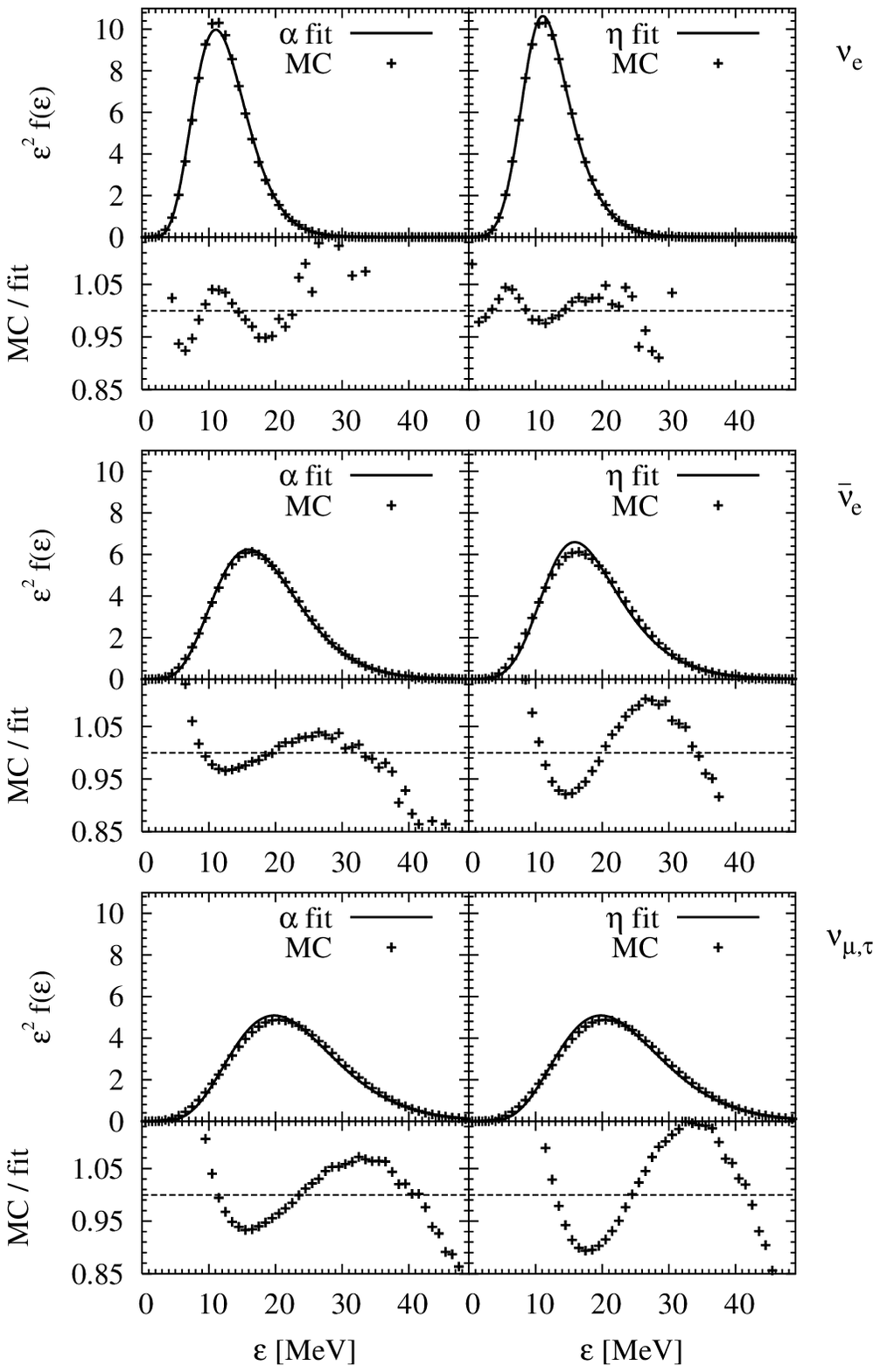}
\caption{\label{fig:3.5PowerLawEtaAlpha} Same as
Figure~\ref{fig:highstatspecs} for the $p=10$, $q=3.5$ power law.}
\end{figure}

\begin{figure}
\columnwidth=12cm
\plotone{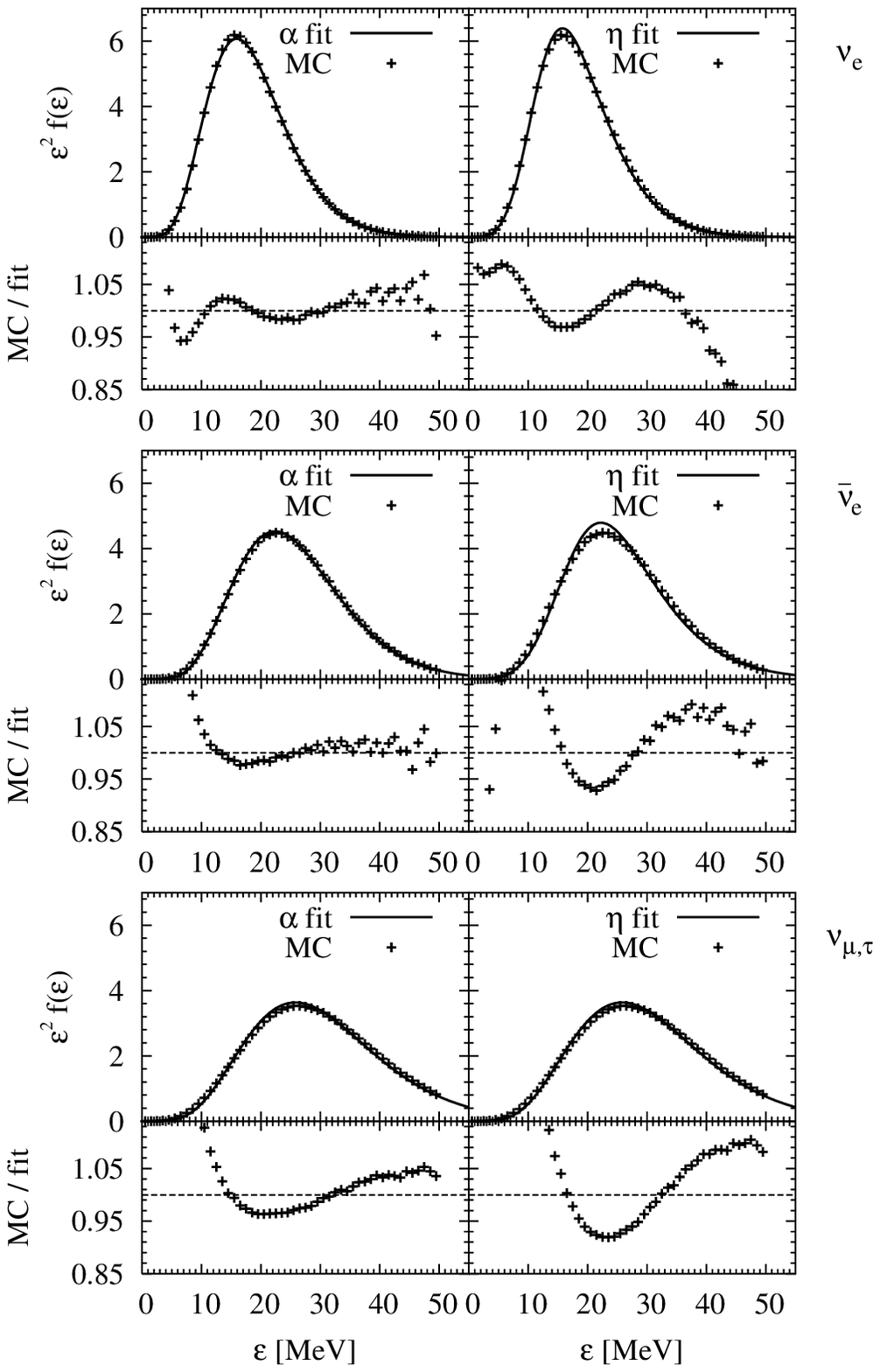}
\caption{\label{fig:2.5PowerLawEtaAlpha} Same as
Figure~\ref{fig:highstatspecs} for the $p=10$, $q=2.5$ power law.}
\end{figure}

We have repeated this exercise for the steep power-law models with
$q=2.5$ and the one with $q=3.5$ and show the results in
Figures~\ref{fig:2.5PowerLawEtaAlpha} and \ref{fig:3.5PowerLawEtaAlpha},
respectively. The quality of the fits is comparable to the previous
example.


\section{Summary}

In comparing the spectra of all neutrino flavors we find a consistent
behavior of all the considered models.  Once all relevant interactions
are included the differences between $\bar\nue$ and $\numu$ spectra
become rather small.  Concerning mean energies from our two realistic
profiles that represent the accretion phase, we always find
$\langle\epsilon_{\nue}\rangle : \langle\epsilon_{\bar\nue}\rangle :
\langle\epsilon_{\nu_\mu}\rangle = 0.8$--$0.9 : 1 : 1.0$--1.09.  These
findings are completely consistent with full hydrodynamic simulations
that include the full set of interactions, i.e., the recent
Garching-group models.  For the luminosities we find good agreement,
too.  The runs commonly show $L_{\nue} \approx L_{\bar\nue}\approx 2 
\, L_{\nu_\mu}$.

For our power-law models we find somewhat stronger hierarchies in the
mean energies.  It is not clear if this will still hold in
self-consistent simulations.  There are currently no self-consistent
models with an accurate treatment of the neutrino transport available
that continued the calculations to the late phases where steep
proto-neutron-star profiles are found.

In our collection of previous literature we found a variety of
different predictions.  The strong hierarchies in mean energies that
were commonly assumed in the literature for calculating
neutrino-oscillation effects are rather outliers.  Also the exact
equipartition of luminosities is not found in elaborate neutrino
transport models.

The standard picture of flavor-dependent SN neutrino spectra needs to
be changed.  Oscillation studies should allow for large differences
in the $\bar\nue$ and $\bar\numu$ luminosities, and for almost equal
mean energies of these flavors.  Results of different self-consistent
simulations show large variations in their absolute values.
Therefore, only concepts that rely on differential effects of
flavor-dependent SN neutrino spectra can have predictive power for
identifying oscillation signatures.

\cleardoublepage

\chapter{Detecting Oscillations of SN Neutrinos}
\label{chap:EarthMatter}

By now, neutrino oscillations are firmly established and the next goal
is to pin down the oscillation parameters that are still unknown,
like the neutrino mass hierarchy and the small mixing angle
$\theta_{13}$.  Taking our findings on neutrino spectra into account
requires new methods for extracting information, but enables us to use
ratios of observables instead of their absolute magnitude.

We present a method for identifying the effect of earth-matter
oscillations in a detected SN neutrino signal by comparing the
time dependence of the energy deposited in two detectors.  By
identifying spectral modulations with a Fourier-transform method the
detection at a single detector is possible, but demands better energy
resolution than the two-detector method.  Finding oscillation
signatures in the signal can pin down the mass hierarchy or restrict
the small mixing angle.

Parts of this chapter were published in a similar form in our
publications\\
A.~S.~Dighe, M.~T.~Keil, and G.~G.~Raffelt,\\
``Detecting the neutrino mass hierarchy with a supernova at IceCube,''
JCAP, in press (2003), hep-ph/0303210, and 
``Identifying earth matter effects on supernova neutrinos at a single
detector,'' hep-ph/0304150.

\section{Oscillations of SN Neutrinos}
More and more accurate measurements have firmly established neutrino
oscillations.  After many years of atmospheric- and solar-neutrino
experiments and also long-baseline experiments (KamLAND and K2K) we
have a good understanding of how neutrinos oscillate (Bahcall,
Gonzales-Garcia, \& Pe\~na-Garay 2003, Fogli et al.\ 2002,
Gonzales-Garcia \& Nir 2002, de~Holanda \& Smirnov 2003). The flavors
$\nue$, $\numu$, and $\nu_\tau$ are in weak interaction eigenstates
and therefore relevant when we talk about interactions of particles.
Each of these weak eigenstates is a non-trivial superposition of three
mass eigenstates $\nu_1$, $\nu_2$, and $\nu_3$,
\be
\pmatrix{\nue\cr \nu_\mu \cr \nu_\tau\cr}
=U\pmatrix{\nu_1\cr \nu_2 \cr \nu_3\cr}\,,
\ee
where $U$ is the leptonic mixing matrix that can be written in the 
canonical form
\be
U=\pmatrix{1&0&0\cr0&c_{23}&s_{23}\cr0&-s_{23}&c_{23}\cr}
\pmatrix{c_{13}&0&e^{i\delta}s_{13}\cr0&1&0\cr
-e^{-i\delta}s_{13}&0&c_{13}\cr}
\pmatrix{c_{12}&s_{12}&0\cr-s_{12}&c_{12}&0\cr0&0&1\cr}\,.
\ee
Here $c_{12}=\cos\theta_{12}$ and $s_{12}=\sin\theta_{12}$ etc., and
$\delta$ is a phase that can lead to CP-violating effects, that are,
however, irrelevant for SN neutrinos.  The three matrices then just
correspond to the rotation matrices in the 2-3, 1-3, and 1-2
planes, respectively.

The mass eigenstates are most convenient for describing neutrino
propagation in space.  If there are three different mass
eigenstates there can be three different masses.  The oscillation,
i.e., the interference of the three eigenstates, depends on the
difference of the squares of their masses.  Usually, the mass-squared
differences are given as $\Delta m^2_{ij} \equiv m^2_i - m^2_j$.  Of
course, only two of them are independent.  Compared to the current
limit on the absolute neutrino mass, experiments find very small
mass-squared differences.  Two of the three masses appear to be rather
degenerate.  The third mass eigenstate can then lie far above or far
below this doublet.  Whether the third mass eigenstate lies above or
below the almost degenerate doublet is referred to as normal or
inverted hierarchy, respectively.  Which hierarchy is realized in
nature is an open question.

The mass squared differences relevant for the atmospheric and solar
neutrino oscillations show the strong hierarchy $\msqa \gg
\msqs$. This hierarchy, combined with the observed smallness of the
angle $\theta_{13}$ at CHOOZ (Apollonio et al.~1999) implies that the
atmospheric neutrino oscillations essentially decouple from the solar
ones and each of these is dominated by only one of the mixing angles.
The atmospheric neutrino oscillations are controlled by $\theta_{23}$
that may well be maximal (45$^\circ$). The solar case is dominated by
$\theta_{12}$, that is large but not maximal.  To a reasonably good
accuracy $\msqa \approx \dmsq_{32}$ and $\msqs \approx \dmsq_{21}$ and
therefore $\dmsq_{31} \approx \dmsq_{32}$.  From a global 3-flavor
analysis of all data one finds the 3$\sigma$  ranges for the mass
differences and mixing angles summarized in
Table~\ref{tab:mixingparameters}.

\begin{deluxetable}{lll}
\tablecaption{\label{tab:mixingparameters}Neutrino mixing parameters}
\tablewidth{0pt}
\tablehead{Observation&Mixing angle&$\Delta m^2$ [meV$^2$]}
\startdata
Sun, KamLAND & $\theta_{12}=$ 27$^\circ$--42$^\circ$&
$\dmsq_{21} =$ 55--190\\
Atmosphere, K2K  & $\theta_{23} =$ 32$^\circ$--60$^\circ$&
$|\dmsq_{32}|=$ 1400--6000\\
CHOOZ & $\theta_{13} {}<14^\circ$ & 
$\dmsq_{31} \approx \Delta m_{32}^2$\\
\enddata
\tablecomments{Neutrino mixing parameters from a
global analysis of all experiments 
(3$\sigma$~ranges) by Gonzalez-Garcia \& Nir (2002).}
\end{deluxetable}

Obviously, in the context of SNe neutrino oscillations can only be
found if flavor-dependent differences in the SN neutrino fluxes and
spectra are present.  We denote the fluxes of $\bar\nue$ and
$\bar\numu$ at earth that would be observable in the absence of
oscillations by $F_\ebar^0$ and $F_{\bar\mu}^0$, respectively.  In the
presence of oscillations a $\bar\nue$ detector actually observes
\be
F_{\ebar}^D(\epsilon)  =  \bar{p}^D(\epsilon) F_{\ebar}^0(\epsilon) + 
\left[ 1-\bar{p}^D(\epsilon) \right] F_{\bar\mu}^0\,,
\label{eq:feDbar}
\ee
where $\bar{p}^D(\epsilon)$ is the $\bar\nue$ survival probability after
propagation through the SN mantle and perhaps part of the earth before
reaching the detector.  Water Cherenkov detectors as well as
scintillation detectors can only detect $\bar\nue$ with a good
efficiency.  Therefore we give the $\bar\nue$ flux at the detector.

When neutrinos travel in matter their oscillatory behavior can change
significantly.  Depending on the neutrino source the path towards a
detector leads through matter surrounding the source, like in stars or
in a SN and might also go through the earth.  This matter has a
significant population of electrons, but other leptons are absent.
By charged-current interactions the medium distinguishes between
electron flavor and other flavors, which implies a potential for
electron neutrinos and anti-neutrinos.  In other words, the medium can
change its refractive index for electron (anti-)neutrinos. Since the
other flavors are unaffected, the oscillatory behavior can change
significantly, depending on the electron density.

A significant modification of the survival probability due to the
propagation through the earth appears only for those combinations of
neutrino mixing parameters shown in Table~\ref{tab:EarthCases}.  The
earth matter effect depends strongly on two parameters, the sign of
$\dmsq_{32}$ and the value of $|\theta_{13}|$ (Dighe \& Smirnov 2000,
Dighe 2001).  The normal hierarchy corresponds to $m_1<m_2<m_3$, i.e.,
$\dmsq_{32}>0$, whereas the inverted hierarchy corresponds to
$m_3<m_1<m_2$, i.e., $\dmsq_{32}<0$.  Note that the presence or
absence of the earth effect discriminates between values of $\sin^2
\theta_{13}$ less or greater than $10^{-3}$, i.e., $\theta_{13}$ less
or larger than about $1.8^\circ$.  Thus, the earth effect is sensitive
to values of $\theta_{13}$ that are much smaller than the current
limit (Table~\ref{tab:mixingparameters}).

\begin{deluxetable}{lll}
\tablecaption{\label{tab:EarthCases}The earth effect in a SN signal.}
\tablewidth{0pt}
\tablehead{13-Mixing&Normal Hierarchy&Inverted Hierarchy}
\startdata
$\sin^2\theta_{13} \lsim 10^{-3}$&$\nue$ and $\bar\nue$
&$\nue$ and $\bar\nue$\\
$\sin^2\theta_{13} \gsim 10^{-3}$&$\bar\nue$&$\nue$ \\
\enddata
\tablecomments{The earth effect appears for the indicated flavors in a
SN signal.}
\end{deluxetable}

Let us consider those scenarios where the mass hierarchy and the value
of $\theta_{13}$ are such that the earth effect appears for
$\bar\nue$.  In such cases the $\bar\nue$ survival probability
$\bar{p}^D(E)$ is given by
\bea
\bar{p}^D &\approx& \cos^2 \theta_{12} - \sin 2\bar{\theta}_{\re 2}^\oplus
~ \sin (2\bar{\theta}_{\re 2}^\oplus - 2\theta_{12}) \nonumber \\
&\times& \sin^2 \left(
\frac{\overline{\dmsq_\oplus}}{10^{-5} {\,\rm eV}^2}\,\frac{ L }{1000 \,\rm
km}\,\frac{12.5\,\rm MeV}{\epsilon} \right)\,,
\label{eq:pbar}
\eea
where the energy dependence of all quantities will always be implicit.
Here $\bar{\theta}_{\re 2}^\oplus$ is the mixing angle between $\bar\nue$
and $\bar{\nu}_2$ in earth matter while $\overline{\dmsq_\oplus}$ is
the mass squared difference between the two anti-neutrino mass
eigenstates $\bar{\nu}_1$ and $\bar{\nu}_2$, $L$~is the distance
traveled through the earth, and $\epsilon$ is the neutrino energy.  We
have assumed a constant matter density inside the earth, which is a
good approximation for $L<10,500$ km, i.e., as long as the neutrinos do
not pass through the core of the earth.

\vfill

\section{Detecting Oscillations With Two Distant Detectors}
\label{sec:ice}

\subsection{The Basic Principle}

How significant the oscillation signal in a detected SN neutrino
spectrum is, basically depends on the difference between the
$\bar\nue$ and $\bar\numu$ fluxes and the distance traveled through
the earth.  There is no exact prediction for the spectra emitted by a
SN, but some features are established.  The 
mean energies of $\bar\nue$ and $\bar\numu$ will be very similar, a
difference of typically 0--20\% should be expected.  At the
same time the number fluxes will be rather different by up to a factor
2 in any direction.  The ratio of fluxes will certainly change with
time, because there are two distinct phases responsible for the
neutrino signal, namely the accretion phase and Kelvin-Helmholtz
cooling of the proto-neutron star (Chapter~\ref{chap:paradigm}).  

With the time dependence of the flux difference it is possible to find
an oscillation signature without a detailed knowledge of the original
SN signal.  Oscillations inside the earth depend on the flux
difference and will therefore also change with time.  Comparing the
time dependence of a neutrino signal of two detectors yields a good
chance for detecting the earth effect.  The setup should be as
follows: one detector sees the SN from above, i.e., without earth
effect, and another detector that neutrinos reach after traveling
through the earth.  The chance for encountering this setup increases
with the separation of the detectors.  Detectors that are separated by
a long distance and will be operational for many decades are the
future IceCube detector in Antarctica (Ahrens et al.~2002b) and, for
example, the Super-Kamiokande detector in Japan.

\subsection{The SN Signal at IceCube}

It has been recognized for a long time that neutrino telescopes
detecting Cherenkov light in ice can detect a SN neutrino burst
because the Cherenkov glow of the ice can be identified as
time-correlated noise among all phototubes (Halzen, Jacobsen, \& Zas
1994, 1996).  This approach has been used by the neutrino telescope
AMANDA to exclude the occurrence of a galactic SN over a recent
observation period (Ahrens et al.~2002a).

The SN neutrinos streaming through the antarctic ice interact
according to $\bar\nue \rp\to \rn \re^+$ and some other less important
reactions. The positrons, in turn, emit Cherenkov light producing a
homogeneous and isotropic glow of the ice.  The optical modules (OMs)
that are frozen into the ice are immersed in this diffuse bath of
photons and pick up a number corresponding to their angular acceptance
and quantum efficiency.  Estimating the event rate of a SN signal
induced by the $\bar\nue \rp\to \rn \re^+$ reaction yields (Dighe et
al.~2003a) 
\be
\label{eq:rateprediction}
\Gamma_{\rm events}=62~{\rm s}^{-1}\,
\frac{L_{\bar\nue}}{10^{52}~\rm erg~s^{-1}}
~\left(\frac{10~\rm kpc}{D}\right)^2\,
f_{\rm flux}\,f_{\rm det} \;,
\ee
where $L_{\bar\nue}$ is the $\bar\nue$ luminosity after flavor
oscillations, and D the distance between SN and detector.  The factors
$f_{\rm flux}$ and $f_{\rm det}$ parameterize the flux and detector
characteristics, respectively.  Both factors are of order 1.  For the
flux we have
\bea
f_{\rm flux}&=&
\frac{15~\rm MeV}{\langle \epsilon_{\bar\nue}\rangle}~\frac{8}{15}
~\frac{\langle \epsilon_{\bar\nue}^3\rangle}{(15~\rm
MeV)^3} \nonumber \\
&=&\frac{(3+\alpha)(2+\alpha)}{(1+\alpha)^2}~\frac{8}{15}
~\frac{\langle \epsilon_{\bar\nue}\rangle^2}{(15~\rm MeV)^2}\;,
\eea
where we used Equation~(\ref{eq:momentsalpharelation}) in the last
line, i.e., we applied our $\alpha$ parameterization.  For $\alpha=3$
and $\langle \epsilon_{\bar\nue}\rangle=15$~MeV the factor is 1.

The factor $f_{\rm det}$ contains the efficiency for converting the
neutrino signal into photons by the ice as well as the efficiency for
detecting these photons.  For our purpose we assume $f_{\rm det}=1$.

In the 4800 OMs of IceCube a SN in our galaxy would yield a total
event number of about $1.5\times 10^6$ photons if we assume the SN
radiates $5\times 10^{52}$~erg.  This rate needs to be compared to the
background counting rate of 300 Hz per OM and thus for a duration of
10~s we expect $1.44\times 10^7$ photons in total.  Assuming Poisson
fluctuations, the uncertainty of this number is $3.8\times10^3$, i.e.,
0.1\% of the SN signal.  Therefore, one can determine the SN signal
with a statistical sub-percent precision, ignoring for now problems of
absolute detector calibration.

\begin{figure}[t]
\columnwidth=11cm
\plotone{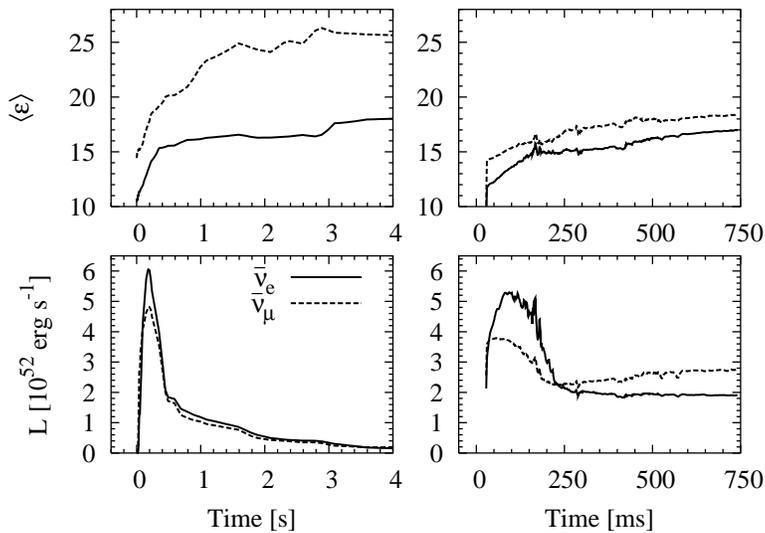}
\caption{\label{fig:simulations}Supernova $\bar\nue$ and $\bar\nu_\mu$
light curves and average energies.  {\em Left:} Livermore
simulation (Totani et al.1999).  {\em Right:} Garching
simulation (Raffelt et al.~2003).}
\end{figure}

In order to illustrate the statistical power of IceCube to observe a
SN signal we use two different numerical SN simulations.  The first
was performed by the Livermore group (Totani et al.~1998) that
involves traditional input physics for $\numu$ interactions and a
flux-limited diffusion scheme for treating neutrino transport. The
great advantage of this simulation is that it covers the full
evolution from infall over the explosion to the Kelvin-Helmholtz
cooling phase of the proto-neutron star.  We show the Livermore
$\bar\nue$ and $\bar\numu$ lightcurves in Figure~\ref{fig:simulations}
(left panels).  For all flavors these curves are also displayed in
Figure~\ref{fig:lllLumi}.

Our second simulation was performed with the Garching code (Rampp \&
Janka 2002).  It includes all relevant neutrino interaction rates,
including nucleon bremsstrahlung, neutrino pair processes, weak
magnetism, nucleon recoils, and nuclear correlation effects.  The
neutrino transport part is based on a Boltzmann solver.  The
neutrino-radiation hydrodynamics program allows one to perform
spherically symmetric as well as multi-dimensional simulations.  The
progenitor model is a 15$\,M_{\odot}$ star with a $1.28\,M_{\odot}$
iron core.  The period from shock formation to 468~ms after bounce was
evolved in two dimensions. The subsequent evolution of the model is
simulated in spherical symmetry.  At 150~ms the explosion sets in,
although a small modification of the Boltzmann transport was necessary
to allow this to happen (Janka et al.~2003). Recall that unmanipulated
full-scale models with an accurate treatment of the microphysics
currently do not obtain explosions
(Section~\ref{sec:SelfconsinstentSims}).  This run will be continued
beyond the current epoch of 750~ms post bounce; we here use the 
preliminary results currently available (Raffelt et al.~2003).  We
show the Garching $\bar\nue$ and $\bar\nu_\mu$ lightcurves in
Figure~\ref{fig:simulations} (right panels).

We take the Livermore simulation to represent traditional predictions
for flavor-dependent SN neutrino fluxes and spectra that were used in
many previous discussions of SN neutrino oscillations.  The Garching
simulation is taken to represent a situation when the $\bar\numu$
interactions are more systematically included so that the
flavor-dependent spectra and fluxes are more similar than had been
assumed previously.  We think it is useful to juxtapose the IceCube
response for both cases.

Another difference is that in Livermore the accretion phase lasts
longer. Since the explosion mechanism is not finally settled, it is
not obvious which case is more realistic.  Moreover, there could be
differences between different SNe. The overall features are certainly
comparable between the two simulations.

\begin{figure}[t]
\columnwidth=11cm
\plotone{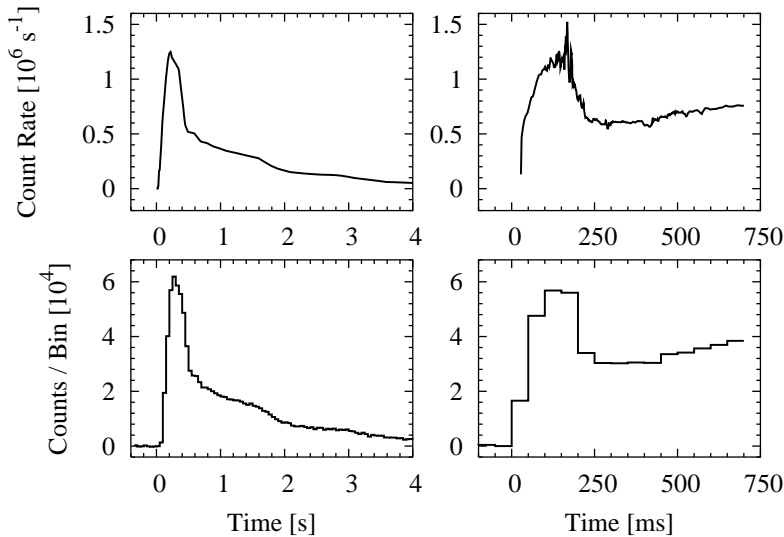}
\caption{\label{fig:SimulationsIceCube}Supernova signal in IceCube
assuming a distance of 10~kpc, based on the Livermore simulation
(left) and the Garching one (right), in both cases ignoring flavor
oscillations.  In the bottom panels we have used 50~ms bins and have
added noise from a background rate of 300~Hz per OM.}
\end{figure}

In Figure~\ref{fig:SimulationsIceCube} we show the expected counting
rates in IceCube as given in Equation~(\ref{eq:rateprediction}) for an
assumed distance of 10~kpc and 4800 OMs for the Livermore (left)
and Garching (right) simulations.  We also show this signal in 50~ms
bins where we have added noise from a background of 300~Hz per OM. The
baseline is at the average background rate so that negative counts
correspond to downward background fluctuations.

One could easily identify the existence and duration of the accretion
phase and thus test the standard delayed-explosion scenario. One could
also measure the overall duration of the cooling phase and thus
exclude the presence of significant exotic energy losses.  Therefore,
many of the particle-physics limits based on the SN~1987A
neutrinos (Raffelt 1999) could be supported with a statistically
serious signal.  If the SN core were to collapse to a black hole after
some time, the sudden turn-off of the neutrino flux could be
identified. In short, when a galactic SN occurs, IceCube is a powerful
stand-alone neutrino detector, providing us with a plethora of
information that is of fundamental astrophysical and particle-physics
interest.

\subsection{The Oscillation Signal at Ice Cube}

In order to calculate the extent of the earth effect for IceCube, we
will assume that the relevant mixing parameters are $\Delta
m^2_{12}=6\times10^{-5}~\rm eV^2$ and $\sin^2(2\theta_{12})=0.9$.  We
further assume that the source spectra are given by the functional
form of our $\alpha$ fit Equation~(\ref{eq:powerlawfit}).  The values
of the parameters $\alpha$ and $\langle \epsilon \rangle$ for both the
$\bar\nue$ and $\bar\numu$ spectra are in general time dependent.

\begin{figure}[b]
\columnwidth=9cm
\plotone{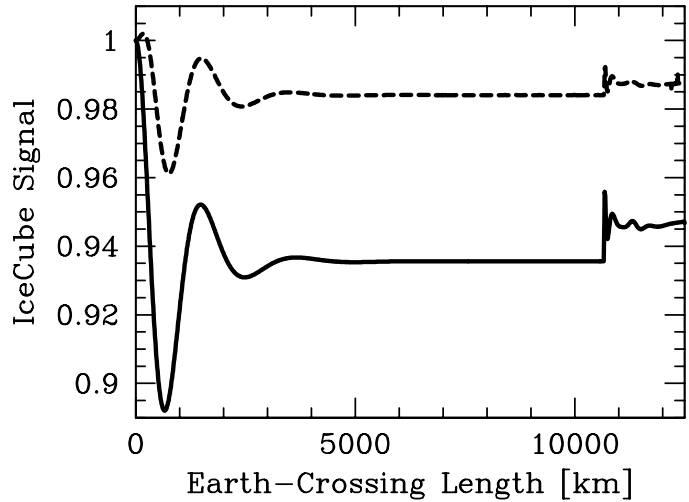}
\caption{\label{fig:exampleEartheffect}Variation of the expected
IceCube signal with neutrino earth crossing length $L$ for the assumed
flux and mixing parameters of Table~\ref{tab:ExampleCases}.  The
signal is normalized to 1 when no earth effect is present, i.e.\ for
$L=0$.  The dashed line is for the case representing the accretion
phase, the solid line for the cooling phase.}
\end{figure}

In Figure~\ref{fig:exampleEartheffect} we show the variation of the
expected IceCube signal with earth-crossing length $L$ for the two
sets of parameters detailed in Table~\ref{tab:ExampleCases}.  The
first could be representative of the accretion phase, the second of
the cooling signal.  We use the two-density approximation for the
earth density profile, where the core has a density of $11.5~\rm
g~cm^{-3}$ and a radius of 3,500~km, while the density of the earth
mantle was taken to be $4.5~\rm g~cm^{-3}$.  We observe that for short
distances, corresponding to near-horizontal neutrino trajectories, the
signal varies strongly with $L$. Between about 3,000 and 10,500~km it
reaches an asymptotic value that we call the ``asymptotic mantle
value.''  For Case~(a), this value corresponds to about 1.5\%
depletion of the signal, whereas for~(b) it corresponds to about 6.5\%
depletion.

\begin{figure}[t]
\columnwidth=11cm
\plotone{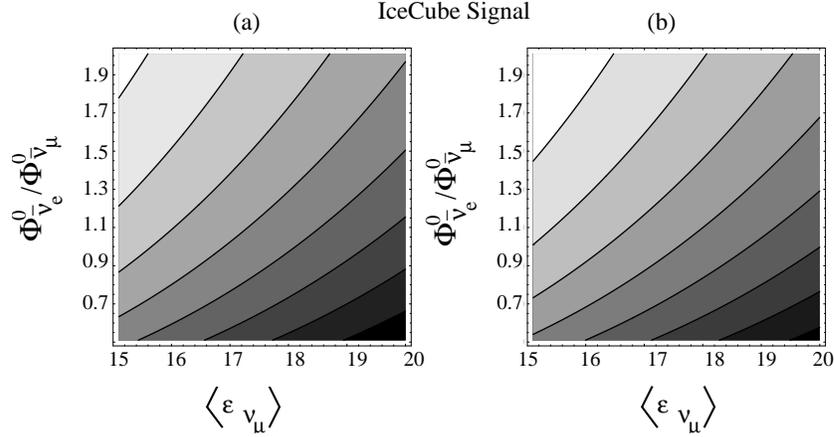}
\caption{\label{fig:cont} Asymptotic IceCube signal modification by
the earth effect.  The fixed flux parameters are (a)~$\langle
\epsilon_{\bar\nue}\rangle=15~\rm MeV$, $\alpha_{\bar\nue}=4.0$, and
$\alpha_{\bar\nu_\mu}=3.0$ and (b)~$\langle \epsilon_{\bar\nue}\rangle=15~\rm
MeV$, $\alpha_{\bar\nue}=\alpha_{\bar\nu_\mu}=3.0$.  The contours are
equally spaced starting from 1.02 (light) in 0.02 decrements to
smaller values (darker).}
\end{figure}

\begin{deluxetable}{llllllll}
\tablecaption{\label{tab:ExampleCases}Flux parameters for two
representative cases.}
\tablewidth{0pt}
\tablehead{
Example&Phase&$\langle \epsilon_{\bar\nue}\rangle$&
$\langle \epsilon_{\bar\nu_\mu}\rangle$&
$\alpha_{\bar\nue}$&$\alpha_{\bar\nu_\mu}$&
$\Phi^0_{\bar\nue}/\Phi^0_{\bar\nu_\mu}$&
Asymptotic\\
&&[MeV]&[MeV]&&&&Earth Effect}
\startdata
(a)&Accretion&15&17&4&3&1.5&$-1.5$\%\\
(b)&Cooling  &15&18&3&3&0.8&$-6.5$\%\\
\enddata
\end{deluxetable}

Beyond an earth-crossing length of about 10,500~km, the neutrinos have
to cross the earth core with another large jump in density.  The core
effects change the asymptotic mantle value by roughly 1\% as can be
seen in Figure~\ref{fig:exampleEartheffect}. We neglect the core effects
in the following analysis, and the ``asymptotic value'' always refers
to the asymptotic mantle value.

For the largest part of the sky the earth effect either appears with
this asymptotic value (``neutrinos coming from below''), or it does not
appear at all (``neutrinos from above''). Therefore, we now focus on
the asymptotic value and study how the signal modification depends on
the assumed flux parameters. In Table~\ref{tab:asymptotic1} we show
the signal modification for $\langle \epsilon_{\bar\nue}\rangle=15~\rm MeV$,
$\alpha_{\bar\nue}=4.0$, and $\alpha_{\bar\nu_\mu}=3.0$ as a function
of $\langle \epsilon_{\bar\nu_\mu}\rangle$ and the flux ratio
$\Phi^0_{\bar\nue}/\Phi^0_{\bar\nu_\mu}$. In
Table~\ref{tab:asymptotic2} we show the same with
$\alpha_{\bar\nue}=\alpha_{\bar\nu_\mu}=3.0$.  The results are shown in
the form of contour plots in Figure~\ref{fig:cont}.

\begin{deluxetable}{lllllll}
\tablecaption{\label{tab:asymptotic1}Asymptotic IceCube signal modification
by the earth effect}
\tablewidth{0pt}
\tablehead{Flux ratio&\multicolumn{6}{c}{$\langle
\epsilon_{\bar\nu_\mu}\rangle$ [MeV]}\\
$\Phi^0_{\bar\nue}/\Phi^0_{\bar\nu_\mu}$& 15&16&17&18&19&20} 
\startdata
2.0 & 1.026 & 1.014 & 1.002 & 0.988 & 0.974 & 0.960 \\
1.9 & 1.023 & 1.011 & 0.999 & 0.985 & 0.971 & 0.956 \\
1.8 & 1.021 & 1.009 & 0.995 & 0.982 & 0.967 & 0.952 \\
1.7 & 1.018 & 1.005 & 0.992 & 0.978 & 0.963 & 0.948 \\
1.6 & 1.015 & 1.002 & 0.988 & 0.974 & 0.959 & 0.944 \\
1.5 & 1.012 & 0.998 & 0.984 & 0.969 & 0.954 & 0.939 \\
1.4 & 1.008 & 0.994 & 0.980 & 0.965 & 0.949 & 0.934 \\
1.3 & 1.004 & 0.990 & 0.975 & 0.960 & 0.944 & 0.928 \\
1.2 & 1.000 & 0.985 & 0.970 & 0.954 & 0.938 & 0.922 \\
1.1 & 0.995 & 0.980 & 0.964 & 0.948 & 0.932 & 0.915 \\
1.0 & 0.989 & 0.974 & 0.957 & 0.941 & 0.925 & 0.908 \\
0.9 & 0.983 & 0.967 & 0.950 & 0.934 & 0.917 & 0.901 \\
0.8 & 0.976 & 0.959 & 0.942 & 0.925 & 0.909 & 0.892 \\
0.7 & 0.967 & 0.950 & 0.933 & 0.916 & 0.899 & 0.883 \\
0.6 & 0.958 & 0.940 & 0.923 & 0.906 & 0.889 & 0.873 \\
0.5 & 0.946 & 0.928 & 0.911 & 0.894 & 0.877 & 0.862 \\
\enddata
\tablecomments{The fixed flux parameters are $\langle
\epsilon_{\bar\nue}\rangle=15~\rm MeV$, $\alpha_{\bar\nue}=4.0$, and
$\alpha_{\bar\nu_\mu}=3.0$.} 
\end{deluxetable}

\begin{deluxetable}{lllllll}
\tablecaption{\label{tab:asymptotic2}
Same as Table~\ref{tab:asymptotic1} with
$\alpha_{\bar\nue}=\alpha_{\bar\nu_\mu}=3.0$.}
\tablewidth{0pt}
\tablehead{Flux ratio&\multicolumn{6}{c}{$\langle
\epsilon_{\bar\nu_\mu}\rangle$ [MeV]}\\
$\Phi^0_{\bar\nue}/\Phi^0_{\bar\nu_\mu}$&
15&16&17&18&19&20}
\startdata
2.0 & 1.036 & 1.024 & 1.012 & 1.000 & 0.986 & 0.972 \\
1.9 & 1.033 & 1.022 & 1.010 & 0.996 & 0.983 & 0.968 \\
1.8 & 1.031 & 1.019 & 1.006 & 0.993 & 0.979 & 0.964 \\
1.7 & 1.028 & 1.016 & 1.003 & 0.989 & 0.975 & 0.960 \\
1.6 & 1.025 & 1.013 & 0.999 & 0.985 & 0.971 & 0.955 \\
1.5 & 1.022 & 1.009 & 0.995 & 0.981 & 0.966 & 0.951 \\
1.4 & 1.019 & 1.005 & 0.991 & 0.976 & 0.961 & 0.945 \\
1.3 & 1.015 & 1.001 & 0.986 & 0.971 & 0.955 & 0.940 \\
1.2 & 1.010 & 0.996 & 0.981 & 0.965 & 0.949 & 0.933 \\
1.1 & 1.006 & 0.991 & 0.975 & 0.959 & 0.943 & 0.927 \\
1.0 & 1.000 & 0.985 & 0.969 & 0.952 & 0.936 & 0.919 \\
0.9 & 0.994 & 0.978 & 0.961 & 0.945 & 0.928 & 0.911 \\
0.8 & 0.986 & 0.970 & 0.953 & 0.936 & 0.919 & 0.903 \\
0.7 & 0.978 & 0.961 & 0.944 & 0.926 & 0.910 & 0.893 \\
0.6 & 0.968 & 0.950 & 0.933 & 0.916 & 0.899 & 0.882 \\
0.5 & 0.956 & 0.938 & 0.920 & 0.903 & 0.886 & 0.870 \\
\enddata
\end{deluxetable}

Even for mildly different fluxes or spectra the signal modification is
several percent, by far exceeding the statistical uncertainty of the
IceCube signal, although the absolute calibration of IceCube
may remain uncertain to within several percent. However, the signal
modification will vary with time during the SN burst. During the early
accretion phase that is expected to last for a few 100~ms and
corresponds to a significant fraction of the overall signal, the
$\bar\nu_\mu$ flux may be almost a factor of 2 smaller than the
$\bar\nue$ flux, but it will be slightly hotter and less
pinched (Raffelt et al.~2003). This corresponds to Case~(a) of
Table~\ref{tab:ExampleCases}; it is evident from
Figure~\ref{fig:exampleEartheffect} and Table~\ref{tab:asymptotic1}
that this implies that the earth effect is very small. During the
Kelvin-Helmholtz cooling phase the flux ratio is reversed with more
$\bar\nu_\mu$ being emitted than $\bar\nue$, but still with the same
hierarchy of energies. This corresponds to Case~(b); in this case the
earth effect could be about 6\%. This time dependence may allow one to
detect the earth effect without a precise absolute detector
calibration.

\begin{figure}
\columnwidth=13cm
\plotone{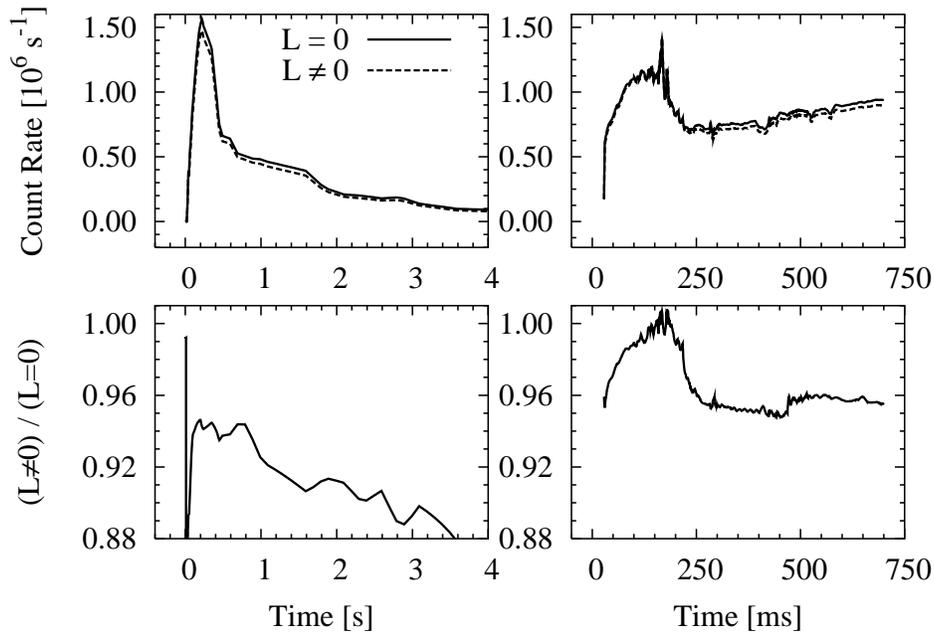}
\caption{\label{fig:percenteffect} Earth effect in IceCube. The upper
panels show the expected counting rate based on the Livermore (left)
and Garching (right) models, including flavor oscillations. The solid
line is without earth effect ($L=0$), the dashed line with asymptotic
earth effect ($L\not=0$). The lower panels show the ratio between
these curves, i.e., the ratio of counting rates with/without earth
effect.}
\end{figure}

In order to illustrate the time dependence of the earth effect we show
in Figure~\ref{fig:percenteffect} the expected counting rate in IceCube
for both the Livermore (left panels) and Garching (right panels)
simulations. In the upper panels we show the expected counting rate
with flavor oscillations in the SN mantle, but no earth effect (solid
lines), or with the asymptotic earth effect (dashed lines) that
obtains for a large earth-crossing path. Naturally the differences are
very small so that we show in the lower panels the ratio of these
curves, i.e., the expected counting rate with/without earth effect as
a function of time for both Livermore and Garching.  While for the
Livermore simulation there is a large earth effect even at early
times, the change from early to late times in both cases is around
4--5\%. Therefore, the most model-independent signature is a time
variation of the earth effect during the SN neutrino signal.

In order to demonstrate the statistical significance of these effects
we integrate the expected signal for both simulations separately for
the accretion phase and the subsequent cooling phase; the results are
shown in Table~\ref{tab:IntegratedRates}. For both simulations the
earth effect itself and its change with time is statistically highly
significant. Due to more schematic input physics the Livermore model
overestimates the differences between $\bar\nue$ and $\bar\numu$
spectra.  However, the relative change of the earth effect during
accretion and cooling is not vastly different between the two
simulations.  Recalling that the absolute detector calibration may be
very uncertain so that one has to rely on the temporal variation of
the earth effect, the difference between Livermore and Garching
becomes much smaller.  We expect that it is quite generic that the
temporal change of the earth effect is a few percent of the overall
counting rate.

\vfill

\begin{deluxetable}{lrrrr}
\tablecaption{\label{tab:IntegratedRates}
IceCube Cherenkov counts for the numerical SN models.}
\tablewidth{0pt}
\tablehead{
&\multicolumn{2}{c}{Livermore}&\multicolumn{2}{c}{Garching}\\
&\multicolumn{1}{c}{Accretion}&\multicolumn{1}{c}{Cooling}
&\multicolumn{1}{c}{Accretion}&\multicolumn{1}{c}{Cooling}}
\startdata
Integration time [s]&0--0.500&0.500--3&0--0.250&0.250--0.700\\
SN Signal [Counts]\\
\quad No Earth Effect        &519,080&818,043&173,085&407,715\\
\quad Asymptotic Earth Effect&488,093&751,137&171,310&390,252\\
\quad Difference             & 30,987&  66,906&  1,775& 17,463\\
\quad Fractional Difference&$-5.97$\%&$-8.18$\%&$-1.03$\%&$-4.28$\%\\
Background [Counts]&720,000&4,320,000&360,000&648,000\\
$\sqrt{\rm Background}$/Signal&0.16\%&0.25\%&0.35\%&0.20\%\\
\enddata
\end{deluxetable}

\subsection{Super- or Hyper-Kamiokande and IceCube}
\label{sec:SKandHK}

One can measure the earth effect in IceCube only in conjunction with
another high-statistics detector.  We do not attempt to simulate in
detail the SN signal in this other detector but simply assume that it
can be measured with a precision about as good as in IceCube.  One
candidate is Super-Kamiokande, a water Cherenkov detector that would
measure around $10^4$ events from a galactic SN at a distance of
10~kpc. Therefore, the statistical precision for the total neutrino
energy deposition in the water is around 1\% and thus worse than in
IceCube. Even though Super-Kamiokande will measure a larger number of
Cherenkov photons than IceCube, a single neutrino event will cause an
entire Cherenkov ring to be measured, i.e., the photons are highly
correlated.  Therefore, in the estimated statistical $\sqrt{N}$
fluctuation of the signal, the fluctuating number $N$ is that of the
detected neutrinos.  If the future Hyper-Kamiokande is built, its
fiducial volume would be about 30 times that of Super-Kamiokande. In
this case the statistical signal precision exceeds that of IceCube for
the equivalent observable.

We denote the equivalent IceCube signal measured by Super- or
Hyper-Kamiokande as $N_{\rm SK}$ and the IceCube signal as $N_{\rm
IC}$.  If the distances traveled by the neutrinos before reaching
these two detectors are different, the earth effect on the neutrino
spectra may be different, which will reflect in the ratio $N_{\rm
SK}/N_{\rm IC}$.  Of course, in the absence of the earth effect this
ratio equals unity by definition.

The geographical position of IceCube with respect to Super- or
Hyper-Kamiokande at a northern latitude of $36.4^\circ$ is well-suited
for the detection of the earth effect through a combination of the
signals.  Using Figure~\ref{fig:exampleEartheffect} we can already
draw some qualitative conclusions about the ratio $N_{\rm SK}/N_{\rm
IC}$.  Clearly, $N_{\rm SK}/N_{\rm IC}=1$ if neutrinos do not travel
through the earth before reaching either detector.  If the distance
traveled by neutrinos through the earth is more that 3,000~km for both
detectors, the earth effects on both $N_{\rm SK}$ and $N_{\rm IC}$ are
nearly equal and their ratio stays around unity.  If the neutrinos
come ``from above'' for SK and ``from below'' for IceCube, or vice
versa, the earth matter effect will shift this ratio from unity.

\begin{figure}[h]
\columnwidth=11cm
\plotone{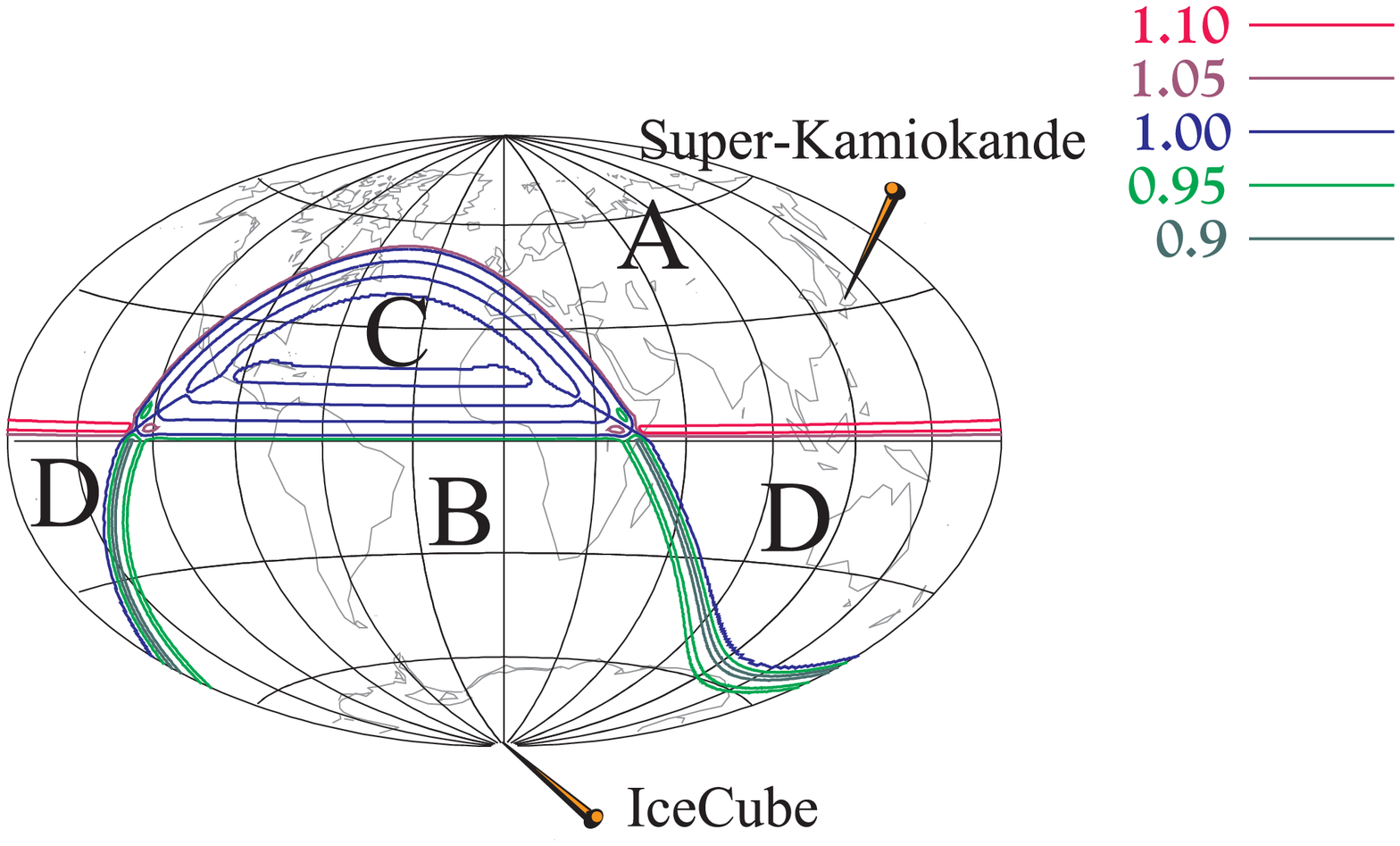}
\caption{\label{fig:Earthmap}Contours of $N_{\rm SK}/N_{\rm IC}$ on
the map of the sky projected on the earth. The regions A, B, C, D are
described in Table~\ref{tab:SkyMapRegions}.}
\end{figure}

In Figure~\ref{fig:Earthmap}, we show contours of $N_{\rm SK}/N_{\rm
IC}$ for the SN position in terms of the location on earth where the
SN is at the zenith. The map is an area preserving Hammer-Aitoff
projection so that the sizes of different regions in the figure gives
a realistic idea of the ``good'' and ``bad'' regions of the sky.  In
order to generate the contours we use the parameters of Case~(b) in
Table~\ref{tab:ExampleCases} so that the asymptotic suppression of the
signal is about 6.5\%. The sky falls into four distinct regions
depending on the direction of the neutrinos relative to either
detector as described in Table~\ref{tab:SkyMapRegions}.  When the
neutrinos come from above for both detectors (Region~D) there is no
earth effect. If they come from below in both (Region~C), the earth
effect is large in both. Depending on the exact distance traveled
through the earth, the event ratio can be large, but generally
fluctuates around~1.  In the other regions where the neutrinos come
from above for one detector and from below for the other (Regions A
and~B) the relative effect is large.

\vfill

\begin{deluxetable}{lllll}
\tablecaption{\label{tab:SkyMapRegions}Regions in Figure~\ref{fig:Earthmap}
for the earth effect in IceCube and Super-Kamiokande.}
\tablewidth{0pt}
\tablehead{
Region&Sky fraction&
\multicolumn{2}{l}{Neutrinos come from}&$N_{\rm SK}/N_{\rm IC}$\\
&&IceCube&Super-K&}
\startdata
A&0.35&below&above&1.070\\
B&0.35&above&below&0.935\\
C&0.15&below&below&Fluctuations around 1 \\
D&0.15&above&above&1\\
\enddata
\end{deluxetable}

\section{Detecting the Earth-Matter Effect at a Single Detector}
\label{sec:fourier}

According to Equation~(\ref{eq:pbar}) the modulations imprinted on the
detected spectra are proportional to $\sin^2(\#/\epsilon)$.
Therefore the prefactor $\#$ should appear as a dominant frequency
in the Fourier transform of the inverse energy spectrum (Dighe et
al.~2003b), because in inverse-energy space the modulations of the SN
neutrino spectrum are nearly equispaced.

The equidistant peaks in the modulation of the inverse-energy spectrum
are a necessary feature of the earth effects. Indeed,
the net $\bar\nue$ flux at the detector may be written using (\ref{eq:feDbar}) 
and (\ref{eq:pbar}) in the form
\be
F_{\ebar}^D =  \sin^2 \theta_{12}  F_{\bar\mu}^0 + 
\cos^2 \theta_{12} F_{\ebar}^0 + \Delta F^0
\bar{A}_\oplus \sin^2(\overline{\dmsq_\oplus} L/\epsilon)~~,
\label{eq:feDbar-y}
\ee
where $\Delta F^0 \equiv (F_{\ebar}^0 - F_{\bar\mu}^0)$ depends
only on the primary neutrino spectra, whereas
$\bar{A}_\oplus \equiv - \sin 2\bar{\theta}_{\re 2}^\oplus
~ \sin (2\bar{\theta}_{\re 2}^\oplus - 2\theta_{12})$ 
depends only on the mixing parameters and is independent of the
primary spectra. 
The last term in (\ref{eq:feDbar-y}) is the
earth oscillation term that contains a frequency $k_\oplus \equiv
2 \overline{\dmsq_\oplus} L$ in $\epsilon^{-1}$, with the coefficient $\Delta F^0
\bar{A}_\oplus$ being a comparatively slowly varying function of $\epsilon^{-1}$.
The first two terms in (\ref{eq:feDbar-y}) are also slowly varying
functions of $\epsilon^{-1}$, and hence contain frequencies in $\epsilon^{-1}$ that are much
smaller than $k_\oplus$.

The frequency $k_\oplus$ is completely independent of the primary
neutrino spectra, and indeed can be determined to a good accuracy from
the knowledge of the solar oscillation parameters, the earth matter 
density, and the direction of the SN.
If this frequency component is isolated from the inverse-energy
spectrum of $\bar\nue$, the earth effects would be identified. 

Due to the energy dependence of the prefactors of the oscillation term
$\Delta F^0 \bar{A}_\oplus$ and of $\dmsq_\oplus$ in
Equation~(\ref{eq:feDbar-y}) the peak in the Fourier-transformed
spectrum gets a certain width around $k_\oplus$.  This peak can be
clearly identified if the earth effect is present.

In an experiment there are two effects obscuring the signal.  One is
the statistical fluctuations of the signal and the other is the
smearing of the modulation signal by the energy resolution of the
detector.  Higher statistics as well as better energy resolution
enhance the oscillation signature.  A prescription for identifying the
prominent peak on top of the background can be found in Dighe et
al.~(2003b).  If the direction of the SN is known we can estimate
$k_\oplus$ and by that simplify the peak identification.

Once the peak is identified this is a  clear signal of earth-matter
effects.  As stated in Section~\ref{sec:SKandHK} an identification of
the earth-matter effect will se\-verely restrict the neutrino mixing
parameter space, since the effects are present only with the
combinations of the neutrino mass hierarchy and the mixing angle
$\theta_{13}$ given in Table~\ref{tab:EarthCases}. In particular, if
$\sin^2 \theta_{13}$ is measured at a laboratory experiment to be
greater than $10^{-3}$, then the earth effects on the $\bar\nue$
spectrum imply the normal mass hierarchy. However, if the earth
effects are not detected, it does not rule out any neutrino mixing
parameters, owing to the current uncertainties in the primary fluxes.

With the position of the peak it is even possible to determine
$\dmsq_\odot$ to an accuracy of a few percent and therefore much 
better than the current limits, that are $\msqs = (5.5$--$19) \times
10^{-5}$ eV$^2$.   The Fourier-transform method will measure
$\dmsq_\odot$ to a precision of 10\%, comparable to what can be
reached by KamLAND (de Gouvea \& Pe\~na-Garay, 2001).

\section{Summary}
\label{concl}

With our new findings for the SN neutrino spectra methods for
identifying oscillation signatures in an observed neutrino spectrum
need to be changed.  We presented two very promising concepts for
detecting the earth-matter effect.  For the IceCube detector in
Antarctica earth matter-effects are present in the signal of a future
galactic SN on the level of a few percent.  If the IceCube signal can
be compared with another high-statistics signal, notably in
Super-Kamiokande or Hyper-Kamiokande, the earth effect becomes clearly
visible as a difference between the detectors. As one is looking for a
signal modification in the range of a few percent, the absolute
detector calibration may not be good enough in one or both of the
instruments. However, for typical numerical SN simulations the effect
is time dependent and most notably differs between the early accretion
phase and the subsequent neutron star cooling phase. Therefore, one
would have to search for a temporal variation of the relative detector
signals of a few percent. The large number of optical modules in
IceCube renders this task statistically possible.  In fact depending
on the differences in flavor-dependent fluxes, the statistical
accuracy of Super-Kamiokande may turn out to be the limiting
factor. This limitation is not significant for Hyper-Kamiokande.

The unique location of IceCube in Antarctica implies that for about
70\% of the sky this detector sees the SN through the earth
when Super- and Hyper-Kamiokande see it from above, or the other way
round, i.e., the chances of a relative signal difference between the
detectors are large. If both detectors were to see the SN from above
or both through the earth, the comparison of the signals would would
not reveal the earth effect.  

Once future detectors like Hyper-Kamiokande or large scintillation
detectors become available a powerful way of identifying earth-matter
effects is through the Fourier transform of the inverse-energy
spectrum.  In the inverse-energy spectrum modulations due to earth
matter will be approximately equispaced and therefore appear as a
single peak in the Fourier transformed spectrum.  With this method one
can unambiguously prove the existence of earth-matter oscillations and
even measure $\msqs$ to a accuracy of a few percent.

Identifying the earth-matter effects in the neutrino signal of a
future galactic SN would restrict the parameter space for neutrino
oscillations severely.  If, in addition, the magnitude of the mixing
angle $\theta_{13}$ can be established to be large in the sense of
$\sin^2\theta_{13}\gsim10^{-3}$ by a long-baseline experiment (Barger 
et al. 2001, Cervera et al.~2000, Freund et al.~2001), it implies the
normal mass hierarchy.  

On the other hand, if $\sin^2\theta_{13}\lsim10^{-3}$ has been
established, the earth effect is unavoidable whatever the hierarchy
is.  Not observing it for such a small $\theta_{13}$ would imply that
the primary SN neutrino fluxes and spectra are more similar than
indicated by state-of-the-art numerical simulations.  For
$\sin^2\theta_{13}\gsim10^{-3}$ not observing the earth-matter effects
does not allow one to draw conclusions, because it can be due the
original SN spectra and fluxes, or due to the neutrino mass hierarchy.

\cleardoublepage


\chapter{Discussion and Summary}
\label{chapter:conclusions}

It has been known for some time that the ``traditional'' set
of $\numu$-matter interactions employed by hydrodynamic SN simulations
was incomplete.  In addition to the ingredients that were known to be
missing, i.e., nucleon bremsstrahlung and nucleon recoil, we showed
that $\nue\bar\nue$ annihilation into $\numu\bar\numu$ is always more
important than the traditional $\re^+\re^-$ annihilation process by a
factor of 2--3.

In a systematic approach we studied the formation of neutrino spectra
and fluxes in a SN core.  Using a Monte Carlo code for neutrino
transport, we varied the microscopic input physics as well as the
underlying static proto-neutron star atmosphere. We used two
background models from self-consistent hydrodynamic simulations, and
several power-law models with varying power-law indices for the
density and temperature and different constant values for the electron
fraction $\Ye$.

The $\nu_\mu$ transport opacity is dominated by neutral-current
scattering on nucleons.  In addition, there are number-changing
processes (nucleon bremsstrahlung, leptonic pair annihilation) and
energy-changing processes ($\nu_\mu\re^\pm$ and $\numu\nue$,
$\numu\bar\nue$ scattering).  Recoil in nucleon scattering allows for
a small energy exchange in each collision.  The $\nu_\mu$ spectra and
fluxes are roughly accounted for if one includes one significant
channel of pair production and one for energy exchange in addition to
$\nu_\mu{\rN}$ scattering.  For example, the traditional set of
microphysics (iso-energetic $\nu_\mu{\rN}$ scattering, $\re^+\re^-$
annihilation, and $\nu_\mu\re^\pm$ scattering) yields crudely comparable
spectra and fluxes to a calculation where pairs are produced by
nucleon bremsstrahlung and energy is exchanged by nucleon recoil. The
overall result is robust to within 30\% against the detailed choice of
microphysics.

However, in view of neutrino oscillations, where flavor-dependent flux
differences are important, state-of-the-art simulations should aim at
a high precision for the fluxes and spectral energies. Therefore, one
needs to include bremsstrahlung, leptonic pair annihilation,
neutrino-electron scattering, and energy transfer in neutrino-nucleon
collisions.  As expected, the traditional $\re^+\re^-$ annihilation
process is always much less important than $\nue\bar\nue$
annihilation. None of the reactions studied here can be neglected
except perhaps the traditional $\re^+\re^-$ annihilation process, and
$\nu_\mu\nue$ and $\nu_\mu\bar\nue$ scattering.
 
The existing treatments of the nuclear-physics aspects of the
$\rN\rN\to\rN\rN\nu\bar\nu$ bremsstrahlung process are rather
schematic. We find, however, that the $\nu_\mu$ fluxes and spectra do
not depend sensitively on the exact strength of the bremsstrahlung
rate.  Therefore, while a more adequate treatment of bremsstrahlung
remains desirable, the final results are unlikely to be much affected.

The transport of $\nu_\mu$ and $\bar\nu_\mu$ is usually treated
identically.  However, weak-magnetism effects render the $\nu_\mu\rN$
and $\bar\nu_\mu\rN$ scattering cross sections somewhat different
(Horowitz 2002), causing a small $\nu_\mu$ chemical potential to build
up.  We find that the differences between the average energies of
$\nu_\mu$ and $\bar\nu_\mu$ are only a few percent and can thus be
neglected for most purposes.

Including all processes works in the direction of making the fluxes
and spectra of $\nu_\mu$ more similar to those of $\bar\nue$ compared
to a calculation with the traditional set of input physics.  During
the accretion phase the neutron-star atmosphere is relatively
expanded, i.e., the density and temperature gradients are relatively
shallow. Our investigation suggests that during this phase
$\langle\epsilon_{\nu_\mu}\rangle$ is only slightly larger than
$\langle\epsilon_{\bar\nue}\rangle$, perhaps by a few percent or 10\%
at most. This result agrees with the first hydrodynamic simulation
including all of the relevant microphysics except $\nue\bar\nue$
annihilation (Accretion-Phase Model II) provided to us by M.~Rampp.
For the luminosities of the different neutrino species one finds
$L_{\bar\nue}\approx L_{\nue}\approx 2\,L_{\nu_\mu}$.  The smallness of
$L_{\nu_\mu}$ is not surprising because the effective radiating
surface is much smaller than for~$\bar\nue$.

During the Kelvin-Helmholtz cooling phase the neutron-star atmosphere
will be more compact, the density and temperature gradients will be
steeper. Therefore, the radiating surfaces for all species will become
more similar. In this situation $L_{\nu_\mu}$ may well become larger
than $L_{\bar\nue}$. However, the relative luminosities depend
sensitively on the electron concentration. Therefore, without a
self-consistent hydrostatic late-time model it is difficult to claim
this luminosity cross-over with confidence.

The ratio of the spectral energies is most sensitive to the
temperature gradient relative to the density gradient. In our
power-law models we used $\rho\propto r^{-p}$ and $T\propto
r^{-q}$. Varying $q/p$ between 0.25 and 0.35 we find that
$\langle\epsilon_{\bar\nue}\rangle:\langle\epsilon_{\nu_\mu}\rangle$
varies between $1:1.10$ and $1:1.22$.  Noting that the upper range for
$q/p$ seems unrealistically large we conclude that even at late times
the spectral differences should be small; 20\% sounds like a safe
upper limit.  However, the power-law models might overestimate the
spread of mean energies, as can be inferred from comparing the shallow
power-law model with the realistic accretion-phase models.  We stress
that there is no physical reason for the inequality
$\langle\epsilon_{\bar\nue}\rangle <
\langle\epsilon_{\nu_\mu}\rangle$.  It may well be that at some point
the mean energies are equal or even cross over.

The statements about the relative neutrino energies in the previous
literature fall into two classes. One group of workers, using the
traditional set of microphysics, found spectral differences between
$\bar\nue$ and $\nu_\mu$ on the 25\% level, a range which largely
agrees with our findings in view of the different microphysics.  Other
papers claim ratios as large as
$\langle\epsilon_{\bar\nue}\rangle:\langle\epsilon_{\nu_\mu}\rangle
=1:1.8$ or even exceeding $1:2$.  We have no explanation for these
latter results. At least within the framework of our simple power-law
models we do not understand which parameter could be reasonably
adjusted to reach such extreme spectral differences.

In a high-statistics neutrino observation of a future galactic SN one
may well be able to discover signatures of flavor oscillations.
However, when studying these questions one has to allow for the
possibility of very small spectral differences, and conversely, for
the possibility of large flux differences. This situation is almost
orthogonal to what has often been assumed in papers studying possible
oscillation signatures. A realistic assessment of the potential of a
future galactic SN to disentangle different neutrino mixing scenarios
should allow for the possibility of very small spectral differences
among the different flavors of anti-neutrinos. The spectral
differences between $\nu_e$ and $\nu_{\mu,\tau}$ are always much
larger, but a large SN neutrino (as opposed to anti-neutrino) detector
does not exist.  

Previous analyses on oscillation effects in the neutrino flux of a SN
not only assumed equipartition of the luminosities and a large hierarchy
in mean energies, but also heavily relied on the absolute magnitude of
model predictions.  Comparing such predictions with results from an
observation has little predictive power due to uncertainties in
the models, e.g., the nuclear equation of state.  Reliable methods
involve only relative flux differences and small relative differences
in energies.  With such rather model independent assumptions one can
unambiguously identify oscillation effects.

We proposed a method for identifying earth-matter effects by comparing
the  SN neutrino signal of two detectors.  One detector records the
signal from above, the other through the earth.  If the energy
deposited per unit volume in both detectors differs, the earth-matter
effect is 
detected.  Surprisingly, the future IceCube detector at the South Pole,
will be able to determine the deposited energy with a precision
comparable to the discussed Hyper-Kamiokande detector.  The unique
location has the advantage that with Super- or Hyper-Kamiokande as a
co-detector the likelihood for encountering the required setup
(exactly one line of sight through the earth) is 70 \%.  If the
earth-matter effect is detected we can infer the hierarchy of neutrino
masses to be normal if $\sin^2\theta_{13}\gsim 10^{-3}$ is
established by other experiments.  For $\sin^2\theta_{13}\lsim
10^{-3}$ the earth effect will always be present.  In any case, a
positive signal allows for strong conclusions whereas the reasons for
a not observing an effect can either be due to smaller flux
differences from the SN or oscillation parameters.

Our second proposed mechanism, namely identifying the earth effect at
a single detector would, together with the first method, enhance the
covered area of the sky to 85 \%.  A detector like the proposed
Hyper-Kamiokande or a future large scintillation detector will record
the neutrinos from a galactic SN with such a high accuracy that the
Fourier transform of the inverse-energy spectrum can reveal
modulations induced by the earth effect.  It would even be possible to
determine the solar mass splitting at the level of accuracy that will
be reached by KamLAND.

These methods have the advantage of being independent of the exact
original fluxes and spectra.  We just assume that a difference between
the $\bar\nue$ and $\bar\numu$ fluxes is present.  For the two-detector
setup we propose to use the time variation of the flux difference as
the observable in order to get rid of systematic uncertainties in the
absolute detector normalization. This time dependence is found in all
numerical simulations.  The physical reason is that neutrino emission
is first powered by the accretion of infalling material and later on
by proto-neutron star cooling.

With this first systematic study of all relevant neutrino-matter
interactions we were able to significantly improve the understanding
of flavor-dependent SN neutrino emission.  Our findings essentially
mark a change of paradigm for oscillation studies involving SN
neutrinos.  The mean energies of $\numu$ and $\bar\nue$ are almost
equal whereas the luminosities differ by up to a factor of 2,
contrary to the previous paradigm.  The new picture allows one to
detect neutrino-oscillation effects, for example with two distant
detectors by comparing the time variation of the deposited energy, or
in a single detector by identifying the modulations of the spectrum
with the help of a Fourier transform.  These concepts rely on robust
features of SN simulations instead of comparing predictions with
measurements, like previous methods.  The detection of a SN by itself
will be a very valuable astrophysics observation and can additionally
contribute to particle physics.

\cleardoublepage

\appendix

\chapter{Abbreviations}

\begin{tabular}{l@{\quad   --- \quad }l}
$\numu$ & unless specified differently, $\numu$ stands for $\numu$,
$\nu_\tau$, $\bar\numu$, and $\bar\nu_\tau$ \\
$\bar\numu$ & unless specified differently, $\bar\numu$ stands for
$\bar\numu$ and $\bar\nu_\tau$ \\
CM & center of momentum \\
LTE & local thermal equilibrium \\
$M_\odot$ & solar mass\\
mfp & mean free path \\
SN & supernova \\
$\Ye$ & electron fraction per baryon \\
$Y_{\rm L}$ & lepton fraction per baryon \\
\end{tabular}

\chapter{Monte Carlo Code}
\setcounter{equation}{0}
\label{app:MonteCarlo}

\section{General Concept}

Our Monte Carlo code is based on that developed by Janka (1987) where
a detailed description of the numerical aspects can be found.  The
code was first applied to calculations of neutrino transport in
SNe by Janka \& Hillebrandt (1989a,b) and Janka (1991).  It
uses Monte Carlo methods to follow the individual destinies of sample
neutrinos (particle ``packages'' with suitably attributed weights to
represent a number of real neutrinos) on their way through the star
from the moment of creation or inflow to their absorption or escape
through the inner or outer boundaries. The considered stellar
background is assumed to be spherically symmetric and static, and the
sample neutrinos are characterized by their weight factors and by
continuous values of energy, radial position, and direction of motion,
represented by the cosine of the angle relative to the radial
direction. The rates of neutrino interactions with particles of the
stellar medium can be evaluated by taking into account Fermion
blocking effects according to the local phase-space distributions of
neutrinos (Janka \& Hillebrandt 1989b).

As background stellar models we use the ones described in
Sec.~\ref{sec:PNSProfiles}. They are defined by radial
profiles of the density $\rho$, temperature $T$, and electron fraction
$\Ye$, i.e.\ the number of electrons per baryon.  The calculations
span the range between some inner radius $R_{\rm in}$ and outer radius
$R_{\rm out}$. These bound the computational domain which is divided
into 30 equally spaced radial zones.  In each zone $\rho$, $T$, and
$\Ye$ are taken to be constant.  $R_{\rm in}$ is chosen at such high
density and temperature that the neutrinos are in LTE in at least the
first radial zone. $R_{\rm out}$ is placed in a region where the
neutrinos essentially stream freely.  At $R_{\rm in}$ neutrinos are
injected isotropically according to LTE.  While a small net flux
across the inner boundary develops, the neutrinos emerging from the
star are generated almost exclusively within our computational
domain. If $R_{\rm in}$ is chosen so deep that the neutrinos are in
LTE, the assumed boundary condition for the flux will not affect the
results.

The stellar medium is assumed to be in thermodynamic equilibrium with
nuclei being completely disintegrated into free nucleons.  Based on
$\rho$, $T$, and $\Ye$ we calculate all the required thermodynamic
quantities, notably the number densities, chemical potentials, and
temperatures of protons, neutrons, electrons, positrons, and the
relevant neutrinos.  Except for runs that include weak magnetism,
the chemical potentials for $\nu_\mu$ and $\nu_\tau$ are taken to be
zero.  Next we compute the interaction rates in each radial zone for
all included processes.  In the simulations discussed in the present
work, fermion phase-space blocking is calculated from the neutrino
equilibrium distributions instead of the computed phase-space
distributions.  This simplification saves a lot of CPU time because
otherwise the rates have to be re-evaluated whenever the distribution
of neutrinos has changed after a transport time step.  The
approximation is justified because phase-space blocking is most
important in regions where neutrinos frequently interact and thus are
close to LTE.  Test runs without this approximation show that the
results are not affected within our numerical accuracy.

At the start of a Monte Carlo run, 800,000 test neutrinos are randomly
distributed in the model according to the local equilibrium
distributions.  Each test neutrino represents a certain number of real
neutrinos.  In this initial setup the number of real neutrinos is
determined by LTE.  Then transport is started.  The time step is fixed
at $\Delta t = 10^{-7}$ s; recall that the interaction rates do not
change.  At the beginning of each step neutrino creation takes place.
The number of test particles that can be created is given by the
number of neutrinos that were lost through the inner and outer
boundaries plus those absorbed by the medium.  Based on $\Delta t$,
the production rates, and the fact that the inner boundary radiates
neutrinos, we calculate the number of neutrinos that are produced in
one time step and distribute them among the available test neutrinos
by attributing suitable weight factors. The sample particles are
created within the medium or injected at the inner boundary in
appropriate proportions.

During a time step the path of each test particle through the stellar
atmosphere is followed by Monte Carlo sampling.  With random numbers
and stored tables of the total interaction rates we decide whether it
flies freely or interacts.  If it interacts it can scatter or it can
be absorbed; in the last case we turn to the next particle.  For
scattering we determine the new momentum and position and the time of
flight before the interaction took place and continue with the process
until the time step is used up.  Particles leaving through the lower
or upper boundaries are eliminated from the transport.

After a certain number of time steps (typically around 15,000) the
neutrino distribution reaches a stationary state and further changes
occur only due to statistical fluctuations. At that stage we start
averaging the output quantities over the next 500 time steps.

\section{Structure of the Code}

\paragraph{Thermodynamic Properties of the Medium}

For each radial zone the number of baryons is computed by the given
medium density.  Using the given temperature $T$ and $Y_\re$ together
with the number density of baryons $n_{\rm B}$ and the fact that
electrons and positrons are in LTE we obtain the chemical potential of
electrons and positrons by numerically solving
Equation~(\ref{eq:ElChemPo}).  With $\mu$ and $T$ the number densities
of electrons and positrons are calculated.  In the same way the
properties of protons and neutrons are determined.

\paragraph{Main Part: The Time-Step Loop}
The main part of the program consists of a loop that processes one
time step after the other.  Each timestep corresponds to $10^{-7}$s of
neutrino transport.  Since we use blocking factors of neutrinos in LTE
as an approximation to the real neutrino occupation, we only calculate
the interaction rates in the first time step and store them in arrays.
From these rates the cumulated probabilities are obtained.  In the
time-step loop specific rates are obtained from the arrays by linear
interpolation.

\paragraph{Injection and Creation of Neutrinos}
In each time step the number of sample neutrinos that were absorbed
or left the model through the inner or outer boundary during the
preceding timestep is now available for injection at the inner
boundary and emission events in the medium.  According to the
likelihood of each event these sample neutrinos are divided into one
group that is injected at the inner boundary and various others that
correspond to the creation mechanisms inside the medium.  With the
same likelihoods the number of real neutrinos that would be produced
in the time step is calculated and equally distributed among the
sample neutrinos as their weights.  Neutrinos injected at the lower
boundary get $R_{\rm in}$ as their radial position, a random energy
according to the LTE distribution in the first cell, and an angle
according to the distribution of neutrinos freely streaming off a
sphere.  The creation mechanisms inside the medium are the charged
current interactions $\re^- \rp \rightarrow \nue \rn$  and $\re^+ \rn
\rightarrow \bar\nue \rp$ for $\nue$ and $\bar\nue$, $\re^+ \re^-
\rightarrow \nu \bar\nu$ for all flavors, and  $\rN\rN \rightarrow
\rN\rN \numu \bar\numu$, $\nue \bar\nue \rightarrow \numu \bar\numu$
for $\numu$.  By Monte Carlo methods we obtain the radial position,
energy, and angle versus the radial direction.

\paragraph{Transport}
After all new neutrinos are created transport starts.  The number of
neutrinos to be transported equals the number of neutrinos that
remained in the star plus the injected and created neutrinos.  The
transport loop then takes one neutrino after the other through the
star until their time is used up or they leave the model.  Neutrinos
that remained inside the model from the preceding timestep have the
whole time $\Delta \rm t = 10^{-7} \rm s$.  Since injected and created
neutrinos appear at some random time their transport time step lasts
only $r \Delta \rm t$ where $r$ is a uniformly distributed random
number between 0 and 1.  Then the time until the first interaction
($T_{IA}$) is obtained with the help of the highest total interaction
rate for the energy of the transported neutrino.  If $T_{IA}$ lies
outside $\Delta \rm t$ we store the neutrino properties after it
freely traveled until the end of $\Delta \rm t$ and jump to the next
one. If the interaction takes place within the available time we have
to decide which interaction takes place.  The interaction rates are
also density dependent, i.e. for most locations inside the star we
underestimated $T_{IA}$ by using the upper limit of the total rate.
Therefore, when deciding which interaction takes place one of the
``interactions'' leaves the neutrino properties unchanged and 
in fact is no interaction at all.  This compensates for the too low
estimate $T_{IA}$ locally.

Before we go to the interaction we first calculate the new position of
the neutrino at the point of interaction.  If it leaves the model 
through the inner or outer boundary on its way to that point it is
eliminated and we jump to the next one.  In case it stays within the
area of interest we use the interaction rates at the interaction point
to decide which process takes place.  Then according to that process
the new energy and angle versus radial direction of the outgoing
neutrino are determined and stored.  If $\Delta \rm t$ is not used up
we determine a new $T_{IA}$ and go on until $\Delta \rm t$ is used
up.  

After all neutrinos used up their time or left the star we go over to
the next time step.

\cleardoublepage


\cleardoublepage

\end{document}